\documentclass[11pt,a4paper]{oxarticle}
\pdfoutput=1

\usepackage[usenames, 
Recompile
13
 dvipsnames]{xcolor}
\usepackage{graphicx, float, array, xspace, amscd, amsthm, amssymb, latexsym, longtable, bbm, fancyhdr,slashed,pifont}
\usepackage[letterpaper, body={6.5in, 9.25in}, top=1.0in, left=1.5in]{geometry}
\usepackage[bbgreekl]{mathbbol}
\usepackage[utf8]{inputenc}
\usepackage[sort&compress, comma, square, numbers]{natbib}
\usepackage[resetlabels]{multibib} 
\newcites{other}{Other Work Not Included Here}
\usepackage{braket}
\usepackage{multirow}
\usepackage{subfig}
\usepackage{bm}
\usepackage{hyperref}
\usepackage{tikz}
\usepackage{tikz-cd}
\usepackage{amsmath}
\usepackage{mathtools}
\usepackage{centernot}

\usepackage{xspace}

\usepackage{tabularx, caption, boldline}
\usepackage{graphicx}
\usepackage{cellspace}
\DeclareSymbolFontAlphabet{\mathbb}{AMSb}
\DeclareSymbolFontAlphabet{\mathbbl}{bbold}

\usepackage{booktabs}
\usepackage{pgfplotstable}
\usepackage{comment}
\usepackage{adjustbox}

\let\SS=\S 

\renewcommand{\S}{\Sigma}

\DeclareFontFamily{OT1}{pzc}{}
\DeclareFontShape{OT1}{pzc}{m}{it}{<-> s * [1.200] pzcmi7t}{}
\DeclareMathAlphabet{\mathpzc}{OT1}{pzc}{m}{it}

\newcommand{\cF}{\mathcal{F}}

\newcommand{\cL}{\mathcal{L}}
\newcommand{\cM}{\mathcal{M}}

\DeclareFontFamily{U}{bbold}{}
\DeclareFontShape{U}{bbold}{m}{n}
{  <-5.5> s*[1.05] bbold5
	<5.5-6.5> s*[1.05] bbold6
	<6.5-7.5> s*[1.05] bbold7
	<7.5-8.5> s*[1.05] bbold8
	<8.5-9.5> s*[1.05] bbold9
	<9.5-11.5> s*[1.05] bbold10
	<11.5-16> s*[1.05] bbold12
	<16-> s*[1.05] bbold17
}{}

\newcommand{\IC}{\mathbbl{C}}

\newcommand{\IN}{\mathbbl{N}}

\newcommand{\IP}{\mathbbl{P}}
\newcommand{\IQ}{\mathbbl{Q}}
\newcommand{\IR}{\mathbbl{R}}

\newcommand{\IW}{\mathbbl{W}}

\newcommand{\IZ}{\mathbbl{Z}}

\font\eightrm=cmr8 at 8pt
\font\csc=cmcsc10

\newcommand{\beq}{\begin{equation}}
\newcommand{\eeq}{\end{equation}}
\newcommand{\beqnn}{\begin{equation*}}
\newcommand{\eeqnn}{\end{equation*}}
\newcommand{\bea}{\begin{eqnarray}}
\newcommand{\eea}{\end{eqnarray}}
\newcommand{\bean}{\begin{eqnarray*}}
	\newcommand{\eean}{\end{eqnarray*}}

\newcommand{\tref}[1]{Table~\ref{#1}}
\newcommand{\sref}[1]{\SS\ref{#1}}

\newcommand{\ii}{\text{i}}

\newcommand{\place}[3]{\vbox to0pt{\kern-\parskip\kern-7pt
		\kern-#2truein\hbox{\kern#1truein #3}
		\vss}\nointerlineskip}

\DeclareFontFamily{U}{wncy}{}
\DeclareFontShape{U}{wncy}{m}{n}{<->wncyr10}{}
\DeclareSymbolFont{mcy}{U}{wncy}{m}{n}
\DeclareMathSymbol{\sha}{\mathord}{mcy}{"58}

\newcommand{\capt}[3]{\parbox{#1}{\renewcommand{\baselinestretch}{1.0}
		\caption{\label{#2}\small\it #3}}}

\newcommand{\1}{\mathbf{1}}

\newcommand{\cicy}[2]{\begin{matrix} #1\end{matrix}\!\left[\begin{matrix}#2 \end{matrix}\right]}

\newcommand{\+}{\phantom{-}}

\renewcommand{\=}{\;=\;}

\makeatletter
\g@addto@macro\bfseries{\boldmath}

\def\blindfootnote{\xdef\@thefnmark{}\@footnotetext}
\makeatother

\newcommand{\fourFthree}[8]{ {}_4F_3\left(\genfrac{}{}{0pt}{0}{#1,\,#2,\,#3,\,#4}{#5,~#6,~#7};~#8\right) }

\newcommand{\Qeq}{\;\stackrel{\mathclap{\normalfont\mbox{?}}}{=}\;}
\newcommand{\QQQeq}{\;\stackrel{\mathclap{\normalfont\mbox{???}}}{=}\;}

\usepackage{makecell}

\newcommand{\MR}{\text{\rotatebox[x=0mm, y=1.25mm]{180}{\reflectbox{\text{R}}}}}

\renewcommand{\baselinestretch}{1.1}
\numberwithin{equation}{section}
\begin{document}
%


\thispagestyle{empty}      
\begin{center}
\null\vskip0.1in
{\Huge New Examples of Abelian D4D2D0 Indices}\\[5pt]
\vskip1cm
{}
{\csc 
Joseph McGovern\\[15pt]
\small{mcgovernjv@gmail.com}\\[1cm]
}
%
%
{School of Mathematics and Statistics\\
University of Melbourne\\
Parkville, VIC 3010\\
Australia\\
Earth}
\vskip100pt
{\bf Abstract}
\end{center}
\vskip-7pt
\begin{minipage}{\textwidth}
\baselineskip=15pt
\noindent
\vskip10pt
We apply the methods of \cite{Alexandrov:2023zjb} to compute generating series of D4D2D0 indices with a single unit of D4 charge for several compact Calabi-Yau threefolds, assuming modularity of these indices. Our examples include a $\IZ_{7}$ quotient of R{\o}dland's pfaffian threefold, a $\IZ_{5}$ quotient of Hosono-Takagi's double quintic symmetroid threefold, the $\IZ_{3}$  quotient of the bicubic intersection in $\IP^{5}$, and the $\IZ_{5}$ quotient of the quintic hypersurface in $\IP^{4}$. For these examples we compute GV invariants to the highest genus that available boundary conditions make possible, and for the case of the quintic quotient alone this is sufficiently many GV invariants for us to make one nontrivial test of the modularity of these indices. As discovered in \cite {Alexandrov:2023zjb}, the assumption of modularity allows us to compute terms in the topological string genus expansion beyond those obtainable with previously understood boundary data. We also consider five multiparameter examples with $h^{1,1}>1$, for which only a single index needs to be computed for modularity to fix the rest. We propose a modification of the formula in \cite{Alexandrov:2022pgd} that incorporates torsion to solve these models. Our new examples are only tractable because they have sufficiently small triple intersection and second Chern numbers, which happens because all of our examples are suitable quotient manifolds. In an appendix we discuss some aspects of quotient threefolds and their Wall data.

\vspace*{10pt}
\end{minipage}
\clearpage


\thispagestyle{empty}  
{\baselineskip=17pt
	\tableofcontents} 
	
	\vfil
	\rightline{\it This is the mystery of the quotient, quotient}
\rightline{\it Upon us all, upon us all, a little rain must fall}
\vskip5pt
\rightline{Led Zeppelin, \textit{The Rain Song}}\thispagestyle{empty}

\clearpage


\setcounter{page}{1}
\renewcommand{\baselinestretch}{1.1}

\section{Introduction}\setcounter{page}{1}
Performing microstate counts for black holes in 4d $\mathcal{N}=2$ string vacua remains an open problem. These counts are related to bound states of D6, D4, D2, and D0 branes in IIA compactifications on a compact Calabi-Yau threefold $Y$, which are counted by generalised Donaldson-Thomas invariants \cite{Joyce:2008pc,Alexandrov:2022pgd}. 

Configurations with no D6 charge are counted by rank-0 generalised DT invariants, also known as D4D2D0 indices. They admit a lift to an M-theory compactification on $Y$ \cite{Maldacena:1997de}. In this M-theory setup the microstate counting problem can be approached by computing the modified elliptic genus \cite{Gaiotto:2006wm,Gaiotto:2007cd} of a 2d $\mathcal{N}=(0,4)$ SCFT, constructed by dimensionally reducing the 6d $\mathcal{N}=(0,2)$ theory on a divisor $\mathcal{D}_{\bm{p}}\subset Y$. The vector $\bm{p}$ gives the D4 charge, and also the homology class $[\mathcal{D}_{\bm{p}}]=p^{i}\epsilon_{i}\in H_{4}(Y,\IZ)$ with $\epsilon_{i}$ a basis of 4-cycles. There is an important distinction between the cases of irreducible and reducible divisors $\mathcal{D}_{\bm{p}}$. In the reducible case, there is a modular anomaly which complicates the analysis \cite{Alexandrov:2018lgp,Alexandrov:2016tnf,Alexandrov:2017qhn}. 

As explained in \cite{Alexandrov:2012au} the D4D2D0 indices, which depend on the K{\"a}hler parameters of $Y$ and vary discontinuously at walls of marginal stability, match with the indices computed by the elliptic genus of the (0,4) SCFT when the complexified K{\"a}hler parameters are sent to $\ii\infty$. In this limit, the values taken by the D4D2D0 indices are sometimes referred to as MSW indices.

Our aim in this paper is to compute new explicit Abelian D4D2D0 indices in this limit, with ``Abelian" specifying that we only consider an irreducible divisor. We are only able to do so by making use of the methods pioneered in \cite{Alexandrov:2023zjb}. One component of their technique is to exploit explicit formulae relating rank 0 generalised DT invariants to Pandharipande-Thomas invariants, due to \cite{Feyzbakhsh:2022ydn}. Next, they invoke the MNOP conjecture \cite{Maulik:2003rzb,Maulik:2004txy} which relates the PT and Gopakumar-Vafa invariants of $Y$. Gopakumar-Vafa invariants for the hypergeometric models studied in \cite{Alexandrov:2023zjb}, smooth complete intersections in weighted projective spaces, were available in light of the solution approach to topological string theory developed in \cite{Huang:2006hq}. As we will soon discuss, once sufficiently many MSW indices are known then all MSW indices for a given irreducible divisor are known. In short, the techniques of \cite{Alexandrov:2023zjb} provide every MSW index for a given irreducible $\mathcal{D}_{\bm{p}}$ once sufficiently many terms in the genus expansion of the topological string free energy are known.

We will more fully explain the following equations in \sref{sect:indices}, and for now we only display what we need to illustrate the basic motivation of this paper. For now it suffices to know that the MSW index is labelled by a positive integer $n$ and a subscript $\bm{\mu}$. The generating function $h_{\bm{p}}(\tau)$ for MSW invariants $\overline{\Omega}_{\bm{p},\bm{\mu}}(\Delta_{\bm{\mu}}-n)$ with D4 charge $\bm{p}$ is vector valued. Components of this vector have a label $\bm{\mu}$. Each component of this vector has a Fourier expansion given by
\begin{equation}\label{eq:MSWintro}
h_{\bm{p},\bm{\mu}}(\tau)\=\sum_{n\geq0} \overline{\Omega}_{\bm{p},\bm{\mu}}(\Delta_{\bm{\mu}}-n)\text{q}^{n-\Delta_{\bm{\mu}}}~,\qquad\text{q}\=\text{e}^{2\pi\ii\tau}~.
\end{equation}
The $\Delta_{\bm{\mu}}$ are certain rational numbers. In the case of an irreducible divisor $\mathcal{D}_{\bm{p}}$ it is conjectured that this generating function transforms as a vector-valued modular form (VVMF) of weight $-1-\frac{b_{2}}{2}$, where $b_{2}$ is the second Betti number of $Y$. Although mathematically open, this conjecture is physically very well justified by realising $h_{1}(\tau)$ in a decomposition of the modified elliptic genus of the MSW SCFT, whereby modular symmetry of the SCFT leads to the conclusion that $h_{\bm{p}}(\tau)$ is a VVMF with this weight \cite{Gaiotto:2006wm}.

This modularity is at the heart of the program for computing these indices, which began with \cite{Gaiotto:2006wm,Gaiotto:2007cd}. While the direct computation of each individual index may be formidable, the space of VVMFs with the correct transformation law is finite dimensional. Consequently one only needs to compute finitely many indices to know them all (that is, for a single unit of D4 charge). 

Note that the set of monomials in each Fourier expansion \eqref{eq:MSWintro} contains finitely many such that $\text{q}$ has a negative exponent, one for every pair $(\bm{\mu},n)$ such that $0<n<\Delta_{\bm{\mu}}$. These are referred to as the polar terms. This is important to consider because the dimensionality of the space of VVMFs with prescribed weight, multiplier system, and polar exponents $\Delta_{\bm{\mu}}$ is less than or equal to the number of these polar terms (the difference is typically small relative to the number of polar terms \cite{Manschot:2008zb}). This means that the more polar terms there are, the more indices one has to compute before modularity fixes the rest. One can see that there will be more polar terms for a given model, and so that model will be more difficult, if that model has\\[-25pt] 
\begin{itemize}
\item A large number of distinct $\bm{\mu}$~.\\[-25pt]
\item Large values of $\Delta_{\bm{\mu}}$~.\\[-25pt]
\end{itemize}
These numbers, which count the difficulty of a given model, can be computed in terms of data of the divisor $\mathcal{D}_{\bm{p}}$ and the threefold $Y$. Recall that amongst the topological data of $Y$ we have the triple intersection and second Chern numbers:
\begin{equation}
\kappa_{ijk}\=\int_{Y}e_{i}\wedge e_{j}\wedge e_{k}~,\qquad c_{2,i}\=\int_{Y}c_{2}(Y)\wedge e_{i}~,
\end{equation}
where $e_{i}$ is a generating set for $H^{2}(Y,\IZ)_{\text{Free}}$ and $c_{2}(Y)$ is the second Chern class of $Y$.

We will always choose our D4 charge vector $\bm{p}$ to be one of $(1),\,(1,0),\,(1,0,0)$ according to whether $b_{2}$ equals 1, 2, or 3. With this choice,\\[-10pt] 
\begin{equation}
\text{the number of distinct }\bm{\mu}\text{ equals }\text{Det}\left[\kappa_{1jk}\right]\quad\text{   and   }\quad\Delta_{\bm{0}}\=\frac{\kappa_{111}+c_{2,1}}{24}~.
\end{equation}
The remaining $\Delta_{\bm{\mu}}$ have different expressions, but are smaller than $\Delta_{\bm{0}}$. It suffices for now to just give $\Delta_{\bm{0}}$. This justifies the rough statement: models with small $\kappa_{ijk}$ and $c_{2,i}$ are easier.

To be more specific, a model is easy if it requires knowledge of the topological string free energy only to a low genus. By this metric, the easiest model studied in \cite{Alexandrov:2023zjb} would be the intersection of two sextic hypersurfaces in $\IW\IP^{5}_{1,1,2,2,3,3}$. This model has $\kappa_{111}=1~,\,c_{2,1}=22$. As a result, there is only one polar term. It turns out that the index $\overline{\Omega}_{\mathbf{p},\bm{0}}(\Delta_{\bm{0}})$, giving the coefficient of the most polar term, can be computed from $\kappa_{ijk}$ and $c_{2,i}$ without knowledge of any GV invariants. Therefore this model is solved almost immediately. We are led to wonder if there are other such simple models out there.

This is more than simply a matter of doing less work. The methods of \cite{Huang:2006hq} do not provide GV invariants to an arbitrarily high genus. One must impose certain boundary conditions at each genus to fix the holomorphic ambiguity, and in practice there are not enough boundary conditions past some model-specific genus. At the time of writing there exists no way to compute GV invariants to an arbitrarily high genus for a compact Calabi-Yau threefold. For each fixed model GV invariants are needed to a sufficiently high genus before the methods of \cite{Alexandrov:2023zjb} can be utilised, so it is worthwhile to find cases where this condition is met. We seek Calabi-Yau threefolds $Y$ such that their topological data is sufficiently small that the set of GV invariants of $Y$ which can be computed using available boundary data is sufficient to compute enough MSW indices for the rest to be fixed by modularity.

A first place to look might be the one parameter threefolds realised as complete intersections in Grassmannians, or the related Pfaffian threefold, for which topological string free energies were computed in \cite{Hosono:2007vf,Haghighat:2008ut}. Unfortunately, the smallest triple intersection number in this set is $\kappa_{111}=14$ for R{\o}dland's Pfaffian threefold \cite{rodland2000pfaffian}, which makes obtaining enough MSW indices unrealistic with available methods. Note that the hypergeometric model $\IP^{7}[2,2,3]$, with $\kappa_{111}=12$, was omitted from the analysis of \cite{Alexandrov:2023zjb} owing to limited knowledge of GV invariants.

There is another model in the literature known to only have a single polar term: the $\IZ_{5}$ quotient of the quintic in $\IP^{4}$, for which the modified elliptic genus was computed in \cite{Gaiotto:2006wm}. This inspires our approach. For a Calabi-Yau $Y$ obtained as the quotient of a simply connected threefold $\widetilde{Y}$ by a freely acting symmetry group $\IZ_{M}$, where $h^{1,1}(Y)=h^{1,1}(\widetilde{Y})$, one has
\begin{equation}\label{eq:WallDivision}
\kappa_{ijk}^{(Y)}\=\frac{1}{M}\kappa_{ijk}^{(\widetilde{Y})}~,\qquad c_{2,i}^{(Y)}\=\frac{1}{M}c_{2,i}^{(\widetilde{Y})}~.
\end{equation}
We explain this relationship between the topological data of $Y$ and $\widetilde{Y}$ in \sref{sect:QuotientData}. We go on in that appendix to discuss some interesting examples of quotient manifolds informed by that discussion.

There is then hope to find amenable examples in the set of quotient manifolds, whose topological data may be small because it is obtained by dividing the cover's data by the order of the quotient group\footnote{This is not the case for general quotients, as we discuss in \sref{sect:QuotientData}.}. After consulting the tables in \cite{Candelas:2016fdy}, we are led to the examples in \tref{tab:Manifolds} (where we use the CICY notation \cite{Candelas:1987kf}). We also consider the Double Quintic Symmetroid model studied in \cite{hosono2016double}, which admits a $\IZ_{5}$ quotient. 

This open problem of fixing the holomorphic ambiguity at higher genera can be addressed with the modularity of MSW invariants, as carried out to great effect in \cite{Alexandrov:2023zjb}. For every $h^{1,1}=1$ model in \tref{tab:Manifolds}, we compute as many topological string free energies as we can using the boundary data specified in \sref{sect:GVtables}, where we list GV invariants. For the $\IP^{4}[5]/\IZ_{5}$ and $\IP^{5}[3,3]/\IZ_{3}$ models, we are able to compute GV invariants beyond the maximal genus those boundary conditions enable, by incorporating data from the MSW generating function and the explicit formulae relating the MSW invariants to GV invariants. We are not the first to compute GV invariants for $\IP^{4}[5]/\IZ_{5}$ and $\IP^{5}[3,3]/\IZ_{3}$, which are mentioned on footnote 12 of \cite{Huang:2006hq} and reference 14 of \cite{Gopakumar:1997dv}\footnote{The $\IP^{4}[5]/\IZ_{5}$ GV invariants were also computed independently by Emanuel Scheidegger in the course of helpful conversations in the BICMR.}. For these models, our computations provide a small number of independent checks on the modularity of MSW invariants.

Regrettably, we are unable to use the Abelian MSW indices to drive the genus expansions for the (1,8) and (1,6) models higher than the genus 5 results we give in \sref{sect:GVtables}. We do compute a handful of invariants at genera beyond 5, but these are insufficient to fix the topological string free energies at these genera. 

The simply connected covers of these models have derived equivalent partner manifolds, respectively the intersection of seven degree 1 hypersurfaces in $Gr(2,7)$ \cite{rodland2000pfaffian,Hori:2006dk,borisov2009pfaffian,Addington:2014sla} and the Reye congruence \cite{Hosono:2011np,hosono2016double}. These respectively have $\IZ_{7}$ and $\IZ_{5}$ quotients, which we assume to be derived equivalent partners of the (1,8) and (1,6) models. It is only by making this assumption that we are able compute the invariants listed in \sref{sect:GVtables}.

\hfill

\begin{minipage}{.45\linewidth}
\begin{table}[H]
\begin{center}
\begin{tabular}{| c || l | c | c |}
\hline ($\phantom{\bigg|}h^{1,1},h^{2,1}\phantom{\bigg|}$) &\hfil Geometry  & Triple Intersection Numbers & \hfil$c_{2}$    \\\hline 
(1,21) & $\IP^{4}[5]/\IZ_{5}\phantom{\bigg|}$ & $\kappa_{111}=1$ & $(10)$  \\[25pt]
(1,25) & $\IP^{5}[3,3]/\IZ_{3}$ & $\kappa_{111}=3$ & $(18)$  \\[25pt]
(1,8) & (Pfaffian in $\IP^{6}$)$/\IZ_{7}$ & $\kappa_{111}=2$ & $(8)$  \\[25pt]
(1,6) & $\binom{\text{Smooth double cover of }}{\text{determinantal quintic in }\IP^{4}}/\IZ_{5}$ & $\kappa_{111}=2$ & $(8)$  \\[25pt]
(2,29) & $\cicy{\IP^{2}\\\IP^{2}}{3\\3}/\IZ_{3}$ & \makecell{ $\kappa_{111}=\kappa_{222}=0$\\$\kappa_{112}=\kappa_{122}=1$} & $\begin{pmatrix}12\\12\end{pmatrix}$ \\[25pt]
(2,20) & $\cicy{\IP^{2}\\\IP^{5}}{1&1&1&0\\1&1&1&3}/\IZ_{3}$ &\makecell{ $\kappa_{111}=0\;\;\;\;\kappa_{112}=1$\\$\kappa_{122}=3\;\;\;\;\kappa_{222}=3$} & $\begin{pmatrix}12\\18\end{pmatrix}$ \\[25pt]
(2,12) & $\cicy{\IP^{4}\\\IP^{4}}{1&1&1&1&1\\1&1&1&1&1}/\IZ_{5}$ & \makecell{ $\kappa_{111}=\kappa_{222}=1$\\$\kappa_{112}=\kappa_{122}=2$} & $\begin{pmatrix}10\\10\end{pmatrix}$ \\[25pt]
(3,18) & $\cicy{\IP^{2}\\\IP^{2}\\\IP^{2}}{1&1&1\\1&1&1\\1&1&1}/\IZ_{3}$ & \makecell{ $\kappa_{111}=\kappa_{222}=\kappa_{333}=0$\\$\kappa_{112}=\kappa_{113}=\kappa_{223}=1$\\$\kappa_{122}=\kappa_{133}=\kappa_{233}=1$\\$\kappa_{123}=2$} & $\begin{pmatrix}12\\12\\12\end{pmatrix}$ \\[35pt]
(3,15) & $\cicy{\IP^{2}\\\IP^{2}\\\IP^{5}}{1&1&1&0&0&0\\0&0&0&1&1&1\\1&1&1&1&1&1}/\IZ_{3}$ & \makecell{$\kappa_{111}=\kappa_{222}=0$\\$\kappa_{112}=\kappa_{122}=\kappa_{113}=\kappa_{223}=1$\\$\kappa_{123}=\kappa_{133}=\kappa_{233}=\kappa_{333}=3$} & $\begin{pmatrix}12\\12\\18\end{pmatrix}$ \\[25pt]
\hline   
\end{tabular} 
\end{center}
\end{table}
\end{minipage}
\vskip-5pt
\begin{table}[H]
\begin{center}
\capt{\textwidth}{tab:Manifolds}{Quotient geometries studied in this paper.}
\end{center}
\end{table}
\vskip-40pt\underline{Outline of paper}

In \sref{sect:Quotients} we review aspects of quotient Calabi-Yaus  that inform us, and in \sref{sect:D4D2D0} we review aspects of \cite{Alexandrov:2023zjb,Feyzbakhsh:2022ydn} which we avail of. These sections contain no new results, are included only in the interest of self-containment, and can be safely skipped by experts. Section \sref{sect:1pexamples} includes some discussion of the specific quotient geometries we study, and their mirrors. Readers solely interested in seeing the Abelian D4D2D0 generating functions can skip to sections \sref{sect:Quinticresults}, \sref{sect:Bicubicresults}, \sref{sect:Rodlandresults}, \sref{sect:HTresults}, and \sref{sect:Multiparameter}.

GV invariants are given in \sref{sect:GVtables}, and to higher degree in an ancillary Mathematica notebook. In \sref{sect:QuotientData} we discuss the Wall data of quotient manifolds. We apply our discussion to address some questions posed in \cite{Hori:2016txh,Gendler:2023ujl} that concern the use of Wall data to distinguish diffeomorphism classes of non-simply connected CY threefolds.

\newpage
\section{Quotients of CY threefolds by freely acting groups}\label{sect:Quotients}
Throughout this paper, $Y$ will be the Calabi-Yau threefold obtained by taking the quotient of another threefold $\widetilde{Y}$ by a freely acting symmetry group $\IZ_{M}$. We will always have a $\IZ_{M}$ action such that $h^{1,1}(Y)=h^{1,1}(\widetilde{Y})$, equivalently $b_{2}(Y)=b_{2}(\widetilde{Y})$. Here we review some details particular to such threefolds which informs our computations.
\subsection{Torsion in the second cohomology}
Many relevant aspects of the homology and cohomology groups of quotient manifolds are discussed in \cite{Aspinwall:1994uj,Batyrev:2005jc,Braun:2007xh}. As we explain in Appendix \sref{sect:QuotientData}, the second integral homology $H_{2}(Y,\IZ)$ is torsion-free. By Poincar{\'e} duality, whereby 
\begin{equation}\label{eq:PD}
H^{k}(Y,\IZ)\cong H_{6-k}(Y,\IZ)~,
\end{equation}
the fourth integral cohomology $H^{4}(Y,\IZ)\cong H_{2}(Y,\IZ)$ will also be torsion free.

Importantly however, for the second cohomology we have
\begin{equation}\label{eq:TorsionH2}
H^{2}(Y,\IZ)\cong\IZ^{b_{2}}\oplus\IZ_{M}
\end{equation}
which has a $\IZ_{M}$ torsion factor. The Poincar{\'e} dual statement is $H_{4}(Y,\IZ)\cong H^{2}(Y,\IZ)$, there is $\IZ_{M}$ torsion in the fourth homology.

The fundamental group $\pi_{1}(Y)$ is $\IZ_{M}$, because $Y$ is the quotient of the simply connected $\widetilde{Y}$ by $\IZ_{M}$. In general $H_{1}(Y,\IZ)$ is the abelianisation of $\pi_{1}(Y)$, so in our case $H_{1}(Y,\IZ)\cong\IZ_{M}$.

As explained in \cite{Batyrev:2005jc}, \eqref{eq:TorsionH2} is guaranteed by the universal coefficient theorem 
\begin{equation}\label{eq:UCT}
\text{Tors}(H_{i}(Y,\IZ))\cong\text{Hom}\left(\text{Tors}(H^{i+1}(Y,\IZ)),\IQ/\IZ\right)~.
\end{equation}
Using \eqref{eq:PD} and \eqref{eq:UCT} one can then demonstrate \eqref{eq:TorsionH2} via
\begin{equation}\label{eq:TorsionZm}\begin{aligned}
\text{Tors}\left(H^{2}(Y,\IZ)\right)&\cong\text{Tors}\left(H_{4}(Y,\IZ)\right)\\
&\cong\text{Hom}\left(\text{Tors}\left(H^{5}(Y,\IZ)\right),\IQ/\IZ\right)\\
&\cong\text{Hom}\left(\text{Tors}\left(H_{1}(Y,\IZ)\right),\IQ/\IZ\right)\\
&\cong\text{Hom}\left(\IZ_{M},\IQ/\IZ\right)\cong\IZ_{M}~.
\end{aligned}
\end{equation}

\subsection{Picard-Fuchs equations and mirror quotients}\label{sect:PFandMirrors}
Computing topological string free energies requires some knowledge of the Calabi-Yau threefold $X$ that is mirror to $Y$. We will denote by $\widetilde{X}$ the mirror threefold of the cover $\widetilde{Y}$. When $Y$ is the (1,21) model $\IP^{4}[5]/\IZ_{5}$, it has long been known that $X=\widetilde{X}/\IZ_{5}$ \cite{Greene:1990ud}. In every $h^{1,1}=1$ example listed in \tref{tab:Manifolds} we can identify a freely acting $\IZ_{M}$ symmetry of the appropriate $\widetilde{X}$ appearing in the literature. 

We now argue that, for the quotients we consider, the Picard-Fuchs equations for both mirrors of $Y$ and $\widetilde{Y}$ must be the same. Note that in the one parameter case ($h^{1,1}(Y)=h^{1,1}(\widetilde{Y})=1$) the genus 0 invariants $n^{(0)}_{k}$ are generated by the Yukawa coupling \cite{Candelas:1990rm} through the Lambert series
\begin{equation}\label{eq:Cttt}
C_{ttt}(q)\=\kappa_{111}+\sum_{k=1}^{\infty}n^{(0)}_{k}\,\frac{q^{k}}{1-q^{k}}~,\qquad q\=\text{e}^{2\pi\ii t}~.
\end{equation}
As explained in \cite{Duco}, the function $C_{ttt}(q)$ puts the Picard-Fuchs operator into a canonical form:
\begin{equation}\label{eq:PFQ}
\mathcal{L}^{PF}\=\theta_{q}^{2}\frac{1}{C_{ttt}(q)}\theta_{q}^{2}~,\qquad \theta_{q}=q\frac{\text{d}}{\text{d}q}~.
\end{equation}
As we explain in Appendix \sref{sect:QuotientData}, the triple intersection numbers share the relation $\kappa_{111}^{(Y)}=\frac{1}{M}\kappa_{111}^{(\widetilde{Y})}$. If we have similarly that 
\begin{equation}\label{eq:requiredrelation}
n^{(0),Y}_{k}\=\frac{1}{M}n^{(0),\widetilde{Y}}_{k},
\end{equation}
then the Yukawa couplings \eqref{eq:Cttt} for $\widetilde{Y}$ and $Y$ will be equal up to an overall multiple of $M$, hence the Picard-Fuchs operators \label{eq:PFQ} for the two mirror geometries will be the same. 

The required relation \eqref{eq:requiredrelation} is certainly true in cases where $n^{(0),Y}_{k}$ equals the count of rational curves of degree $k$ on $Y$, when all such curves are smooth. This follows because such a curve has no unramified covers, and by assumption $\mathbb{Z}_{M}$ acts freely on $\widetilde{Y}$, hence the preimage of each such curve is $M$ distinct copies of the curve in $\widetilde{Y}$, implying\footnote{The same degree $k$ appears on both sides of \eqref{eq:requiredrelation}, this follows from our discussion in \sref{sect:QuotientData} which closely follows the arguments of \cite{Aspinwall:1994uj} (where a similar argument for such a relation \eqref{eq:requiredrelation} appears). For more general quotient groups than $\mathbb{Z}_{M}$, the degrees of the invariants on the LHS and RHS of \eqref{eq:requiredrelation} can differ (as in the example of \cite{Aspinwall:1994uj}).} \eqref{eq:requiredrelation}. 

The relation between $n_{k}^{(0)}$ and curve counts is not so simple, as explained in \cite[\SS2.1]{Katz:1999xq}. To be sure that the required relation \eqref{eq:requiredrelation} does hold, we appeal to the GLSM results of \cite{Jockers:2012dk}. There, it is shown that the genus-0 invariants of $n^{(0),\widetilde{Y}}_{k}$ can be read off of the sphere partition function $Z_{S^{2}}^{\widetilde{Y}}(\,\widetilde{t}\,,\,\overline{\widetilde{t}}\,)$ of a GLSM with some gauge group $H$, whose large volume phase flows to a sigma model on $\widetilde{Y}$. To perform the same computation for $n^{(0),Y}_{k}$, one should repeat this computation but replace the gauge group $H$ with $H\times \mathbb{Z}_{M}$. The computation of the sphere partition function goes through almost exactly the same for both models, however one must divide by the order of the Weyl group of the gauge group, which differs in each model by a factor of $M$. This leads to
\begin{equation}\label{eq:ZS2rel}
Z_{S^{2}}^{Y}(\,t\,,\,\overline{t}\,)=\frac{1}{M}Z_{S^{2}}^{\widetilde{Y}}(\,\widetilde{t}\,,\,\overline{\widetilde{t}}\,).
\end{equation}
Following through with the prescription for computing genus-0 invariants outlined in \cite{Jockers:2012dk}, and noting that $t=\widetilde{t}$ for our quotients (as discussed in \sref{sect:QuotientData}), one has that \eqref{eq:ZS2rel} implies \eqref{eq:requiredrelation}. This argument also provides the relation \eqref{eq:WallDivision} that we discuss further in \sref{sect:QuotientData}, since the triple intersection and second-Chern numbers can similarly be read off from $Z_{S^{2}}$. This argument also gives an equality between the Frobenius-basis periods of $\widetilde{Y}$ and $Y$, which gives another demonstration that the mirrors of $Y$ and $\widetilde{Y}$ have the same Picard-Fuchs operators.


From transposing the Hodge diamond, the Euler characteristic of $X$ will be minus the Euler characteristic of $Y$. This means that $\chi(X)$ must equal $\chi(\widetilde{X})/M$, since $\chi(Y)=\chi(\widetilde{Y})/M$. Since $\chi(\widetilde{X})/M$ is the Euler characteristic of $\widetilde{X}/\IZ_{M}$, which also has the same PF equation (because the fundamental period of $\widetilde{X}$ and $X$ are identical), we proceed on the assumption that the mirror of $Y$ is $X=\widetilde{X}/\IZ_{M}$ when we compute GV invariants. Treating this problem more rigorously for the $(1,6)$ model could require a more involved understanding of issues discussed in \sref{sect:HTgeometry}. Treating the mirror of the (1,8) geometry  more carefully might proceed along the lines of \cite{Batyrev:1998kx}.

\subsection{The genus 1 topological string free energy}
The reason that we needed to identify a mirror geometry (rather than merely a mirror PF operator) in \sref{sect:PFandMirrors} is that the nature of the singularities that $X$ can acquire for certain values of the complex strcuture moduli provide important boundary data for higher genus computations.

The BCOV result for the genus 1 B-model topological string free energy is \cite{Bershadsky:1993cx,Bershadsky:1993ta} 
\begin{equation}\label{eq:Bgenus1}
\cF^{(1)}(\varphi)\=-\frac{1}{2}\left(3+h^{1,1}(Y)-\frac{\chi(Y)}{12}\right)\log\left( \varpi_{0}\right)-\frac{1}{2}\log\left(\frac{\text{d}t}{\text{d}\varphi}\right)-\frac{c_{2}+12}{24}\log(\varphi)-\frac{1}{12}\sum_{i}|G_{i}|\log(\Delta_{i})~.
\end{equation}
$X$ will become singular for certain moduli $\varphi$ that solve a polynomial equation $\Delta(\varphi)=0$. The polynomial $\Delta$ is the discriminant of $X$. Let $\Delta$ be a product of factors $\Delta_{i}$ that are irreducible over $\IQ$. For $\varphi$ a root of $\Delta_{i}$, $X$ will become singular because a Lens space of the form $S^{3}/G_{i}$ shrinks, with $G_{i}$ being a group specific to the factor $\Delta_{i}$. The appearance of the $|G_{i}|$ in \eqref{eq:Bgenus1} was discovered in \cite{Gopakumar:1997dv}, and explained by the fact that as the $S^{3}/G_{i}$ shrinks a number $|G_{i}|$ of hypermultiplets become massless. 

Although it will not happen for the examples presently under study, we mention that it was observed in \cite{Candelas:2019llw} that at a root of some $\Delta_{i}$ multiple distinct Lens spaces may shrink. The proposal of \cite{Candelas:2019llw} was that the number of such collapsing Lens spaces should appear in the genus one free energy, multiplying the coefficients of $\log(\Delta_{i})$ in \eqref{eq:Bgenus1}. This is consistent with the coefficient of $\log(\Delta_{i})$ being minus the number of massless hypermultiplets divided by 12, as per \cite{Gopakumar:1997dv}.

\newpage
\section{GV-PT-D4D2D0 relations}\label{sect:D4D2D0}

\subsection{The GV and BCOV formulae}
The Gromov-Witten invariants $N^{(g)}_{\beta}$ of the Calabi-Yau threefold $Y$ are computed by the all-genus A-model topological string free energy \cite{Hori:2003ic}. The Gopakumar-Vafa invariants $n^{(g)}_{\beta}$ are then obtained by a multicover formula \cite{Gopakumar:1998ii,Gopakumar:1998jq}, with the integrality explained in physical terms as giving counts of BPS particles in an M-theory compactification on $Y$ (so necessarily integral). A mathematical proof of integrality of the $n^{(g)}_{\beta}$, as defined by this multicovering relation in terms of Gromov-Witten invariants, was given in \cite{Ionel:2018ebm}. 

The Gopakumar-Vafa  formula reads
\begin{equation}\label{eq:GVformula}
\begin{aligned}
F^{\text{All Genus}}(\mathbf{t},\lambda)&\=\sum_{g=0}^{\infty}\lambda^{2g-2}F^{(g)}(\mathbf{t})\\[5pt]
&\=\lambda^{-2}c(\mathbf{t})+l(\mathbf{t})+\sum_{g=0}^{\infty}\lambda^{2g-2}\sum_{\beta\in H_{2}(Y,\IZ)}N^{(g)}_{\beta}\text{q}^{\beta\cdot\mathbf{t}}\\[5pt]
&\=\lambda^{-2}c(\mathbf{t})+l(\mathbf{t})+\sum_{g=0}^{\infty}\sum_{\beta\in H_{2}(Y,\IZ)}\sum_{m=1}^{\infty}n^{(g)}_{\beta}\frac{1}{m}\left(2\sin \frac{m\lambda}{2}\right)^{2g-2}\text{q}^{m \beta}~.
\end{aligned}\end{equation}
$c(\mathbf{t})$ and $l(\mathbf{t})$ are the cubic and linear polynomials in $\mathbf{t}$ that respectively appear in the genus 0 and genus 1 free energies. For a homology class $\beta$, $\text{q}^{m\beta}$ denotes $\exp\left(2\pi\ii m\sum_{i=1}^{h^{1,1}}\beta_{i} t^{i}\right)$. The genus 0 $\lambda^{-2}$ term, and the associated multicovering formula, were obtained in \cite{Candelas:1990rm}. A refined version of this formula incorporating discrete M-theory charges (which do not arise in this paper) has been provided in \cite{Schimannek:2021pau,Katz:2022lyl}.

In order to obtain the integers $n^{(g)}_{\beta}$, one uses mirror symmetry. This has been carried out to great effect in \cite{Huang:2006hq,Hosono:2007vf,Haghighat:2008ut}, and we do not offer any new insights here not already contained in those papers. The B-model and A-model free energies $\cF^{(g)}$ and $F^{(g)}(\mathbf{t})$ at a fixed genus are related by
\begin{equation}\label{eq:BtoA_FreeEnergy}
F^{(g)}(\mathbf{t})=\varpi_{0}(\varphi)^{2g-2}\mathcal{F}^{(g)}(\varphi)\bigg\vert_{\varphi=\varphi(\mathbf{t})}.
\end{equation}
$\varpi_{0}$ is the holomorphic series solution of the Picard-Fuchs system of $X$ in the solution expansion about a point of Maximal Unipotent Monodromy (MUM), with leading term 1 (more generally this will be $\varphi^{\alpha}$, where $\alpha$ is the distinct root of the indicial equation at the MUM point). 

In the vicinity of a MUM point, a basis of $b_{3}(X)$ solutions can be found such that one has leading term 1, $h^{1,1}(Y)$ have logarithmic leading behaviour, $h^{1,1}(Y)$ have log-squared behaviour, and a final solution has log-cubed asymptotics. The mirror map, which relates the complexified K\"ahler parameters of $Y$ to the complex structure parameters $\varphi$ of $X$, reads
\begin{equation}
t^{i}=\frac{1}{2\pi\ii}\frac{\varpi_{1,i}(\varphi)}{\varpi_{0}(\varphi)}~.
\end{equation}
$\varpi_{1,i}(\varphi)$ is the single-log solution to the Picard-Fuchs system, with leading behaviour ${\varpi_{0}\log(\varphi_{i})+O(\varphi)}$ at the MUM point.

There is a unique (up to scale) holomorphic $(3,0)$ form $\Omega$ on $X$. The complex structure moduli space $\cM$ of $X$ has a K{\"a}hler structure, with K{\"a}hler potential $K$ given by
\begin{equation}
\text{e}^{-K}\=\ii\int_{X}\Omega\wedge\overline{\Omega}~.
\end{equation}
This provides a K{\"a}hler connection $\partial_{\varphi^{i}}K$ on a line bundle $\cL$ over $\cM$. One also has a Levi-Civita metric connection $\Gamma^{i}_{jk}$ from the metric $G_{i\bar{j}}=\partial_{i}\partial_{\bar{j}}K$.

The Yukawa coupling is a symmetric 3-tensor on $\cM$ valued in $\cL^{2}$, given by
\begin{equation}
C_{ijk}\=\int_{X}\Omega\wedge\partial_{\varphi^{i}}\partial_{\varphi^{j}}\partial_{\varphi^{k}}\Omega~,
\end{equation}
which obeys $D_{i}C_{jkl}=D_{j}C_{ikl}$, $\partial_{\bar{i}}C_{jkl}=0$.

The $S_{4}$ permutation symmetry on the indices of $D_{i}C_{jkl}$ allows one to express $C_{ijk}$ using a prepotential, the genus 0 B-model free energy $\mathcal{F}^{(0)}$, which is a section of $\cL^{2}$. One has $C_{ijk}=D_{i}D_{j}D_{k}\cF^{(0)}$. 

While the genus 0 result is holomorphic, only depending on $\varphi$ and not $\overline{\varphi}$, the holomorphic anomaly of Bershdasky-Cecotti-Ooguri-Vafa \cite{Bershadsky:1993cx,Bershadsky:1993ta} leads to a $\overline{\varphi}$ dependence in the higher genus $\mathcal{F}^{(g)}(\varphi,\overline{\varphi})$, which are sections of $\cL^{2-2g}$. This anholomorphic dependence is not seen in \eqref{eq:BtoA_FreeEnergy} because one takes the topological limit (introduced in \cite{Bershadsky:1993ta}, see also the discussion in \cite{Haghighat:2008ut}):
\begin{equation}
\mathcal{F}^{(g)}(\varphi)=\mathcal{F}^{(g)}(\varphi,\overline{\varphi})\bigg\vert_{\overline{\varphi}\mapsto\infty}~.
\end{equation}
The genus 1 holomorphic anomaly equation is given in \cite{Bershadsky:1993cx,Bershadsky:1993ta}, 
\begin{equation}
\partial_{\bar{i}}\partial_{j}\cF^{(1)}=\frac{1}{2}\overline{C}^{nm}_{\bar{i}}C_{jmn}+\left(1-\chi(Y)/24\right)G_{\bar{i}j}~.
\end{equation} 
The solution is
\begin{equation}
\mathcal{F}^{(1)}(\varphi,\overline{\varphi})\=\frac{1}{2}\left(3+h^{1,1}(Y)-\frac{\chi(Y)}{12}\right)K-\frac{1}{2}\log\left(\text{det}(G)\right)-\log\left|\prod_{c}\Delta_{c}^{\frac{|G_{c}|}{12}}\,\prod_{i=1}^{h^{1,1}(Y)}(\varphi^{i})^{\frac{1}{2}+\frac{c_{2,i}}{24}}\right|^{2}~.
\end{equation}
The third term above is the genus 1 ambiguity, fixed by imposing boundary data at the hyperconifold and MUM point singularities. In the topological limit, in the one-parameter case, this becomes \eqref{eq:Bgenus1}.

The Holomorphic Anomaly Equations (HAE) express, for $g\geq2$, the anholomorphic derivatives of the free energy $\mathcal{F}^{(g)}$ in terms of free energies $\mathcal{F}^{(h)}$ at lower genera, $h<g$:
\begin{equation}\label{eq:HAE}
\frac{\partial}{\partial \overline{\varphi^{k}}}\,\mathcal{F}^{(g)}(\varphi,\overline{\varphi})\=\frac{1}{2}\overline{C_{kij}}G^{i\overline{i}}G^{j\overline{j}}\text{e}^{2K(\varphi,\overline{\varphi})}\left[D_{i}D_{j}\mathcal{F}^{(g-1)}(\varphi,\overline{\varphi})+\sum_{r=1}^{g-1}\left(D_{i}\mathcal{F}^{(r)}(\varphi,\overline{\varphi})\right)\left(D_{j}\mathcal{F}^{(g-r)}(\varphi,\overline{\varphi})\right)\right]~.
\end{equation}

If the $\mathcal{F}^{(h<g)}$ are known, then the above equation manifestly determines $\mathcal{F}^{(g)}$ up to the addition of a holomorphic function $f^{(g)}(\varphi)$. 

The most computationally practical way to proceed is to use the polynomial structure of these free energies, as explained in \cite{Yamaguchi:2004bt,Alim:2007qj,Alim:2012gq}. One computes at each genus a polynomial $P^{(g)}$ of bounded degree in certain propagator functions $S^{ij},S^{i},S$ (see for instance \cite{Alim:2007qj}), and the free energy is the sum of this polynomial and the holomorphic ambiguity: $\mathcal{F}^{(g)}=P^{(g)}+f^{(g)}$. The recursion relation above becomes a recursive set of PDEs that define this polynomial $P^{(g)}$ of the propagators.

The holomorphic ambiguity is a rational function (as is necessary for meromorphicity on the moduli space), and by considering the possible singularities of $\cF^{(g)}$ it becomes clear that $f^{(g)}$'s denominator can be completely fixed, and the growth at infinity bounded. This means that at each genus there is a finite set of numbers to fix in order to completely determine $f^{(g)}$, the coefficients of the numerator polynomial of this rational function.

\subsection{Known constraints on the holomorphic ambiguity}
One set of results we wish to stress in this paper, which exactly follows the analysis for simply connected hypergeometric models in \cite{Alexandrov:2023zjb}, is that conjectural modularity of Maldacena-Strominger-Witten invariants provides new constraints on the holomorphic ambiguity $f^{(g)}$. This allows for explicit topological string expansions to a higher genus than previously possible. In order for these results to be evident, we here discuss previously established constraints on the holomorphic ambiguity that we have used to compute the tables in Appendix \sref{sect:GVtables}. We discuss these without including the prospect of incorporating the data obtained by directly obtaining some curve counts using relative Hilbert schemes as was done in \cite{Huang:2006hq} for complete intersections in weighted projective spaces. 

For one-parameter hypergeometric models, the determination of $f^{(g)}(\varphi)$ (and so $\mathcal{F}^{(g)}$ and $F^{(g)}$) was carried out to a high genus by Huang-Klemm-Quackenbush in \cite{Huang:2006hq}. This problem was revisited by Hosono-Konishi \cite{Hosono:2007vf} for the Grassmannian/Pfaffian pair of models discussed by R{\o}dland \cite{rodland2000pfaffian}, and then for all one-parameter Grassmannian models by Klemm-Haghighat \cite{Haghighat:2008ut}. Here we collect the salient results for the fixing of the ambiguity. From here forwards, we will specialise to the one-parameter case $h^{1,1}(Y)=h^{2,1}(X)=1$ with $\mathbf{t}=t^{1}=t$.

At a conifold point, about which the mirror coordinate $t_{c}=k_{c}\varphi+O(\varphi^{2})$ for some constant $k_{c}$, the Schwinger-Gap computation of \cite{Huang:2006hq} provides the behaviour
\begin{equation}\label{eq:gap}
F^{(g)}(t_{c})=|G_{c}|\frac{(-1)^{g-1}B_{2g}}{2g(2g-2)t_{c}^{2g-2}}+O(1)~.
\end{equation}
$B_{n}$ is a Bernoulli number. $|G_{c}|$ are the same numbers appearing as exponents in \eqref{eq:Bgenus1}, giving the number of hypermultiplets that become massless at the conifold. We obtain this $k_{c}$ for each of our examples $Y$ by taking the $k_{c}$ for the covers $\widetilde{Y}$, from the works \cite{Huang:2006hq,Hosono:2007vf,Hosono:2011np}, and dividing by the order of the quotient group. This is the correct normalisation, based on the arguments of \cite{Huang:2006hq} that relate $k_{c}$ to the mass of a shrinking D3 brane in IIB: Since our quotient acts freely, the shrinking D3 brane in our examples are obtained as free quotients of the vanishing cycles on the covers, and so this mass is divided by the order of the group.

The point at infinity, the origin in $\tilde{\varphi}$-space where $\widetilde{\varphi}=\frac{1}{\beta\varphi}$ for some constant $\beta$, will be a singularity of the Picard-Fuchs operator. The B-model free energy is subject to K\"ahler gauge transformations that can add or remove singularities, but the A-model free energy is regular at $\widetilde{\varphi}=0$ for the models we study (see the discussion on C-points in \cite{Huang:2006hq} for cases where $F^{(g)}$ is singular at infinity). Poles of $f^{(g)}$ at infinity can and must cancel with poles of the propagators $S^{ij},\,S^{i},\,S$, and in the original $\varphi$ coordinate these specific poles appear as polynomial terms in $f^{(g)}$. There is a bound on the degree of this polynomial, coming from the prefactor $\varpi_{0}^{2g-2}$ in \eqref{eq:BtoA_FreeEnergy}. Without making a change of K\"ahler gauge, the expansion of $\varpi_{0}$ about infinity has leading term $\tilde{\varphi}^{a}$, where $a$ is the smallest entry of the Riemann symbol of the Picard-Fuchs operator at $\infty$.

Finally, as we have mentioned, the construction of the propagators may introduce ``fake" singularities. $f^{(g)}$ must cancel these. In practice one expects to encounter this problem when one has an apparent singularity, which is a singularity of the Picard-Fuchs operator but not a root of the geometric discriminant polynomial $\Delta$. The highest pole order we encounter at these apparent singularities is $3g-3$, just as in \cite{Hosono:2007vf}.

We will write again the discriminant $\Delta(\varphi)$ as $\prod_{i}\Delta_{i}(\varphi)$, where $\Delta_{i}$ are irreducible over $\IQ$. We allow for an apparent singularity at the root of the linear polynomial $\Delta_{App}(\varphi)$. Moreover, we from here on assume that $\varphi=0$ is a MUM point with all indices in the Riemann symbol equalling 0.

Based on the considerations we have discussed above, the holomorphic ambiguity takes the form
\begin{equation}\label{eq:OurAmb}
\begin{aligned}
&f^{(g)}(\varphi)\=\\[5pt]
&\sum_{j=0}^{\lfloor{(2g-2)a\rfloor}}b_{j}\varphi^{j}+\underbrace{\sum_{\Delta_{i}}\left[\sum_{j=0}^{2g-2}\frac{\sum_{k=1}^{\text{deg}(\Delta_{i})-1}b_{i,j,k}\varphi^{k}}{\Delta_{i}(\varphi)^{j}}\right]}_{\text{Fixed by gap condition}}+\underbrace{\sum_{j=0}^{3g-3}\frac{b_{App,j}}{\Delta_{App}(\varphi)^{j}}}_{\text{Fixed by regularity at } \Delta_{App}=0}+\underbrace{\sum_{j=\lfloor(2g-2)a\rfloor+1}^{g-1}B_{j}\varphi^{j}}_{\text{Fixed by regularity at }\infty}~.
\end{aligned}\end{equation} 
If there is no apparent singularity, then we discard the third term above. The fourth term arises only as a result of choices made when constructing the propagators\footnote{To be more specific, one constructs the propagators $S^{ij},S^{i},S$ as in \cite{Alim:2007qj} (equations 14, 16,17,18). There is freedom to choose a subset of the propagator ambiguities $s^{l}_{ij},h^{jk}_{i},h^{j}_{i},h_{i},h_{ij}$ therein. We make the choice $h^{11}_{1}=0$, $h^{1}_{1}=0$ globally, and $s^{1}_{11}=0$ in a patch containing $\varphi=0$, whereupon specific rational functions $h_{1}$ and $h_{11}$ are forced upon us. This choice leads to the fourth term in \eqref{eq:OurAmb}, other choices may lead to further new terms that can always be removed upon considering regularity (or indeed remove the fourth term altogether). Note that $s^{l}_{ij}$ transforms under general coordinate transformations in the same manner as $\Gamma^{l}_{ij}$, and not as a tensor (see \cite{Huang:2015sta}, equation B.5), and so cannot be taken to vanish globally with our prior choices.}, and is immediately fixed by regularity at every genus so that it causes no conceptual obstacle.

The gap condition \eqref{eq:gap} (once $k_{c}$ is known), and regularity at $\Delta_{App}(\varphi)=0$ and infinity, are sufficient to completely fix the second, third, and fourth terms in \eqref{eq:OurAmb}. It is therefore best to discuss the remaining problem in determining $f^{(g)}$ as the problem  of determining the polynomial
\begin{equation}\label{eq:AmbPoly}
b(\varphi)=\sum_{j=1}^{\lfloor{(2g-2)a\rfloor}}b_{j}\varphi^{j}~.
\end{equation}
The constant term $b_{0}$ can always be computed without difficulty, because the constant term in $F^{(g)}(t)$'s expansion about a MUM point is known \cite{Gopakumar:1998ii,Marino:1998pg,Faber:1998gsw}. Moreover, for any fixed degree $Q$ (the positive integer giving the homology class in $H_{2}(Y,\IZ)$) there is  a maximal genus $g_{\text{max}}(Q)$ such that $n^{(g)}_{Q}\neq0$ (this finiteness conjecture is now a theorem in light of \cite{doan2021gopakumar}). By combining the MNOP conjecture \cite{Maulik:2003rzb,Maulik:2004txy} with new results on Wall-crossing for rank-1 Donaldson-Thomas invariants, assuming what was referred to in \cite{Alexandrov:2023zjb} as the BMT inequality \cite{Bayer2011BridgelandSC,Bayer2014TheSO}, the authors of \cite{Alexandrov:2023zjb} have argued that a bound is given by
\begin{equation}\label{eq:Castelnuovo}
g_{\text{max}}(Q)\leq\begin{cases}\lfloor\frac{Q^{2}}{2\kappa_{111}}+\frac{Q}{2}\rfloor+1~,\quad Q\geq \kappa_{111}\\[5pt]
\lfloor\frac{2Q^{2}}{3\kappa_{111}}+\frac{Q}{3}\rfloor+1~,\quad 0<Q<\kappa_{111}\end{cases}~.
\end{equation}
We assume in this work that this BMT inequality holds for the geometries we study, so that the above bound can be applied. Furthermore, we must make this assumption in order to apply the results of \cite{Alexandrov:2023zjb}, Appendix A when we compute D4D2D0 indices.

From \eqref{eq:Castelnuovo}, one can obtain a number $Q_{\text{Castelnuovo}}(g)$ such that for $Q\leq Q_{\text{Castelnuovo}}(g)$, $n^{g}_{Q}$ vanishes as a consequence of \eqref{eq:Castelnuovo}. We stress that there may be degrees $Q>Q_{\text{Castelnuovo}}(g)$ such that $n^{(g)}_{Q}$ is zero without being implied to vanish by \eqref{eq:Castelnuovo}, which in some examples is observed to occur for $Q$ close to $Q_{\text{Castelnuovo}}(g)$. 

So with a MUM point, we are able to place some constraints on the $b_{j}$ in \eqref{eq:AmbPoly}. However, one must note that the number of Castelnuovo zeroes grows with the genus $g$ asympotically as $\sqrt{2\kappa_{111}g}$. The number of indeterminate $b_{j}$ is $\lfloor2g-2\rfloor a$, which always outstrips the conditions provided by the Castelnuovo zeroes as the genus $g$ is increased. 

It may happen (as for the (1,6) and (1,8) models) that $\varphi=\infty$ is also MUM point, from which a different set of invariants $n^{(g)}_{Q}$ can be computed. In these cases, the additional MUM point provides additional Castelnuovo zeroes and one more constant term condition (in the expansion about infinity).

\subsection{The GV-PT correspondence}
The relations between MSW and GV invariants that we wish to study and exploit in this paper are stated in terms of Pandharipande-Thomas invariants $\text{PT}(Q,n)$ \cite{Pandharipande:2007kc} and Donaldson-Thomas invariants $\text{DT}(Q,n)$ \cite{Thomas:1998uj} of the Calabi-Yau threefold $Y$. We will take a blindered approach to these rich sets of invariants, only displaying their conjectured relation to GV invariants. The generating functions for these invariants are
\begin{equation}
Z_{\text{DT}}(y,\text{q})\=\sum_{Q,n}\text{DT}(Q,n)y^{Q}\text{q}^{n}~,\qquad Z_{\text{PT}}(y,\text{q})\=\sum_{Q,n}\text{PT}(Q,n)y^{Q}\text{q}^{n}~.
\end{equation}
The DT/PT relation, conjectured in \cite{Pandharipande:2007kc} and proven in \cite{toda2010curve,bridgeland2011hall}, yields an equivalence between these two sets of invariants. The relation uses the MacMahon function $M(\text{q})$, and reads
\begin{equation}
Z_{\text{PT}}(y,\text{q})\=M(-\text{q})^{\chi(Y)}Z_{\text{DT}}(y,\text{q})~,\qquad M(\text{q})=\prod_{k>0}(1-\text{q}^{k})^{-k}~.
\end{equation}
Maulik, Nekrasov, Okounkov, and Pandharipande conjecture an equivalence between Donaldson-Thomas and Gromov-Witten theories \cite{Maulik:2003rzb,Maulik:2004txy}. This produces functional identities that relate the topological string free energy with either $Z_{\text{DT}}$ or $Z_{\text{PT}}$. Namely,
\begin{equation}
\exp\left(F_{\text{Reduced}}(t,\lambda)\right)\=Z_{\text{PT}}\left(\text{e}^{2\pi\ii t},-\text{e}^{\ii\lambda}\right)~,
\end{equation}
wherein $F_{\text{Reduced}}$ denotes $F^{\text{All Genus}}$ from \eqref{eq:GVformula} with the polynomial pieces $c(t),l(t)$ and constant contributions ($\beta=0$) discarded. To obtain PT invariants from a set of GV invariants (themselves obtained by solving for the $F^{(g)}$), one writes this relation in the plethystic form due to \cite{Pandharipande:2011jz}:
\begin{equation}
Z_{\text{PT}}(y,\text{q})=\text{PE}\left[\sum_{Q>0}\sum_{g=0}^{g_{\text{max}}(Q)}(-1)^{g-1}n^{(g)}_{Q}(1-x)^{2g-2}x^{1-g}y^{Q}\right](-\text{q},y)~.
\end{equation}
The plethystic exponential is defined by
\begin{equation}
\text{PE}[f(x,y)](X,Y)=\exp\left(\sum_{k=1}^{\infty}\frac{1}{k}f\left(x^{k},y^{k}\right)\right)\bigg\vert_{(x,y)=(X,Y)}~.
\end{equation}

\subsection{D4D2D0 indices}\label{sect:indices}
We will be deliberately brief in our treatment in this subsection, as these matters have already been covered in \cite{Alexandrov:2023zjb}, whose presentation we follow. We cover a bare minimum so that our results in \sref{sect:1pexamples} and \sref{sect:Multiparameter} can be more self-contained.

For an object $E\in\text{D}^{\text{b}}\text{coh}(Y)$, there is a Mukai vector $\gamma(E)\=\text{Ch}(E)\sqrt{\text{Td}(Y)}$. As in \cite{Alexandrov:2010ca} we use a basis $(1,e_{i},e^{i},e_{Y})$ of $H^{\text{Even}}(Y,\IZ)_{\text{Free}}$. Here $e_{j}\wedge e^{i}=\delta^{i}_{j}e_{Y}$ and $\int_{Y}e_{Y}=1$. We expand $\gamma(E)$ following \cite{Alexandrov:2010ca}:
\begin{equation}\label{eq:MukaiVector}
\gamma(E)\=\text{Ch}(E)\sqrt{\text{Td}(Y)}\=p^{0}+p^{i}e_{i}-q_{i}e^{i}+q_{0}e_{Y}~.
\end{equation}
We will neglect the possibility of turning on torsional D4 charge, so that the divisor's homology class $[\mathcal{D}_{\bm{p}}]$ includes torsion pieces. In \cite{Gaiotto:2006wm} the five such choices that could be made for $\IP^{4}[5]/\IZ_{5}$ were argued to give identical sets of indices, explaining the factor of 5 in their equation (4.1) (however, see our discussion following our equation \eqref{eq:QuinticZ5abelian}). 

The components in the above expansion are subject to quantisation conditions:
\begin{equation}
p^{0}\in\IZ~,\qquad p^{i}\in\IZ~, \qquad q_{i}\in\IZ+\frac{1}{2}\kappa_{ijk}p^{j}p^{k}-\frac{1}{24}c_{2,i}p^{0}~,\qquad q_{0}\in\IZ-\frac{1}{24}c_{2,i}p^{i}~.
\end{equation}
The integers $p^{i}~,\,1\leq i\leq h^{1,1}(Y),$ define a divisor $\mathcal{D}_{\mathbf{p}}\subset Y$. The divisor $\mathcal{D}_{\mathbf{p}}$ is the Poincar{\'e} dual of $p^{i}e_{i}$. There is a quadratic form on the lattice $\Lambda=H_{4}(Y,\IZ)=H^{2}(Y,\IZ)$ given by the matrix
\begin{equation}
\kappa_{ij}\=\kappa_{ijk}p^{k}~,
\end{equation}
and the inverse matrix $\kappa^{ij}$ provides a quadratic form on the dual lattice $\Lambda^{*}=H_{2}(Y,\IZ)=H^{4}(Y,\IZ)$. The form $\kappa_{ij}$ embeds $\Lambda$ into $\Lambda^{*}$ as a sublattice, and so there is a lattice quotient $\Lambda^{*}/\Lambda$. This quotient contains 
\begin{equation}
|\Lambda^{*}/\Lambda|\=\text{Det}\left[\kappa_{ijk}p^{k}\right]
\end{equation}
elements, each a $b_{2}$-vector with integer entries.

We will not in this paper provide a complete definition of generalised Donaldson-Thomas invariants \cite{Joyce:2008pc}. The generalised rational Donaldson Thomas invariant $\overline{\Omega}(p^{0},p^{i},q_{i},q_{0};\mathbf{t})=\overline{\Omega}(\gamma;\mathbf{t})$ is locally constant with respect to variations of the K{\"a}hler moduli $\mathbf{t}$ but can change discontinuously at walls of marginal stability. There is a conjectural integer refinement \cite{Joyce:2008pc}, 
\begin{equation}\label{eq:DTrefinement}
\Omega(\gamma;\mathbf{t})\=\sum_{k|\gamma}\frac{\mu(k)}{k^{2}}\overline{\Omega}(\gamma;\mathbf{t})~,
\end{equation}
which again jumps at walls of marginal stability in $\mathbf{t}$-space. $\Omega(\gamma;\mathbf{t})$ and $\overline{\Omega}(\gamma;\mathbf{t})$ coincide when $\gamma$ is primitive. \cite{Alexandrov:2010ca} explains that, apart from the case of $b_{2}(Y)=1$, this moduli dependence persists in the large volume region of moduli space. 

Now consider the rank 0 case $p^{0}=0$, for which the generalised DT invariants are also termed D4D2D0 indices. By going to the large volume attractor point (a procedure introduced in \cite{deBoer:2008fk}), the MSW invariants (terminology introduced in \cite{Alexandrov:2012au}) are obtained. These are given as
\begin{equation}
\overline{\Omega}_{\infty}(0,p^{i},q_{i},q_{0})\=\lim_{\lambda\rightarrow\infty}\overline{\Omega}(0,p^{i},q_{i},q_{0};-\kappa^{ij}q_{j}+\ii\lambda p^{i})~.
\end{equation}
Importantly, these MSW indices do not depend on any moduli. Moreover, as explained in \cite{Alexandrov:2023zjb} they coincide with the Gieseker index in the case that $b_{2}(Y)=1$.


There is an autoequivalence on $\text{D}^{\text{b}}\text{coh}(Y)$, termed spectral flow, sending an object $E$ to $E\otimes\mathcal{O}_{Y}(k^{i}e_{i})$ with each $k^{i}\in\IZ$. The components of the Mukai vector $\gamma(E)$ transform under spectral flow as
\begin{equation}
\begin{aligned}
p^{0}&\mapsto p^{0}~,\\[5pt]
p^{i}&\mapsto p^{i}+k^{i}p^{0}~,\\[5pt]
q_{i}&\mapsto q_{i}-\kappa_{ijk}k^{j}p^{k}-\frac{1}{2}\kappa_{ijk}k^{j}k^{k}p^{0}~,\\[5pt]
q_{0}&\mapsto q_{0}-k^{i} q_{i}+\frac{1}{2}\kappa_{ijk}k^{i}k^{j}p^{k}+\frac{1}{6}\kappa_{ijk}k^{i}k^{j}k^{k}p^{0}~.
\end{aligned}
\end{equation}
In the case $p^{0}=0$, spectral flow leaves  $p^{i}$ invariant. Also, in this case 
\begin{equation}\label{eq:q0def}
\hat{q}_{0}\coloneq q_{0}-\frac{1}{2}\kappa^{ij}q_{i}q_{j}
\end{equation}
is invariant under spectral flow. Furthermore, the class of the vector $\bm{\mu}$ with components
\begin{equation}
\mu_{i}\coloneq q_{i}-\frac{1}{2}\kappa_{ijk}p^{j}p^{k}
\end{equation} 
in the lattice quotient $\Lambda^{*}/\Lambda$ is invariant. These vectors $\bm{\mu}$, representatives of spectral-flow equivalence classes, are called glue vectors. They number $|\Lambda^{*}/\Lambda|$.

Having these spectral flow invariants means that there is a large redundancy in the labelling, and notation is streamlined by writing
\begin{equation}
\overline{\Omega}_{\mathbf{p},\bm{\mu}}(\hat{q}_{0})=\overline{\Omega}_{\infty}(0,p^{i},q_{i},q_{0})~.
\end{equation}
Any Mukai vector components $q_{i},q_{0}$ that produce the spectral flow invariants $\mu_{i}, \hat{q}_{0}$ can be used on the right hand side. Note that the MSW index is constant along orbits of the spectral flow action on the Mukai vector components.

There are two important numbers associated to the divisor $\mathcal{D}_{\mathbf{p}}$. These are $\chi(\mathcal{D}_{\mathbf{p}})$ (the Euler characteristic) and $\chi(\mathcal{O}(\mathcal{D}_{\mathbf{p}}))$ (the arithmetic genus plus 1). They are given by
\begin{equation}
\chi(\mathcal{D}_{\mathbf{p}})\=\kappa_{ijk}p^{i}p^{j}p^{k}+c_{2,i}p^{i}~,\qquad\chi_{\mathcal{D}_{\mathbf{p}}}\=\chi(\mathcal{O}(\mathcal{D}_{\mathbf{p}}))\=\frac{1}{6}\kappa_{ijk}p^{i}p^{j}p^{k}+\frac{1}{12}c_{2,i}p^{i}~.
\end{equation}
Note that these quantities are both integers. It was proven in \cite{Toda:2011aa} (Corollary 3.3), in the case $\text{Pic}(Y)\cong\IZ$, that $\Omega_{\mathbf{p},\mu}(\hat{q}_{0})$ vanishes unless
\begin{equation}\label{eq:q0max}
\hat{q}_{0}\;\leq\; \hat{q}_{0}^{\text{max}}\=\frac{1}{24}\chi(\mathcal{D}_{\mathbf{p}})~.
\end{equation}
From the quantisation conditions, the $\hat{q}_{0}$ defined in \eqref{eq:q0def} with a fixed $\bm{\mu}$ is such that
\begin{equation}\label{eq:q0integer}
\frac{\chi(\mathcal{D}_{\bm{p}})}{24}-\frac{1}{2}\kappa^{ij}\mu_{i}\mu_{i}-\frac{1}{2}\mu_{i}p^{i}-\hat{q}_{0}\in\IZ~.
\end{equation}
In light of \eqref{eq:q0max} and \eqref{eq:q0integer}, we have that $\overline{\Omega}_{\bm{p},\bm{\mu}}\left(\hat{q}_{0}\right)$ can only be nonvanishing for
\begin{equation}\label{eq:Deltamu}
\hat{q}_{0}=\Delta_{\bm{\mu}}-n~,\qquad n\in\IN_{0}~,\qquad\Delta_{\bm{\mu}}\=\frac{\chi(\mathcal{D}_{\bm{p}})}{24}-\text{Fr}\left(\frac{1}{2}\kappa^{ij}\mu_{i}\mu_{i}+\frac{1}{2}\mu_{i}p^{i}\right)~,
\end{equation}
with $\text{Fr}(x)=x-\lfloor x\rfloor$.

The MSW indices will be collected into the generating function
\begin{equation}\label{eq:GenSeries}
\begin{aligned}
h_{\bm{p},\bm{\mu}}(\tau)&\=\sum_{\+\+\hat{q}_{0}\leq\hat{q}_{0}^{\text{max}}}\overline{\Omega}_{\bm{p},\bm{\mu}}\left(\hat{q}_{0}\right)\text{q}^{-\hat{q}_{0}}\\[5pt]
&\=\sum_{\+\+n\geq0\+\+}\overline{\Omega}_{\bm{p},\bm{\mu}}\left(\Delta_{\bm{\mu}}-n\right)\text{q}^{-\Delta_{\bm{\mu}}+n}~,\qquad\text{q}\=\text{e}^{2\pi\ii\tau}~.
\end{aligned}\end{equation}
There is an additional symmetry, $\bm{\mu}\mapsto-\bm{\mu}$, explained in \cite{Alexandrov:2023zjb} in terms of the derived duality $E\mapsto E^{\vee}$. This symmetry can also be seen as a consequence of the modularity transformations detailed in the following subsection.

\subsection{Modularity of MSW invariants}
MSW indices are so named for the role that they play in the Maldacena-Strominger-Witten SCFT \cite{Maldacena:1997de}, the two-dimensional $\mathcal{N}=(0,4)$ SCFT constructed by reducing the M5 brane worldvolume theory on the divisor $\mathcal{D}_{\bm{p}}$. The MSW invariants appear as Fourier coefficients of the modified elliptic genus of the MSW SCFT, as explained in \cite{Alexandrov:2012au}. 

Now, in the absence of subtle complications one can expect the modular symmetry of the SCFT to be reflected in the elliptic genus, so that the MSW invariants are the Fourier coefficients of a modular form. In \cite{Gaiotto:2006wm,Gaiotto:2007cd} the first examples of these modified elliptic genera were computed by exploiting this modularity. In fact, the bound \eqref{eq:q0max} on $\hat{q}_{0}$ such that the MSW index is nonvanishing follows physically from the unitarity bound, so that there is no state in the SCFT with $L_{0}<0$ \cite{Alexandrov:2012au}.

There are in fact substantial complications that arise when the divisor $\mathcal{D}_{\bm{p}}$ is reducible. In this case, the generating series \eqref{eq:GenSeries} for MSW invariants transforms as a weakly holomorphic vector valued mock modular form, as discovered in \cite{Alexandrov:2016tnf,Alexandrov:2017qhn,Alexandrov:2018lgp}. For explicit examples, one can see the work \cite{Alexandrov:2023ltz}. For progress on finding the generating functions in the mock modular case, see \cite{Alexandrov:2024wla}. In the $\mathcal{N}=4$ case, the utility of mock modular forms for black hole counting problems was established in \cite{Dabholkar:2012nd}.

In this work, we will only ever consider the case of an irreducible divisor. In this case the only nonzero term in the sum \eqref{eq:DTrefinement} is $k=1$, so we have equality between the rational and refined Donaldson-Thomas invariants: $\Omega(\gamma;\mathbf{t})=\overline{\Omega}(\gamma;\mathbf{t})$. We produce a number of new examples to accompany those of \cite{Alexandrov:2023zjb,Alexandrov:2022pgd}, and follow their presentation of the conjectured modularity of MSW indices. Namely, for an irreducible divisor $\mathcal{D}_{\bm{p}}$ the generating function \eqref{eq:GenSeries} should transform as a weakly holomorphic vector valued modular form of weight $-1-\frac{b_{2}}{2}$.

This means that $h_{\bm{p},\bm{\mu}}(\tau)$ in \eqref{eq:GenSeries}, which is a function from the upper half plane to $\mathbb{C}^{|\Lambda^{*}/\Lambda|}$, has a specific transformation law under modular transformations of the parameter $\tau$. The modular group $PSL(2,\IZ)$ is the set of matrices 
\begin{equation}
\gamma\=\begin{pmatrix}
a&b\\c&d
\end{pmatrix}
\end{equation}
with integer entries, unit determinant, and $\pm\gamma$ identified. This acts on the upper half plane by fractional linear transformations, 
\begin{equation}
\gamma:\tau\mapsto\frac{a\tau+b}{c\tau+d}~.
\end{equation}
The transformation law for $h_{\bm{p},\bm{\mu}}(\tau)$ is 
\begin{equation}
h_{\bm{p},\bm{\mu}}(\gamma(\tau))\=(c\tau+d)^{-1-\frac{b_{2}}{2}}\sum_{\bm{\nu}\in\Lambda^{*}/\Lambda}M(\gamma)_{\bm{\mu}}^{\+\bm{\nu}}h_{\bm{p},\bm{\nu}}(\tau)~,
\end{equation}
where $M:PSL(2,\IZ)\mapsto GL(|\Lambda^{*}/\Lambda|,\IC)$ is a particular representation of $PSL(2,\IZ)$. As discussed in \cite{Alexandrov:2019rth,Alexandrov:2016enp}, this representation is chosen so that the MSW generating function transforms with multiplier system $M_{\eta}^{c_{2}\cdot\bm{p}}\times \overline{M_{\theta}}$, where $M_{\theta}$ is the multiplier system for the Siegel-Narain theta functions discussed therein (and one should keep in mind across different papers the conventions on $\theta$ versus $\overline{\theta}$).

Since $SL(2,\IZ)$ is generated by the matrices
\begin{equation}
S\=\begin{pmatrix}
0&-1\\1&\+0
\end{pmatrix}~,\qquad T\=\begin{pmatrix}
1&1\\0&1
\end{pmatrix}~,
\end{equation}
it suffices to provide the matrices $M(S)$ and $M(T)$. These are \cite{Alexandrov:2022pgd}
\begin{equation}\begin{aligned}\label{eq:MultiplierSystem}
M(T)_{\bm{\mu}}^{\+\bm{\nu}}&\=\text{e}^{\pi\ii\big(\left(\mu_{i}+\frac{1}{2}\kappa_{ik}p^{k}\right)\kappa^{ij}\left(\mu_{j}+\frac{1}{2}\kappa_{jl}p^{l}\right)+\frac{1}{12}c_{2,i}p^{i}\big)}\delta_{\bm{\mu}}^{\+\bm{\nu}}~,\\[5pt]
M(S)_{\bm{\mu}}^{\+\bm{\nu}}&\=\frac{(-1)^{\chi_{\mathcal{D}_{\mathbf{p}}}}}{\sqrt{|\Lambda^{*}/\Lambda|}}\text{e}^{(b_{2}-2)\frac{\pi\ii}{4}-2\pi\ii\,\mu_{i}\kappa^{ij}\nu_{j}}~.
\end{aligned}\end{equation}
We hope it causes no confusion that $\bm{\mu},\bm{\nu}$ are themselves vectors with components $\mu_{i},\nu_{i}$, which label components of the matrices $M_{\bm{\mu}}^{\+\bm{\nu}}$ and vector $h_{1,\bm{\mu}}$. The 1 subscript in $h_{1,\bm{\mu}}$ indicates that we are considering cases with a single unit of D4 charge (so an irreducible divisor).

A vector valued modular form of negative weight cannot be bounded as $\tau\mapsto\ii\infty$, there must be at least one nonzero polar term (see the discussion in \cite{Manschot:2008zb}). Therefore, the space of vector valued modular forms with a fixed negative weight and bounded order of poles is finite dimensional. The dimension of this space is bounded above by the number of polar terms (terms in the Fourier expansion with negative power of $\text{q}$). Importantly the dimension may be strictly less than the number of polar terms, as discussed in \cite{Manschot:2007ha}. The difference is explained by the existence of certain cusp forms, with each such cusp form imposing one linear constraint on the set of polar terms so that they admit completion to a modular form. An explicit dimension formula is given in \cite{Manschot:2008zb} (equation 3.9), making use of results in \cite{SkoruppaThesis} (equation 6, page 100).

\subsubsection{Basis of vector-valued modular forms}
We shall make use of the basis constructed in \cite{Alexandrov:2022pgd}, subsection 3.2. 

One begins with the theta series (we use the same conventions for these as in  \cite{Alexandrov:2022pgd}, equation (3.4))
\begin{equation}
\vartheta_{\alpha}^{(m,p)}(\tau,z)\=\sum_{k\in\IZ+\frac{\alpha}{m}+\frac{p}{2}}(-1)^{mpk}\text{q}^{\frac{m}{2}k^{2}}\text{e}^{2\pi\ii mkz}~.
\end{equation}
See that $\vartheta_{\alpha}^{(m,p)}(\tau,z)=\vartheta_{\alpha+m}^{(m,p)}(\tau,z)=\vartheta_{-\alpha}^{(m,p)}(\tau,z)$. Note that we have used $\alpha$ as a subscript instead of $\mu$ (as in \cite{Alexandrov:2022pgd}) because we wish to consider multiparameter cases, so that our glue vector $\bm{\mu}$ may have multiple components. 

Now consider the functions
\begin{equation}
\theta_{\alpha}^{(K)}(\tau)\=\begin{cases}\vartheta_{\alpha}^{(K,1)}(\tau,0)~,&K\text{ even},\\[5pt]-\frac{1}{2\pi}\partial_{z}\vartheta_{\alpha}^{(K,1)}(\tau,0)&K\text{ odd}.\end{cases}
\end{equation}
These are the $r=1$ case of the functions given in \cite{Alexandrov:2022pgd}, equation (3.7). The vectors $\left(\theta_{\alpha}^{(K)}(\tau)\right)_{0\leq\alpha<K}$ are vector valued modular forms of weight $1/2+(K\text{ Mod }2)$. They transform with multiplier system
\begin{equation}
M^{(K)}(T)_{\alpha}^{\+\beta}\=\text{e}^{\frac{\pi\ii}{K}(\alpha+\frac{K}{2})^{2}}\delta_{\alpha}^{\+\beta}~,\qquad M^{(K)}(S)_{\alpha}^{\+\beta}\=\frac{\text{e}^{-\frac{\pi\ii}{2}K}}{\sqrt{\ii K}}\text{e}^{-2\pi\ii\frac{\alpha\beta}{K}}~.
\end{equation}

The Serre derivative $D$ acts on a modular form $f$ of weight $w$ by
\begin{equation}
D[f]=(q\partial_{q}f)-\frac{w}{12}E_{2}f~,
\end{equation}
where $E_{2}$ is the normalised Eisenstein series of weight 2. $D[f]$ is a modular form of weight $w+2$.

In the one-parameter case (with irreducible divisor) $\Lambda^{*}/\Lambda\cong\IZ_{\kappa_{111}}$, and the glue vectors $\bm{\mu}$ all have one component, integers $\mu$ in the range $0\leq\mu<\kappa_{111}$.  We shall follow \cite{Alexandrov:2022pgd} exactly, and express our generating series of MSW invariants as combinations of functions 
\begin{equation}\label{eq:basis}
\frac{P_{\ell}\left(E_{4}(\tau),E_{6}(\tau)\right)D^{\ell}\left[\theta_{\mu}^{(\kappa_{111})}(\tau)\right]}{\eta(\tau)^{4\kappa_{111}+c_{2,1}}}~.
\end{equation}
Here $P_{\ell}$ is a quasihomogeneous polynomial of normalised Eisenstein series $E_{4}(\tau)$ and $E_{6}(\tau)$, chosen such that $P_{\ell}$ is a modular form of weight $2\kappa_{111}+\frac{1}{2}c_{2,1}-2-2\ell-(\kappa_{111}\text{ Mod }2)$. In the case $\kappa_{111}=1$, there is a simplification as $\theta^{(1)}_{0}(\tau)\=\eta(\tau)^{3}$ \cite{Alexandrov:2022pgd}.

The functions \eqref{eq:basis} provide vector-valued modular forms of weight $-\frac{1}{2}$. They transform with the $b_{2}=1$ case of the multiplier system \eqref{eq:MultiplierSystem}. It was shown in \cite{Alexandrov:2022pgd} that there are enough choices of $\ell$ and $P_{\ell}$ so that a complete basis of weight $-1/2$ VVMFs with the correct multiplier system and polar exponents $\Delta_{\mu}$ can be found among the functions \eqref{eq:basis}.

In our specific multiparameter examples we have either $b_{2}=2$ or $b_{2}=3$. Our vectors $\bm{p}$ will respectively be either $(1,0)$ or $(1,0,0)$. We observe that the multiplier systems \eqref{eq:MultiplierSystem} for our examples match with those of $\eta(\tau)^{-4|\Lambda^{*}/\Lambda|-c_{2,1}-1+b_{2}}D^{\ell}\left[\theta_{\alpha(\bm{\mu})}^{(|\Lambda^{*}/\Lambda|)}(\tau)\right]$, where $\alpha$ maps the glue vectors $\bm{\mu}$ into $\IZ_{|\Lambda^{*}/\Lambda|}$. Moreover, this combination of theta and eta functions has the correct polar exponents $\Delta_{\bm{\mu}}$, and we can express the MSW generating functions as a sum of these functions multiplied by suitable powers of $E_{4}(\tau)$ and $E_{6}(\tau)$.

In the course of studying our multiparameter examples the Rademacher expansion, extended to VVMFs in \cite{Dijkgraaf:2000fq} and further detailed in \cite{Manschot:2007ha,Manschot:2008ria} has proved very useful. The multiplier system \eqref{eq:MultiplierSystem}, weight $-1-\frac{b_{2}}{2}$, polar exponents \eqref{eq:Deltamu}, and polar terms (provided by \eqref{eq:TheoremF1}, with our examples only having a single polar term) are sufficient to uniquely fix a VVMF. This data can be inserted into the Rademacher expansion, equation (A.3) of \cite{Manschot:2007ha}, to produce the non-polar terms. In practice one only sums a finite number of terms in the infinite sum and observes this to be close to the true integer value. This method reproduces the modular forms that we display in our results. Rademacher expansions were instrumental in the reproduction of $\mathcal{N}=8$ microstate counts from the gravitational path integral in \cite{Iliesiu:2022kny}.

\subsection{Feyzbakhsh's explicit formulae}

\subsubsection{First theorem}
The first of the explicit results that we shall make use of appears as Theorem 4 in Appendix A of \cite{Alexandrov:2023zjb}. This is a stronger version of Theorem 1.1 of \cite{Feyzbakhsh:2022ydn}. To begin with we will only consider $h^{1,1}(Y)=1$, so that our D4 charge vector has a single component equalling $p$ (which should be set to one for the irreducible case). The result provides an expression for $\overline{\Omega}_{p,\mu}\left(\hat{q}_{0}\right)$, subject to a very strong constraint on the reduced D0-brane charge $\hat{q}_{0}$. In fact this constraint is so strong that in our examples we are only able to apply the theorem for a single reduced D0 charge, $\hat{q}_{0}^{\text{max}}=\chi(\mathcal{D}_{p})/24$, with residual D2 charge $\mu=0$. In light of this we will not present the theorem in full, but state the sole result that we make use of:
\begin{equation}\label{eq:TheoremF1}
\overline{\Omega}_{p,0}\left(\frac{\chi(\mathcal{D}_{p})}{24}\right)\=\left(\# H^{2}(Y,\IZ)_{\text{Torsion}}\right)^{2}(-1)^{1+\chi_{\mathcal{D}_{p}}}\chi_{\mathcal{D}_{p}}~.
\end{equation}
This special case of the theorem was presented as equation (4.7) in \cite{Alexandrov:2023zjb}, but there the factor $\left(\# H^{2}(Y,\IZ)_{\text{Torsion}}\right)^{2}$ was equal to 1 as all examples of $Y$ therein were simply connected, with torsion-free second cohomologies.

Our multiparameter examples with $h^{1,1}>1$ will be solved if we can compute the single polar term in each case. Since (apart from the (2,12) model) these examples do not satisfy Assumption $(\star)$ of \cite{Alexandrov:2023zjb}'s Appendix A, we cannot rigorously apply \eqref{eq:TheoremF1} with $p$ replaced by a vector $\bm{p}$. In line with the AGMP ansatz to be discussed in \sref{sect:AGMP}, we propose to do just this, and apply \eqref{eq:TheoremF1} (with $p\mapsto\bm{p}$) in our multiparameter examples to compute these single polar terms. We are taking the ansatz proposed by the authors of \cite{Alexandrov:2022pgd} for this most polar term, and multiplying by $\left(\# H^{2}(Y,\IZ)_{\text{Torsion}}\right)^{2}$. In every example known so far this works because for those most polar terms the AGMP ansatz matches with the theorem from \cite{Alexandrov:2023zjb}. Although this remains open for these multiparameter cases, it is our best guess. In case it is wrong, our generating functions will still be correct up to scale, assuming modularity.

\subsubsection{Second theorem}
For all bar one of our $h^{1,1}=1$ examples, the theorem \eqref{eq:TheoremF1} is sufficient for us to determine the generating functions for Abelian D4D2D0 indices if we assume their modularity. To test the modularity of D4D2D0 indices and apply it to the problem of constraining the holomorphic ambiguity, and to study the remaining case not solved by \eqref{eq:TheoremF1}, we will make use of a different formula. In this discussion we will have $h^{1,1}(Y)=1$, so that the D4 charge vector has a single component equal to 1. We only apply this result to our $h^{1,1}=1$ cases.

The authors of \cite{Alexandrov:2023zjb} made use of Theorem 1 in their Appendix A. This builds on Theorem 1.2 in \cite{Feyzbakhsh:2022ydn}. Proposition 2 in that same Appendix A generalises this theorem, in particular allowing for torsion in $H^{2}(Y,\IZ)$. In the $h^{1,1}=1$ cases we are interested in, the summation range in that proposition works out to be the same as the range defined for Theorem 1. This means that although we are considering manifolds with torsion in $H^{2}(Y,\IZ)$, the ultimate expression we use will be the same as \cite{Alexandrov:2023zjb}, equation (4.19).

Their result involves the function
\begin{equation}
f:\IR^{+}\mapsto\IR~,\qquad f(x)\=\begin{cases}
x+\frac{1}{2} & \text{if }\; 0<x<1~, \\[5pt]
\sqrt{2x+\frac{1}{4}} & \text{if }\; 1\leq x<\frac{15}{8}~,\\[5pt]
\frac{2}{3}x+\frac{3}{4} & \text{if }\; \frac{15}{8}\leq x<\frac{9}{4}~,\\[5pt]
\frac{1}{3}x+\frac{3}{2} & \text{if }\; \frac{9}{4}\leq x<3~,\\[5pt]
\frac{1}{2}x+1 & \text{if }\; 3\leq x~.
\end{cases}
\end{equation}
Additionally there are notations
\begin{equation}\label{eq:primesnotation}
\begin{aligned}
\chi(Q',m')&\=m-m'+Q+Q'-\chi_{\mathcal{D}_{1}}~,\\[5pt]
\hat{q}_{0}'&\=m'-m-\frac{1}{2\kappa_{111}}(Q'-Q)^{2}-\frac{1}{2}(Q+Q')+\frac{\chi(\mathcal{D}_{1})}{24}~. 
\end{aligned}
\end{equation}
Now suppose that 
\begin{equation}\label{eq:F2conditions}
(Q,m)\in\IZ_{+}\times\IZ~,\qquad f\left(\frac{Q}{\kappa_{111}}\right)<-\frac{3m}{2Q}~.
\end{equation}
Then there are relations \cite{Alexandrov:2023zjb}
\begin{equation}\label{eq:TheoremF2}
\text{PT}(Q,m)=\sum_{(Q',m')}(-1)^{\chi(Q',m')}\chi(Q',m')\text{PT}(Q',m')\Omega_{1,Q-Q'}\left(\hat{q}_{0}'\right)~,
\end{equation}
where the sum runs over pairs $(Q',m')$ that satisfy
\begin{equation}\label{eq:sumrange}
\begin{aligned}
0\leq& \;Q'\;\leq Q+\kappa_{111}\left(\frac{1}{2}+\frac{3m}{2Q}\right)~,\\[5pt]
-\frac{Q'^{2}}{2\kappa_{111}}-\frac{Q'}{2}\leq& \;m'\; \leq m+\frac{1}{2\kappa_{111}}(Q-Q')^{2}+\frac{1}{2}(Q+Q')~.
\end{aligned}
\end{equation}
The formula \eqref{eq:TheoremF2} appeared in \cite{Alexandrov:2023zjb} as equation (4.12). As explained there, the pairs $(Q',m')$ summed over only involve $Q'<Q$ with strict inequality. For each $Q'$, there are finitely many $m'$ allowed by \eqref{eq:sumrange}. This means that \eqref{eq:TheoremF2} is recursive in $Q$.

Following \cite{Alexandrov:2023zjb}, note that $\text{PT}(0,0)=1$. So one can hope to invert \eqref{eq:TheoremF2} to obtain a recursive expression for $\Omega_{1,\mu}(\hat{q}_{0})$. However, the integers $Q=\mu,\,m=-\hat{q}_{0}+\frac{\chi(\mathcal{D}_{1})}{24}-\frac{1}{2\kappa_{111}}Q^{2}-\frac{1}{2}Q$ may not satisfy \eqref{eq:F2conditions} so that \eqref{eq:TheoremF2} cannot be used. The way around this problem, due to \cite{Alexandrov:2023zjb}, is to abuse spectral flow. There exists some $k_{\text{min}}\in\IZ_{+}$ such that for all integers $k\geq k_{\text{min}}$ the pairs
\begin{equation}\label{eq:Qkmk}
(Q_{k}\,,\,m_{k})=\left(\,\mu+k \kappa_{111}\,,\,\frac{\chi(\mathcal{D}_{1})}{24}-\hat{q}_{0}-\frac{1}{2\kappa_{111}}Q_{k}^{2}-\frac{1}{2}Q_{k}\,\right)
\end{equation} 
satisfy \eqref{eq:F2conditions}. This is a spectral flow transformation, and the D4D2D0 index $\Omega_{1,\mu}(\hat{q}_{0}))$ will be the same for $(Q_{k},m_{k})$ as it is for the original $(Q,m)$. Thus, for each $k\geq k_{\text{min}}$ one can write the Abelian D4D2D0 index as\footnote{By way of explanation, as in \cite{Alexandrov:2023zjb} we seek a useful expression for $\Omega_{1,\mu}\left(\hat{q}_{0}\right)\equiv\Omega_{1,\mu}\left(\Delta_{\mu}-m\right)$ and wish to use \eqref{eq:TheoremF2}. However, it is not guaranteed that $(Q,m)\equiv(\mu,-\hat{q}_{0}+\Delta_{\mu})$ satisfies the conditions \eqref{eq:F2conditions}. If one replaces $(Q,m)$ with $(Q_{k},m_{k})$ as in \eqref{eq:Qkmk}, for a $k$ such that \eqref{eq:F2conditions} is satisfied, then one gets an instance of \eqref{eq:TheoremF2} reading $\text{PT}(Q_{k},m_{k})=(-1)^{\chi(0,0)}\chi(0,0)\text{PT}(0,0)\Omega_{1,Q_{k}}\left(\Delta_{Q}-m_{k}\right)+\text{corrections}$. This equation can be rearranged to give $\Omega_{1,\mu}(\hat{q}_{0}))$, which does not appear in the corrections. Note that $\text{PT}(0,0)=1$, and $\Omega_{1,Q_{k}}\left(\Delta_{Q_{k}}-m_{k}\right)=\Omega_{1,\mu}\left(\Delta_{\mu}-m\right)$ by spectral-flow invariance of generalised DT invariants. This rearrangement gives $\Omega_{1,\mu}(\hat{q}_{0}))=(-1)^{\chi(0,0)}\chi(0,0)^{-1}\left[\text{PT}(Q_{k},m_{k})-\text{corrections}\right]$, which is exactly equation \eqref{eq:TheoremF2r}.}
\begin{equation}\label{eq:TheoremF2r}
\Omega_{1,\mu}\left(\hat{q}_{0}\right)\=\frac{(-1)^{m_{k}+Q_{k}-\chi_{\mathcal{D}_{1}}}}{m_{k}+Q_{k}-\chi_{\mathcal{D}_{1}}}\left[\text{PT}(Q_{k},m_{k})-\sum_{(Q',m')\neq(0,0)}(-1)^{\chi(Q',m')}\chi(Q',m')\text{PT}(Q',m')\Omega_{1,Q_{k}-Q'}(\hat{q}_{0}')\right]~.
\end{equation}
The $\hat{q}_{0}'$ in \eqref{eq:TheoremF2r} are computed from \eqref{eq:primesnotation} with $(Q,m)$ replaced by $(Q_{k},m_{k})$. Similarly, the $\chi(Q',m')$ in \eqref{eq:TheoremF2r} are computed as in \eqref{eq:primesnotation} but with $(Q,m)$ replaced by $(Q_{k},m_{k})$. Note that after suitably applying \eqref{eq:TheoremF2} and specialising \eqref{eq:sumrange}, the summation range in \eqref{eq:TheoremF2r} runs over pairs $(Q',m')$ satisfying the following inequalities:
\begin{equation}\label{eq:sumrangek}
\begin{aligned}
0<& \;Q'\;\leq Q_{k}+\kappa_{111}\left(\frac{1}{2}+\frac{3m_{k}}{2Q_{k}}\right)~,\\[5pt]
-\frac{Q'^{2}}{2\kappa_{111}}-\frac{Q'}{2}\leq& \;m'\; \leq m_{k}+\frac{1}{2\kappa_{111}}(Q_{k}-Q')^{2}+\frac{1}{2}(Q_{k}+Q')~.
\end{aligned}
\end{equation}

\subsection{The AGMP Ansatz}\label{sect:AGMP}
The authors of \cite{Alexandrov:2022pgd} provide a physically-motivated formula for the polar terms in $h_{1,\bm{\mu}}(\tau)$. Although there are cases where this is known to be incorrect (see the note added in page 6,  \cite{Alexandrov:2022pgd} v2), it produces the correct result in many cases. As discussed in \cite{Alexandrov:2023zjb}, a better understanding of the range of validity of this expression would elucidate on proposed relations between the polar terms and bound states of $\text{D6}$-$\overline{\text{D6}}$-branes \cite{Denef:2007vg}. 

The AGMP Ansatz reads, in the case of a simply connected threefold with $b_{2}=1$ (so that the D4 charge $\bm{p}$ is given by an integer $p$),
\begin{equation}\label{eq:AGMP}
\overline{\Omega}_{p,\mu}\left(\hat{q}_{0}\right)\=(-1)^{p\mu+n+\chi_{\mathcal{D}_{p}}}(p\mu+n-\chi_{\mathcal{D}_{p}})\text{DT}(\mu,n)~,
\end{equation}
where $0\leq\mu\leq\frac{1}{2}p\kappa_{111}$ is taken, otherwise one uses the $\mu\leftrightarrow p\kappa_{111}-\mu$ symmetry. Here $n$ is the integer from \eqref{eq:Deltamu}. \eqref{eq:AGMP} is equation (5.20) in \cite{Alexandrov:2022pgd}, and equation (4.8) in \cite{Alexandrov:2023zjb}. This can be obtained by truncating the sum provided by the rigorous theorem 4 of \cite{Alexandrov:2023zjb} (appendix A) and ignoring the assumptions of said theorem. Since that sum includes an overall factor $(\#H^{2}(Y,\IZ)_{\text{Torsion}})^{2}$, we guess that in the non-simply connected case one should modify \eqref{eq:AGMP} to 
\begin{equation}\label{eq:AGMPtorsion}
\overline{\Omega}_{p,\mu}\left(\hat{q}_{0}\right)\=(\#H^{2}(Y,\IZ)_{\text{Torsion}})^{2}(-1)^{p\mu+n+\chi_{\mathcal{D}_{p}}}(p\mu+n-\chi_{\mathcal{D}_{p}})\text{DT}(\mu,n)~,
\end{equation}
and then study whether this formula fails or succeeds.

We raise this for two reasons. The first is that for $\IP^{5}[3,3]/\IZ_{3}$, the GV invariants that we compute in \sref{sect:GVtables} are insufficient for us to use \eqref{eq:TheoremF2r} with any valid spectral flow parameter $k\geq k_{\text{min}}$. We go on to use invalid values $k<k_{\text{min}}$ to arrive at a modular form, and note that this naive result is in agreement with \eqref{eq:AGMPtorsion}.

The second reason is that the (1,8) and (1,6) models in our \tref{tab:Manifolds} are quotients of manifolds that famously have derived equivalent partner manifolds. Subject to the assumption that both quotients, of two derived equivalent manifolds by the same group, are derived equivalent we obtain the GV invariants presented in \sref{sect:GVtables}. It would be interesting to compute D4D2D0 indices for the $\IZ_{7}$ quotient of the codimension 7 complete intersection Calabi-Yau in $\text{Gr}(2,7)$, and the $\IZ_{5}$ quotient of the Reye congruence. We do not have sufficiently many GV invariants to utilise \eqref{eq:TheoremF2r} in these cases, so must resort to trying \eqref{eq:AGMPtorsion} and \eqref{eq:AGMP}. This does not work. The AGMP ansatz predicts sets of polar terms that cannot be completed to modular forms. 

We defer further study of these models to the future, and here only list the polar terms provided by \eqref{eq:AGMPtorsion} before we move on to our more successful examples. 

The first case of these problematic models is the $\IZ_{7}$ quotient of the (1,1,1,1,1,1,1) intersection in $Gr(2,7)$ \cite{rodland2000pfaffian,Hosono:2007vf,Haghighat:2008ut}, which has relevant topological data $\kappa_{111}=6,\,c_{2,1}=12,H^{2}(Y,\IZ)=\IZ\oplus\IZ_{7}$. 

The second case is the $\IZ_{5}$ quotient of the Reye congruence \cite{Hosono:2011np,hosono2016double,Candelas:2008wb}, which has $\kappa_{111}=7$, $c_{2,1}=10$, $H^{2}(Y,\IZ)=\IZ\oplus\IZ_{10}$.

Equation \eqref{eq:AGMPtorsion} leads to
\begin{equation}\label{eq:bad}
\text{First case: }h_{1}^{\text{polar}}(\tau)\=\begin{pmatrix}
-98q^{-3/4}\\882q^{-1/6}\\343q^{-5/12}\\0\\343q^{-5/12}\\882q^{-1/6}
\end{pmatrix}~,\qquad\text{Second case: } h_{1}^{\text{polar}}(\tau)\=\begin{pmatrix}
-200q^{-17/24}\\1600q^{-23/168}\\500q^{-71/168}\\0\\0\\500q^{-71/168}\\1600q^{-23/168}
\end{pmatrix}~.
\end{equation}
Using \eqref{eq:AGMP} instead only scales these vectors, and we still cannot obtain modular forms with the correct transformation properties that have the above polar parts. This is a similar situation as for the models $X_{4,2}$, $X_{3,2,2}$, and $X_{2,2,2,2}$ in Appendix C of \cite{Alexandrov:2022pgd}.

\newpage

\section{One-parameter examples}\label{sect:1pexamples}
\subsection{$\IZ_{5}$ quotient of the quintic threefold, the (1,21) model}
The quintic threefold $\IP^{4}[5]$ is given by the vanishing of a degree $5$ polynomial in $\IP^{4}$, and we will take $\IP^{4}$ to have homogeneous coordinates $\left[y_{0}:y_{1}:y_{2}:y_{3}:y_{4}\right]$. It has been known since \cite{Greene:1990ud} that there is a 21-dimensional locus inside the 101-dimensional complex structure moduli space of $\IP^{4}[5]$ such that the $\IZ_{5}$ action on $\IP^{4}$ generated by
\begin{equation}
\left[y_{0}:y_{1}:y_{2}:y_{3}:y_{4}\right]\mapsto \left[y_{0}:\text{e}^{\frac{2\pi\ii}{5}}y_{1}:\text{e}^{\frac{2\pi\ii}{5}\cdot2}y_{2}:\text{e}^{\frac{2\pi\ii}{5}\cdot3}y_{3}:\text{e}^{\frac{2\pi\ii}{5}\cdot4}y_{4}\right]
\end{equation}
descends to a free $\IZ_{5}$ action on $\IP^{4}[5]$. One could also make a linear change of coordinates so that the $\IZ_{5}$ action is instead generated by $y_{i}\mapsto y_{i+1}$.

\subsubsection{The mirror}
Following \cite{Hosono:1994ax,batyrev1995generalized}, the mirror Laurent polynomial equation with complex structure parameter $\varphi$ is
\begin{equation}\label{eq:mirrorquintic}
x_{1}+x_{2}+x_{3}+x_{4}+\frac{\varphi}{x_{1}x_{2}x_{3}x_{4}}\=1~.
\end{equation}
The mirror of $\IP^{4}[5]$ is a subvariety of the toric variety labelled $\IP_{\Delta^{*}}$ in \cite{Hosono:1994ax}. As we only require two things from the mirror, period integrals and an analysis of the $\IZ_{5}$ fixed points, it is sufficient for our purposes to have the coordinates $x_{i}$ live in a dense torus $(\IC^{*})^{4}\subset\IP_{\Delta^{*}}$, whereupon the locus of \eqref{eq:mirrorquintic} is birational to the true smooth mirror. To recover the perhaps more familiar equation for the mirror quintic used in \cite{Candelas:1990rm} one makes a replacement $x_{i}\mapsto\varphi^{1/5}x_{i}^{5}/(x_{1}x_{2}x_{3}x_{4})$, homogenises to  introduce $x_{5}$, and then identifies $\psi_{\text{there}}=\frac{1}{5}\varphi_{\text{here}}^{-1/5}$.

The discriminant of \eqref{eq:mirrorquintic} is well known, $\Delta=1-5^{5}\varphi$. The Picard-Fuchs operator has been known since \cite{Candelas:1990rm} to be a generalised hypergeometric equation with fundamental solution 

\begin{equation}
\varpi_{0}(\varphi)\=\fourFthree{1/5}{2/5}{3/5}{4/5}{1}{1}{1}{5^{5}\varphi}.
\end{equation}
The computation of higher genus invariants $n^{(g)}_{k}$ then proceeds almost exactly as in \cite{Huang:2006hq}, with orbifolds like the one to be discussed soon already mentioned in that paper as extra results. The major substantial difference is that when one takes a $\IZ_{5}$ quotient, the conifold point is replaced by a hyperconifold point. Instead of a shrinking $S^{3}$ there is a shrinking Lens space $S^{3}/\IZ_{5}$, and this is taken into account when computing the genus one free energy. This change cascades into all of the higher genus free energies.

It has been known since \cite{Greene:1990ud,Aspinwall:1990xe}, where the mirror quintic threefold was originally constructed, that the mirror of the quintic's $\IZ_{5}$ quotient is the $\IZ_{5}$ quotient of the mirror. The equation \eqref{eq:mirrorquintic} has a $\IZ_{5}$ symmetry generated by $x_{1}\mapsto x_{2}\mapsto x_{3}\mapsto x_{4}\mapsto \varphi/(x_{1}x_{2}x_{3}x_{4})$. There is a more symmetric presentation of \eqref{eq:mirrorquintic}, obtained by introducing a new coordinate $x_{5}$ and an additional equation as follows:
\begin{equation}\label{eq:mirrorquintic2}
\begin{aligned}
x_{1}+x_{2}+x_{3}+x_{4}+x_{5}&\=1~,\\[5pt]
x_{1}x_{2}x_{3}x_{4}x_{5}&\=\varphi~.
\end{aligned}
\end{equation}
The $\IZ_{5}$ symmetry is now $x_{i}\mapsto x_{i+1}$. At any fixed point all five $x_{i}$ are equal by assumption, and then the first equation of \eqref{eq:mirrorquintic2} provides $x_{i}=\frac{1}{5}$. Then the second equation of \eqref{eq:mirrorquintic2} can only be solved for one value of $\varphi$, which coincides with the discriminant locus $\varphi=5^{-5}$. 

This means that when we compute the genus one free energy as in \eqref{eq:Bgenus1}, we will have ${|G_{1}|=5}$.

\subsubsection{Abelian D4D2D0 indices}\label{sect:Quinticresults}
The quotient manifold $\IP^{4}[5]/\IZ_{5}$ has topological data
\begin{equation}
\kappa_{111}\=1~,\qquad c_{2}\=10~,\qquad \chi=-40~.
\end{equation}
Since $\kappa_{111}=1$, the VVMF $h_{1,\mu}$ generating Abelian D4D2D0 invariants will have rank 1. One computes
\begin{equation}\begin{aligned}
\chi_{\mathcal{D}_{1}}&\=\frac{1}{6}\kappa_{111}+\frac{1}{12}c_{2}\=1~,\qquad \chi(\mathcal{D}_{1})\=\kappa_{111}+c_{2}=11~, \\[5pt]
&\Delta_{0}\= \frac{\chi(\mathcal{D}_{1})}{24}=\frac{11}{24}~.
\end{aligned}\end{equation}

As consequence of formula \eqref{eq:TorsionZm}, $\text{Tors}\left(H^{2}\left(\IP^{4}[5]/\IZ_{5}\right)\right)\cong\IZ_{5}$. We can then use \eqref{eq:TheoremF1} to compute
\begin{equation}
\Omega_{1,0}\left(\frac{11}{24}\right)\=25~.
\end{equation}

Since $h_{1,\mu}$ has only one component ($\mu=0$), and only a single term in the $\text{q}$-series for this component has a negative exponent, the above computation completely fixes $h_{1,0}$. This example has already been addressed in \cite{Gaiotto:2006wm} (section 4), where it was realised that the solution must take the form $h_{1,0}=C \eta^{-11}E_{4}$, with $\eta$ the Dedekind eta function, $E_{4}$ the weight-4 Eisenstein series, and $C$ an integer. 

We obtain
\begin{equation}\label{eq:QuinticZ5abelian}
h_{1,0}(\tau)\=25\eta(\tau)^{-11}E_{4}(\tau)\=25\text{q}^{-11/24}\left(\underline{\,1\,}+251\text{q}+4877\text{q}^{2}+49378\text{q}^{3}+360005\text{q}^{4}+\,...\right)
\end{equation}
The prefactor of $25$ differs to the result of \cite{Gaiotto:2006wm}, wherein the prefactor was taken to be $5$. Now, there the authors were aiming to compute degeneracies for all D4D2D0 bound states, and so they summed over the torsion classes in $H^{2}(\IP^{4}[5]/\IZ_{5},\IZ)$ to arrive at their factor of five. We do not do this, but there is still a discrepancy as if we did sum over the torsion classes in this manner then our prefactor would be 125.

\subsubsection{Interplay with GV invariants}
In Appendix \sref{sect:GVtables}, \tref{tab:QuinticZ5Numbers}, we provide GV invariants up to genus 10. This is the highest genus we can reach with the boundary conditions described therein. 

We will now attempt to recompute $\Omega_{1,0}\left(\frac{11}{24}\right)=25$ using the other explicit formula \eqref{eq:TheoremF2r}, for which we require these GV invariants. The smallest value of the spectral flow parameter $k$ such that the pair $(Q_{k},m_{k})$ of \eqref{eq:Qkmk} satisfy the inequality \eqref{eq:F2conditions} is $k_{\text{min}}=2$.

We will also consider the recomputation of $\Omega_{1,0}\left(\frac{11}{24}-1\right)=25*251$. For $\mu=0,\hat{q}_{0}=\hat{q}_{0}^{\text{max}}-1$ the minimal allowed value of $k$ is $k_{\text{min}}=4$.

$\underline{\mu=0,\;\hat{q}_{0}=\hat{q}_{0}^{\text{max}},\;k=k_{min}=2\,:}$

For $k=k_{\text{min}}=2$, we have $(Q_{k},m_{k})=(2,-3)$. The only pair in the range defined by \eqref{eq:sumrangek} is $(Q',m')=(0,1)$. Since $\text{PT}(0,1)=0$, the sum in \eqref{eq:TheoremF2r} contributes nothing and so \eqref{eq:TheoremF2r} provides us with
\begin{equation}
\Omega_{1,0}\left(\frac{11}{24}\right)\=\frac{(-1)^{m_{k}+Q_{k}-\chi_{\mathcal{D}_{1}}}}{m_{k}+Q_{k}-\chi_{\mathcal{D}_{1}}}\text{PT}(Q_{k},m_{k})\bigg\vert_{(Q_{k},m_{k})=(2,-3)}\=-\frac{n^{(4)}_{2}}{2}~.
\end{equation}
We have used the PT-GV correspondence in the final equality. Since we have already independently computed $n^{(4)}_{2}=-50$ in \tref{tab:QuinticZ5Numbers}, we see that the explicit formula \eqref{eq:TheoremF2r} is confirmed.

$\underline{\mu=0,\;\hat{q}_{0}=\hat{q}_{0}^{\text{max}},\;k=k_{min}+1=3\,:}$

Now we compute the same D4D2D0 invariant as before, but with a different value for the spectral flow parameter $k$. This time $(Q_{k},m_{k})=(3,-6)$. Once again the only pair in the range defined by \eqref{eq:sumrangek} is $(Q',m')=(0,1)$, and $\text{PT}(0,1)=0$, so the sum in \eqref{eq:TheoremF2r} contributes nothing. From \eqref{eq:TheoremF2r} we read
\begin{equation}
\Omega_{1,0}\left(\frac{11}{24}\right)\=\frac{(-1)^{m_{k}+Q_{k}-\chi_{\mathcal{D}_{1}}}}{m_{k}+Q_{k}-\chi_{\mathcal{D}_{1}}}\text{PT}(Q_{k},m_{k})\bigg\vert_{(Q_{k},m_{k})=(3,-6)}\=-\frac{n^{(7)}_{3}}{4}~.
\end{equation}
From \tref{tab:QuinticZ5Numbers}, we already have $n^{(7)}_{3}=-100$ and we once again confirm that the prediction from \eqref{eq:TheoremF2r} is correct.

$\underline{\mu=0,\;\hat{q}_{0}=\hat{q}_{0}^{\text{max}},\;k=k_{min}+2=4\,:}$

Now we have $(Q_{k},m_{k})=(4,-10)$. There are two pairs in the range defined by \eqref{eq:sumrangek}, those being $(Q',m')=(0,1)$ and $(0,2)$. Since $\text{PT}(0,2)=\text{PT}(0,1)=0$, the sum in \eqref{eq:TheoremF2r} contributes nothing. From \eqref{eq:TheoremF2r} we read off
\begin{equation}\label{eq:Qg11}
\Omega_{1,0}\left(\frac{11}{24}\right)\=\frac{(-1)^{m_{k}+Q_{k}-\chi_{\mathcal{D}_{1}}}}{m_{k}+Q_{k}-\chi_{\mathcal{D}_{1}}}\text{PT}(Q_{k},m_{k})\bigg\vert_{(Q_{k},m_{k})=(4,-10)}\=\frac{n^{(11)}_{4}}{7}~.
\end{equation}
This number does not appear in our tables \sref{sect:GVtables} as we required one additional datum to compute the GV invariants at genus 11. By taking \eqref{eq:Qg11} as input we can solve the topological string to genera 11,12,13. We display some new invariants in \tref{tab:QuinticGenus11}.

\begin{minipage}{.45\linewidth}
\begin{table}[H]
\small{\begin{tabular}{| c || l | l | l |}
\hline $\phantom{\bigg|}k\phantom{\bigg|}$ &\hfil $n^{(11)}_{k}$ & \hfil $n^{(12)}_{k}$ &\hfil $n^{(13)}_{k}$   \\\hline 
1 & 0 & 0 & 0 \\
2 & 0 & 0 & 0 \\
3 & 0 & 0 & 0 \\
4 & 175 & 0 & 0 \\
5 & -7169430652 & 794737725 & -61263800 \\
6 & 2602101884428630 & -231085360252560 & 59862993108300 \\
7 & 87967441882643127859850 & 4756262670014883289150 & 
111922686673158134550 \\
8 & 151493784464416577906799492925 & 32156743263310689530717111260 & 
4228191354832779283415651050 \\
\hline   
\end{tabular} 
}
\end{table}
\end{minipage}
\vskip-5pt
\begin{table}[H]
\begin{center}
\capt{\textwidth}{tab:QuinticGenus11}{GV invariants for the quintic's $\IZ_{5}$ quotient, obtained by combining \eqref{eq:Qg11} with the results of \sref{sect:GVtables}.}
\end{center}
\end{table}

$\underline{\mu=0,\;\hat{q}_{0}=\hat{q}_{0}^{\text{max}}-1,\;k=k_{\text{min}}=4\,:}$

We now turn to recomputing a different D4D2D0 invariant, the first of our computations to test the modularity of \eqref{eq:QuinticZ5abelian} (whereby all terms in the $\text{q}$-series are fixed once we fix the first term). For this case, we have $(Q_{k},m_{k})=(4,-9)$. The contributing pairs $(Q',m')$ are $(0,1)$, $(0,2)$, $(0,3)$, $(1,-1)$, and $(1,0)$. The first three cannot contribute because $\text{PT}(0,1)=\text{PT}(0,2)=\text{PT}(0,3)=0$. However, $\text{PT}(1,-1)=n^{(2)}_{1}$ and $\text{PT}(1,0)=n^{(1)}_{2}+20n^{(2)}_{1}$ are both nonzero. Nonetheless, these pairs also do not contribute because the corresponding values of $\hat{q}'_{0}$ violate \eqref{eq:q0max}. 

Specifically, for $(Q',m')=(1,-1)$ we find that $\hat{q}'_{0}=\hat{q}^{\text{max}}_{0}+1$. For $(Q',m')=(1,0)$ we find $\hat{q}'_{0}=\hat{q}^{\text{max}}_{0}+2$. Therefore, the sum in \eqref{eq:TheoremF2r} once again does not contribute and we read off
\begin{equation}\label{eq:Qg11again}
\Omega_{1,0}\left(-\frac{13}{24}\right)\=\frac{(-1)^{m_{k}+Q_{k}-\chi_{\mathcal{D}_{1}}}}{m_{k}+Q_{k}-\chi_{\mathcal{D}_{1}}}\text{PT}(Q_{k},m_{k})\bigg\vert_{(Q_{k},m_{k})=(4,-9)}\=-\frac{1}{6}\left(n^{(10)}_{4}+20n^{(11)}_{4}\right)~.
\end{equation}
We are pleased to report that, with $n^{(10)}_{4}=-41150$ from \tref{tab:QuinticZ5Numbers} and $n^{(11)}_{4}=175$ from \eqref{eq:Qg11}, the tentative relation \eqref{eq:Qg11again} does indeed hold: $-\frac{1}{6}(-41150+20*175)=6275=25*251$.

\underline{Additional GV invariants assuming \eqref{eq:TheoremF2r} :}

By recomputing $\Omega_{1,0}\left(\frac{11}{24}\right),\,\Omega_{1,0}\left(-\frac{13}{24}\right),\,\Omega_{1,0}\left(-\frac{37}{24}\right)$ using $k=5$, which respectively is $k_{\text{min}}+3,\,k_{\text{min}}+1,k_{\text{min}}$, we are able to obtain the following GV invariants:
\begin{equation}\label{eq:Qg14}
n^{(14)}_{5}\=2965700~,\qquad n^{(15)}_{5}\=-71000~,\qquad n^{(16)}_{5}\=275~.
\end{equation}
This provides us with enough data to expand up to genus 15, as displayed in \tref{tab:QuinticGenus14}. After incorporating \eqref{eq:Qg14} we are still in need of one more datum to obtain further genus 16 invariants.

\begin{minipage}{.45\linewidth}
\begin{table}[H]
\small{\begin{tabular}{| c || l | l | l |}
\hline $\phantom{\bigg|}k\phantom{\bigg|}$ &\hfil $n^{(14)}_{k}$ & \hfil $n^{(15)}_{k}$    \\\hline 
1 & 0 & 0 \\
2 & 0 & 0 \\
3 & 0 & 0 \\
4 & 0 & 0 \\
5 & 2965700 & -71000 \\
6 & -12576522370080 & 2158870171160 \\
7 & 3256013529576897075 & -813245152733660750 \\
8 & 330378094976934638810586210 & 14511004385732885931249005 \\
9 & 2543092010804637552209780798490390 & 
421954236680996731171378302165400 \\
10 & 3250497633874077193629894835512058573790 & 
1301446688552380479335402521670946275490 \\
\hline   
\end{tabular} 
}
\end{table}
\end{minipage}
\vskip-5pt
\begin{table}[H]
\begin{center}
\capt{\textwidth}{tab:QuinticGenus14}{GV invariants for the quintic's $\IZ_{5}$ quotient, obtained by combining \eqref{eq:Qg11} and \eqref{eq:Qg14} with the results of \sref{sect:GVtables}.}
\end{center}
\end{table}

\underline{Testing \eqref{eq:TheoremF2r} with spectral flow parameter $k<k_{\text{min}}$ :}

An important observation of \cite{Alexandrov:2023zjb} is that in some of their examples the explicit formula \eqref{eq:TheoremF2r} was able to correctly relate GV invariants with the indices read off from $h_{1,\mu}$, even if they used a value $k$ for the spectral flow parameter such that $(Q_{k},m_{k})$ did not satisfy the inequality \eqref{eq:F2conditions}. 

Consider $\Omega_{1,0}\left(\frac{11}{24}\right),\,\Omega_{1,0}\left(-\frac{13}{24}\right),$ and $\Omega_{1,0}\left(-\frac{37}{24}\right)$ as read off from \eqref{eq:QuinticZ5abelian}. These three numbers are $25,\,6275,$ and $121925$. When we attempt to recompute these numbers using $k=k_{\text{min}}-1$, we obtain respectively $10,\,6248+\frac{1}{3},$ and $122096$. These are all incorrect, although tantalisingly close.

\subsection{$\IZ_{3}$ quotient of the bicubic threefold, the (1,25) model}
Now we turn to $\widetilde{Y}\cong\IP^{5}[3,3]$, the complete intersection of two cubic hypersurfaces in $\IP^{5}$. As explained in \cite{Candelas:2008wb}, members of this family can be found with freely acting $\IZ_{3}$ and $\IZ_{3}\times\IZ_{3}$ symmetries. The generators of $\IZ_{3}\times\IZ_{3}$ act on the homogeneous coordinates $y_{i}$ of $\IP^{5}$ by the actions
\begin{equation}
\begin{aligned}
[y_{1}:y_{2}:y_{3}:y_{4}:y_{5}:y_{6}]&\mapsto[y_{2}:y_{3}:y_{1}:y_{4}:y_{5}:y_{6}]~,\\[5pt]
[y_{1}:y_{2}:y_{3}:y_{4}:y_{5}:y_{6}]&\mapsto[y_{1}:\zeta y_{2}:\zeta^{2}y_{3}:y_{4}:\zeta y_{5}:\zeta^{2}y_{6}]~,\qquad \zeta\=\text{e}^{2\pi\ii/3}~.
\end{aligned}
\end{equation}
Note that if we fix one of these actions, we can then change coordinates so that the $\IZ_{3}$ has the other action action on the new coordinates. We pick either of the above two $\IZ_{3}$ actions and here consider the quotient $Y\cong\IP^{5}[3,3]/\IZ_{3}$.

\subsubsection{The mirror}
Following \cite{Hosono:1994ax,batyrev1995generalized}, the mirror $\widetilde{X}$ of $\IP^{5}[3,3]$ is a complete intersection in the toric variety $\IP_{\Delta^{*}}$ from \cite{Hosono:1994ax}. The Laurent polynomial equations defining this intersection are
\begin{equation}\label{eq:mirrorbicubic1}
\begin{aligned}
x_{1}+x_{2}+x_{3}&\=1~,\\[5pt]
x_{4}+x_{5}+\frac{\varphi}{x_{1}x_{2}x_{3}x_{4}x_{5}}&\=1~.
\end{aligned}
\end{equation}
The Picard-Fuchs operator is a generalised hypergeometric operator annihilating the fundamental period
\begin{equation}
\varpi_{0}(\varphi)\=\fourFthree{1/3}{1/3}{2/3}{2/3}{1}{1}{1}{3^{6}\varphi}~.
\end{equation}
The discriminant is 
\begin{equation}
\Delta=1-729\varphi~.
\end{equation}
The equations \eqref{eq:mirrorbicubic1} have a $\IZ_{3}$ symmetry generated by simultaneously applying $x_{1}\mapsto x_{2} \mapsto x_{3}\mapsto x_{1}$ and $x_{4}\mapsto x_{5}\mapsto \varphi/(x_{1}x_{2}x_{3}x_{4}x_{5})$. We will once again introduce an extra coordinate $x_{6}$, so that we can write \eqref{eq:mirrorbicubic1} as the intersection of three hypersurfaces
\begin{equation}\label{eq:mirrorbicubic2}
\begin{aligned}
x_{1}+x_{2}+x_{3}&\=1~,\\[5pt]
x_{4}+x_{5}+x_{6}&\=1~,\\[5pt]
x_{1}x_{2}x_{3}x_{4}x_{5}x_{6}&\=\varphi~.
\end{aligned}
\end{equation}
Now our $\IZ_{3}$ action is generated by the simultaneous $x_{1}\mapsto x_{2}\mapsto x_{3}\mapsto x_{1}$, $x_{4}\mapsto x_{5}\mapsto x_{6}\mapsto x_{4}$. We take $X\cong\widetilde{X}/\IZ_{3}$, and turn to studying fixed points. At a fixed point we must have $x_{1}=x_{2}=x_{3}$ and $x_{4}=x_{5}=x_{6}$. From the first two equations in \eqref{eq:mirrorbicubic2}, we must have all $x_{i}=1/3$. The third equation in \eqref{eq:mirrorbicubic2} can only be solved if $\varphi=3^{-6}$, which solves $\Delta=0$. 

This means that when we compute the genus one free energy, we take $|G_{1}|=3$ in \eqref{eq:Bgenus1}.

\subsubsection{Abelian D4D2D0 indices}\label{sect:Bicubicresults}
The topological data of the quotient manifold $\IP^{5}[3,3]/\IZ_{3}$ is
\begin{equation}
\kappa_{111}\=3~,\qquad c_{2}\=18~,\qquad \chi=-48~.
\end{equation}
One computes
\begin{equation}\begin{aligned}
\chi_{\mathcal{D}_{1}}&\=\frac{1}{6}\kappa_{111}+\frac{1}{12}c_{2}\=2~,\qquad \chi(\mathcal{D}_{1})\=\kappa_{111}+c_{2}=21~, \\[5pt]
&\Delta_{0}\= \frac{\chi(\mathcal{D}_{1})}{24}=\frac{7}{8}~,\\[5pt]
&\Delta_{1}\= \frac{\chi(\mathcal{D}_{1})}{24}-\text{Fr}\left(\frac{1^{2}}{2\kappa_{111}}+\frac{1}{2}\right)\=\frac{5}{24}~.
\end{aligned}\end{equation}
From \eqref{eq:TorsionZm}, $\text{Tors}\left(H^{2}\left(\IP^{5}[3,3]/\IZ_{3}\right)\right)\cong\IZ_{3}$. Then \eqref{eq:TheoremF1} provides
\begin{equation}\label{eq:BicubicFirst}
\Omega_{1,0}\left(\frac{7}{8}\right)\=-18~.
\end{equation}
Since $\kappa_{111}=3$, the VVMF $h_{1,\mu}$ generating Abelian D4D2D0 invariants will have rank 2. Each component $h_{1,0}$ and $h_{1,1}=h_{1,2}$ has a single polar term, and we must compute two terms in either $\text{q}$-series in order to fix the entire $h_{1,\mu}$. We have no way of rigorously doing this.

Anyway, note that to compute $\Omega_{1,0}\left(\frac{7}{8}\right)$ using \eqref{eq:TheoremF2r}, $k_{\text{min}}=2$. If we attempt to use $k=2$, then \eqref{eq:TheoremF2r} provides $\Omega_{1,0}\left(\frac{7}{8}\right)=n^{(10)}_{6}/5$. If we take the suggested GV invariant $n^{(10)}_{6}=-90$ as input data then we can solve the holomorphic anomaly equations up to genus $g=11$, and are left in need of extra data still to solve at $g=12$. 

We do not have sufficiently high genus GV invariants to compute any other terms in $h_{1,\mu}$ using \eqref{eq:TheoremF2r} with $k\geq k_{\text{min}}$. But we can try to compute two entries using $k=k_{\text{min}}-1$. We tentatively find
\begin{equation}\label{eq:BicubicCheating}
\begin{aligned}
\text{using }k=2:&\quad\Omega_{1,0}\left(\frac{7}{8}-1\right)\;\Qeq\;-\frac{1}{4}\left(n^{(9)}_{6}+18n^{(10)}_{6}\right)\=1566\=9*174~,\\[5pt]
\text{using }k=1:&\quad\Omega_{1,1}\left(\frac{5}{24}\right)\+\;\;\Qeq\; -\frac{1}{2}n^{(5)}_{4}\=486\=9*54~.
\end{aligned}
\end{equation}
Moreover, $\Omega_{1,1}\left(\frac{5}{24}\right)=486$ agrees with the modified AGMP ansatz \eqref{eq:AGMPtorsion}. We have had to use the value of $n^{(10)}_{6}$ previously predicted by using \eqref{eq:TheoremF2r}.  

After fixing $\Omega_{1,0}\left(\frac{7}{8}\right)$ using \eqref{eq:BicubicFirst} we only had to fix a single additional entry. We find that the two highly speculative computations in \eqref{eq:BicubicCheating} are consistent, in that if we assume either one then the other is implied by the resulting expansion for $h_{1,\mu}$. Based on this quasimiracle we aggressively conjecture that
\begin{equation}\label{eq:BicubicZ3abelian}
\begin{aligned}
h_{1,\mu}&\;\Qeq\;\frac{9}{\eta(\tau)^{30}}\left(\left(\frac{E_{6}(\tau)^{2}}{24}+\frac{E_{4}(\tau)^{3}}{8}\right)\theta^{(3)}_{\mu}(\tau)+2E_{4}(\tau)E_{6}(\tau)D\left[\theta^{(3)}_{\mu}(\tau)\right]\right)~,\\[10pt]
h_{1,0}(\tau)&\;\Qeq\;9\text{q}^{-7/8}\left(\underline{\,-2\,}+174\text{q}+119052\text{q}^{2}+5318746\text{q}^{3}+117995460\text{q}^{4}+\,...\right)~,\\[5pt]
h_{1,1}(\tau)&\;\Qeq\;=9\text{q}^{-5/24}\left(\underline{\,54\,}+26838\text{q}+1669194\text{q}^{2}+44349552\text{q}^{3}+738224496\text{q}^{4}+\,...\right)~.
\end{aligned}
\end{equation}
We have not displayed $h_{1,2}(\tau)=h_{1,1}(\tau)$.

We do not find any other entries of \eqref{eq:BicubicZ3abelian} to be computed correctly using \eqref{eq:TheoremF2r} with $k<k_{\text{min}}$ besides \eqref{eq:BicubicCheating}. By recomputing $\Omega_{1,1}\left(\frac{5}{24}\right)$ using $k=k_{\text{min}}=2$ we predict $n^{(12)}_{7}=-2916$, which provides new input so that we can solve for the topological string partition function at genus 12. The new GV invariants that we are able to compute using this input from \eqref{eq:BicubicZ3abelian} are listed in~\tref{tab:BicubicGenus12}.

\begin{minipage}{.45\linewidth}
\begin{table}[H]
\footnotesize{\begin{tabular}{| c || l | l | l |}
\hline \!\!\!$k$\!\!\! &\hfil Speculative $n^{(10)}_{k}\phantom{\bigg|}$ &\hfil Speculative $n^{(11)}_{k}\phantom{\bigg|}$ &\hfil Speculative $n^{(12)}_{k}\phantom{\bigg|}$   \\\hline 
\!\!\!1\!\!\! &\!\!\!\!\! 0 &\!\!\!\!\! 0 &\!\!\!\!\! 0 \\
\!\!\!2\!\!\! &\!\!\!\!\! 0 &\!\!\!\!\! 0 &\!\!\!\!\! 0 \\
\!\!\!3\!\!\! &\!\!\!\!\! 0 &\!\!\!\!\! 0 &\!\!\!\!\! 0 \\
\!\!\!4\!\!\! &\!\!\!\!\! 0 &\!\!\!\!\! 0 &\!\!\!\!\! 0 \\
\!\!\!5\!\!\! &\!\!\!\!\! 0 &\!\!\!\!\! 0 &\!\!\!\!\! 0 \\
\!\!\!6\!\!\! &\!\!\!\!\! -90 &\!\!\!\!\! 0 &\!\!\!\!\! 0 \\
\!\!\!7\!\!\! &\!\!\!\!\! -84613626 &\!\!\!\!\! 1275750 &\!\!\!\!\! -2916 \\
\!\!\!8\!\!\! &\!\!\!\!\! 9171367649964 &\!\!\!\!\! -91871426772 &\!\!\!\!\! 4618959012 \\
\!\!\!9\!\!\! &\!\!\!\!\! 13003585821138309318 &\!\!\!\!\! 153778390444153740 &\!\!\!\!\! 365228012942442 \\
\!\!\!10\!\!\! &\!\!\!\!\! 1415928995638950548200644 &\!\!\!\!\! 76752945272483301322845 &\!\!\!\!\! 2017959285916872796176 \\
\!\!\!11\!\!\! &\!\!\!\!\! 44581192598784631364029390923 &\!\!\!\!\! 6134256441374176923994189962 &\!\!\!\!\! 508977170467926887932328988 \\
\!\!\!12\!\!\! &\!\!\!\!\! 669514061987901745772649256357758 &\!\!\!\!\! 
180383720682931916283051956277939 &\!\!\!\!\! 32610350325425780729130502024368 
\\
\!\!\!13\!\!\! &\!\!\!\!\! 6077335091750164066412936523391376463\!\!\! &\!\!\!\!\! 2776349986355483474800145118100488378\!\!\! &\!\!\!\!\! 906404671066163665506921981409451202\!\!\! \\
\hline   
\end{tabular} 
}
\end{table}
\end{minipage}
\vskip-5pt
\begin{table}[H]
\begin{center}
\capt{\textwidth}{tab:BicubicGenus12}{Speculative GV invariants for the $\IZ_{3}$ quotient of the bicubic, assuming validity~of~\eqref{eq:TheoremF2r}~and~\eqref{eq:BicubicZ3abelian}.}
\end{center}
\end{table}
With further applications of \eqref{eq:TheoremF2r} to recompute entries of \eqref{eq:BicubicZ3abelian}, we can obtain further predictions
\begin{equation}
\begin{gathered}
n^{(19)}_{9}\Qeq-198~,\qquad n^{(19)}_{10}\Qeq-12478031532~,\\[5pt] n^{(20)}_{10}\Qeq 286364808~,\qquad n^{(21)}_{10}\Qeq-3163860~,\qquad n^{(22)}_{10}\Qeq 6318~.
\end{gathered}
\end{equation}

\subsection{$\IZ_{7}$ quotient of R{\o}dland's pfaffian threefold, the (1,8) model}
Let $A_{i},\,0\leq i\leq6,$ be a set of seven antisymmetric matrices. With $x_{i}$ giving homogeneous coordinates on $\IP^{6}$, introduce the matrix $N_{A}=\sum_{i=0}^{6}x_{i}A_{i}$. The subvariety of $\IP^{6}$ defined as the locus of $x_{i}$ where $\text{Rank}(N_{A})\leq4$ is, for generic $A_{i}$, a smooth Calabi-Yau threefold which in this section we denote $\widetilde{Y}$. This construction is due to R{\o}dland, who also provided a candidate mirror construction. Note that $\widetilde{Y}$ is not a complete intersection (instead it is the noncomplete intersection of the vanishing loci of $N_{A}$'s $6\times6$ Pfaffians), and so the machinery of toric geometry is not readily available for constructing the mirror.

\subsubsection{The quotient and the mirror}

There is a non-freely acting $\IZ_{7}\times\IZ_{7}$ symmetry of $\IP^{6}$, generated by $x_{i}\mapsto x_{i+1}$ and $x_{i}\mapsto w^{i}x_{i}$ with $w$ a seventh root of unity. The tables of \cite{Candelas:2016fdy} include a $\IZ_{7}$ quotient of $\widetilde{Y}$, which in this section will be labelled by $Y$. Ideally, we would display a choice of matrices $A_{i}$ such that the locus $\text{Rank}(N_{A})\leq4$ is both smooth and admits one of the $\IZ_{7}$ as a freely acting symmetry. We opt for a more circuitous argument that such a choice exists, that works with a choice such that $\IZ_{7}\times\IZ_{7}$ acts freely but the $\text{Rank}(N_{A})\leq4$ locus is not smooth. Having done this, we will go on to show that R{\o}dland's proposal for the mirror is compatible with us taking the $\IZ_{7}$ quotient. 

\underline{Existence of a smooth quotient $Y$:}

Following \cite{rodland2000pfaffian}, let $E\cong\IC^{7}$ and consider $E \wedge E$, which can be identified with the set of $7\times7$ skew-symmetric matrices $N$. Each $N$ has 21 independent components that, working projectively, furnish $\IP(E\wedge E)\cong\IP^{20}$. R{\o}dland considers the locus in $\IP(E\wedge E)$ where $N$ has rank $\leq4$, which defines the Pfaffian variety, the non-complete intersection in $\IP^{20}$ given by the vanishing of the Pfaffians of the seven $6\times6$ diagonal minors of $N$. Intersecting the Pfaffian variety with a generic 6-plane $\mathbf{P}^{6}$ in $\IP^{20}$ provides a Calabi-Yau threefold $\widetilde{Y}$, which \cite{rodland2000pfaffian} denoted $X_{A}$.

R{\o}dland identifies an $R\cong\IZ_{7}\times\IZ_{7}$ action on this $\IP(E\wedge E)$. If\footnote{He we use lowercase $e$ to align with \cite{rodland2000pfaffian}, these are not cohomology generators as in the rest of this paper.} $e_{j}$ is a basis of $E$, the $R$ action has generators $\sigma:e_{j}\mapsto e_{j+1}$ and $\tau:e_{j}\mapsto \exp(\frac{2\pi\ii}{7}j)e_{j}$. $H_{\sigma}$ and $H_{\tau}$ will denote the subgroups generated by $\sigma$ and $\tau$. The different choices of linear embeddings of the $\IP^{6}$ in $\IP^{20}$  constitute the complex structure moduli space of $\widetilde{Y}$, and it was shown that $h^{2,1}(\widetilde{Y})=50$. Within the 50 dimensional space of complex structures of $\widetilde{Y}$ is a $\IP^{2}$ subvariety where the $R$ actions on $\IP(E\wedge E)$ and $\IP^{6}$ are compatible, so that $R$ generically acts freely on $\widetilde{Y}$. On this $\IP^{2}$, $\widetilde{Y}$ has 49 double points and is not smooth. $R$ acts freely and transitively on the set of 49 double points. We will use $\widetilde{Y}_{\text{sing}}$ to display this singular family fibred over the $\IP^{2}$. We can take the quotient by $H_{\sigma}$ to obtain a variety $Y_{\text{sing}}$.

We will argue that $Y_{\text{sing}}$ admits a smooth deformation by appealing to a theorem of Friedman \cite{Friedman1986SimultaneousRO}, which was proven by alternative topological methods in \cite{rollenske2009smoothing} (Theorem 1.2 therein). This states that a Calabi-Yau threefold $X$ with ordinary double point singularities admits a smoothing iff there exists a relation $\sum_{k}\delta_{k}[C_{k}]=0$ in $H_{2}(X^{+},\IR)$ with each $\delta_{k}\neq0$, where $X^{+}$ is a small resolution of $X$ obtained by replacing each ODP singularity $p_{k}$ with a smooth rational (-1,-1) curve $C_{k}$ (such an $X^{+}$ exists). We therefore have a small resolution\footnote{It is indeed known that there is a $h^{1,1}=2$ CY3 resolution $\widetilde{Y}^{+}$ of $\widetilde{Y}_{\text{sing}}$, as this also appears in the tables of \cite{Candelas:2016fdy}.} $\widetilde{Y}^{+}$ with some relation between the homology classes of the exceptional curves, but these curves are exchanged freely and transitively by the $R$ action and so we also have a free $R$ action on $\widetilde{Y}^{+}$. If we quotient $\widetilde{Y}^{+}$ by $\IZ_{7}$ then we get $Y_{\text{sing}}^{+}$, which is a small resolution of $Y_{\text{sing}}$. The homology relations for $\widetilde{Y}^{+}$ lead to homology relations between the exceptional curves in $Y^{+}$, which therefore by Friedman's theorem admits a smoothing $Y$. Note that this argument breaks down if we quotient by $\IZ_{7}\times\IZ_{7}$ because then there is only one exceptional curve, and so the necessary nontrivial homology relation cannot be found in $H_{2}(\widetilde{Y}^{+}/(\IZ_{7}\times\IZ_{7}),\IR)$. 

Rather than count the number of $\IZ_{7}$ invariant deformations of $\widetilde{Y}_{\text{sing}}$ to obtain $h^{2,1}(Y)$, we resort to using the Euler characteristic: $h^{2,1}(Y)=h^{1,1}(Y)-(\chi(\widetilde{Y})/7)/2=8$.

\underline{Towards a mirror construction:}

R{\o}dland \cite{rodland2000pfaffian} goes on to argue that a mirror $\widetilde{X}$ to $\widetilde{Y}$ can be constructed by going to a line in this $\IP^{2}$ where $\widetilde{Y}^{sing}$ acquires seven fixed points under $H_{\tau}$, then quotienting by $H_{\tau}$, and then resolving the singularities. We shall explain this, and emphasise that R{\o}dland's mirror construction is compatible with the $H_{\sigma}$ quotient, and so propose that the mirror $X$ of $Y$ is $\widetilde{X}/\IZ_{7}$. We leave it open to construct the mirror of the quotient more carefully, perhaps along the lines of \cite{Batyrev:1998kx}. 

In R{\o}dland's construction $[y_{1}:y_{2}:y_{3}]$ are homogeneous coordinates for the $\IP^{2}$ parameter spaces of $\widetilde{Y}_{\text{sing}}$ and $y_{3}=0$ gives a line (with parameter $y=y_{2}/y_{1}$) on which the seven fixed points of $H_{\tau}$ in $\mathbf{P}^{6}$ lie on $\widetilde{Y}$, and these fixed points are also ODP singularities. Note that for generic points on the line $y_{3}=0$, $H_{\sigma}$ acts freely. On this line $\widetilde{Y}_{\text{sing}}$ has (49+7) double points. The quotient $\widetilde{Y}_{\text{sing}}/H_{\tau}$ has 14 singularities: 7 double points (the image of the previous 49) and 7 further orbifold singularities. These are resolved to produce $\widetilde{X}$. Now, if we quotient by $H_{\sigma}$ we are left with two singularities, one ODP and one $\IZ_{7}$ orbifold. Resolving these produces $\widetilde{X}/\IZ_{7}$, which we take to be our $X$.

\underline{Higher genus considerations: }

The importance of identifying $X$ is that we must work out what kind of Lens space shrinks at the conifold singularities in order to proceed with higher genus computations. For R{\o}dland's model, $\widetilde{Y}_{\text{sing}}$ acquires seven more ordinary double points when the line $y_{3}=0$ inside $\IP^{2}$ intersects one of the other lines where $H_{\sigma}$ does not act freely, and these seven singular points are fixed under the $H_{\sigma}$ action. This was found in \cite{rodland2000pfaffian} to occur for $y^{21}-289y^{14}-58y^{7}+1=0$. Now, after taking the $H_{\tau}$ quotient in order to construct $\widetilde{X}$, these seven singularities are identified into a single singularity so that $\widetilde{X}$ has one ODP fixed under the $H_{\sigma}$ action when $\Delta(y^7)\equiv y^{21}-289y^{14}-58y^{7}+1=0$. This means that $X$ has a $\IZ_{7}$-orbifold singularity for these values of $y$, and so our genus one free energy \eqref{eq:Bgenus1} will have $|G_{1}|=7$.

The Picard-Fuchs operator for $\widetilde{X}$ and $X$ is given in \cite{rodland2000pfaffian}, the mirror complex structure coordinate is $\varphi=y^{7}$. This has singularities at the three hyperconifold points where $\Delta(\varphi)=0$, an apparent singularity at $\varphi=1$, and two MUM points at $\varphi=0$ and $\varphi=\infty$. The higher genus problem for this operator, in the context of $\widetilde{Y}$, was addressed by Hosono and Konishi in \cite{Hosono:2007vf}, which provides the necessary practical methods (which we make use of) to impose regularity at the apparent singularity and impose the conifold gap condition when the discriminant locus $\Delta$ is irreducible. Where $\widetilde{X}$ is concerned, the additional MUM point at infinity is associated with the Pfaffian variety's derived equivalent partner, which is the intersection of seven degree-1 hypersurfaces in the Grassmannian $\text{Gr}(2,7)$. We will assume that in our computations the MUM point $\varphi=\infty$ is similarly associated to a $\IZ_{7}$ quotient of this intersection in $\text{Gr}(2,7)$. Our genus 1 computation, independently of this assumption, produces the same $c_{2}$ as our assumption in the expansion about $\varphi=\infty$. This additional geometry provides us more data to constrain the holomorphic ambiguity: the constant term in the expansion about infinity and also Castelnuovo vanishing of the GV invariants read off from the expansions about infinity (these are in addition to those that we have at $\varphi=0$). 

Once again following \cite{rodland2000pfaffian}, take $\IP(E^{\vee}\wedge E^{\vee})\cong\IP^{20}$. Inside this lives the Grassmannian $\text{Gr}(2,7)$, and also the $\IP^{13}$ dual to our original $\IP^{6}$. The intersection $\text{Gr}(2,7)\cap\IP^{13}$ is another Calabi-Yau threefold, proved to be derived equivalent to $\widetilde{Y}$ in \cite{borisov2009pfaffian} (with a different proof in \cite{Addington:2014sla} that follows more closely the ``physics proof" of \cite{Hori:2006dk}). The group $R\cong H_{\sigma}\times H_{\tau}\cong\IZ_{7}\times\IZ_{7}$ also acts on $\IP(E^{\vee}\wedge E^{\vee})$ and $\IP^{13}$, so we have again a freely acting smooth $\IZ_{7}$ quotient $\left(\text{Gr}(2,7)\cap\IP^{13}\right)/\IZ_{7}$. It may be interesting to seek a proof of $\text{D}^{b}\text{coh}\left[\left(\text{Gr}(2,7)\cap\IP^{13}\right)/\IZ_{7}\right]\cong\text{D}^{b}\text{coh}\left[\widetilde{Y}/\IZ_{7}\right]$~.

\subsubsection{Abelian D4D2D0 indices}\label{sect:Rodlandresults}
The topological data for this example is 
\begin{equation}
\kappa_{111}\=2~,\qquad c_{2}\=8~,\qquad \chi=-14~.
\end{equation}
We go on to compute
\begin{equation}\begin{aligned}
\chi_{\mathcal{D}_{1}}&\=\frac{1}{6}\kappa_{111}+\frac{1}{12}c_{2}\=1~,\qquad \chi(\mathcal{D}_{1})\=\kappa_{111}+c_{2}=10~, \\[5pt]
&\Delta_{0}\= \frac{\chi(\mathcal{D}_{1})}{24}=\frac{5}{12}~,\\[5pt]
&\Delta_{1}\= \frac{\chi(\mathcal{D}_{1})}{24}-\text{Fr}\left(\frac{1^{2}}{2\kappa_{111}}+\frac{1}{2}\right)\=-\frac{1}{3}~.
\end{aligned}\end{equation}
From \eqref{eq:TorsionZm}, $\text{Tors}\left(H^{2}\right)\cong\IZ_{7}$. Then \eqref{eq:TheoremF1} provides
\begin{equation}\label{eq:RodlandFirst}
\Omega_{1,0}\left(\frac{5}{12}\right)\=49~.
\end{equation}
Since $\kappa_{111}=2$, the VVMF $h_{1,\mu}$ will have rank two. However, based on the above calculations $h_{1,1}$ has no polar terms and $h_{1,0}$ has one polar term. Therefore, the problem is one-dimensional and so completely solved by the above application of the theorem \eqref{eq:TheoremF1}. We obtain the result
\begin{equation}\label{eq:RPfaffianZ7abelian}
\begin{aligned}
h_{1,\mu}(\tau)&\=\frac{49}{\eta(\tau)^{16}}\left(-\frac{E_{6}(\tau)}{12}\theta^{(2)}_{\mu}(\tau)-2E_{4}(\tau)D\left[\theta^{(2)}_{\mu}(\tau)\right]\right)~,\\[10pt]
h_{1,0}(\tau)&\=49\text{q}^{-5/12}\left(\underline{\,1\,}+136\text{q}+2081\text{q}^{2}+18152\text{q}^{3}+117028\text{q}^{4}+\,...\right)~,\\[5pt]
h_{1,1}(\tau)&\=49\text{q}^{1/3}\left(56+1136\text{q}+10912\text{q}^{2}+75072\text{q}^{3}+414304 \text{q}^{4}+\,...\right)~.
\end{aligned}
\end{equation}

\subsubsection{Interplay with GV invariants}
We are able to compute GV invariants for this model up to genus 5, as tabulated in \tref{tab:RodlandZ7PfaffianNumbers}. Unfortunately, this is not a sufficiently high-genus set of invariants for us to make any checks of \eqref{eq:TheoremF2r} with $k\geq k_{min}$. We can attempt to compute $\Omega_{1,1}\left(-\frac{1}{3}\right)$ with $k=k_{\text{min}-1}=1$ but this gives the incorrect result $2870$, which differs to $2744=49*56$.

If we attempt to use \eqref{eq:TheoremF2r} to compute $\Omega_{1,0}\left(\frac{5}{12}\right)$ and $\Omega_{1,1}\left(-\frac{1}{3}\right)$, in both cases using $k=k_{\text{min}}=2$, then \eqref{eq:RPfaffianZ7abelian} predicts
\begin{equation}
n^{(7)}_{4}\=147~,\qquad n^{(9)}_{5}\=-10976~.
\end{equation}
We now attempt something extremely dubious, and use \eqref{eq:TheoremF2r} with $k=k_{\text{min}}-1=2$ to attempt to compute $\Omega_{1,0}\left(\frac{5}{12}-1\right)$. Together with \eqref{eq:RPfaffianZ7abelian} this leads us to
\begin{equation}
49*136\=\Omega_{1,0}\left(\frac{5}{12}-1\right)\;\QQQeq\;-\frac{1}{2}\left(n^{(6)}_{4}+12n^{(7)}_{4}\right)\implies n^{(6)}_{4}\;\QQQeq\;-15092~.
\end{equation}
This is quite possibly the wrong value for $n^{(6)}_{4}$. We required one additional GV invariant at genus 6 (beyond those predicted to be zero by the Castelnuovo bound) in order to solve for the topological string free energy. This extremely speculative result could be used as input data in this way, but the resulting integers read off may not the correct GV invariants, as far as we know presently.

We can use \eqref{eq:TheoremF2r} to compute $\Omega_{1,0}\left(\frac{5}{12}\right)$ and $\Omega_{1,0}\left(\frac{5}{12}-1\right)$ using respectively $k=k_{\text{min}}+1=3$ and $k=k_{\text{min}}=2$. Then \eqref{eq:RPfaffianZ7abelian} predicts
\begin{equation}
n^{(12)}_{6}\=-48216~,\qquad n^{(13)}_{6}\=343~.
\end{equation}

If we attempt to use \eqref{eq:TheoremF2r} to compute $\Omega_{1,1}\left(-\frac{1}{3}\right)$, $\Omega_{1,1}\left(-\frac{1}{3}-1\right)$, and $\Omega_{1,1}\left(-\frac{1}{3}-2\right)$ with respectively $k=k_{\text{min}}+1=3$, $k=k_{\text{min}}=3$, and $k=k_{\text{min}}=3$, then we are led to
\begin{equation}
n^{(14)}_{7}\=26182268~,\qquad n^{(15)}_{7}\=-1186192~,\qquad n^{(16)}_{7}\=24696~.
\end{equation}

\subsection{$\IZ_{5}$ quotient of Hosono-Takagi's double quintic symmetroid threefold, the (1,6) model}
\subsubsection{Some geometry}\label{sect:HTgeometry}
Hosono and Takagi have considered a number of different threefolds related by quotient maps, mirror symmetry, and derived equivalence. We identify a $\IZ_{5}$ quotient of one of their geometries, previously undiscussed to our knowledge. For this we compute Abelian D4D2D0 indices. Here we briefly outline the constructions in \cite{Hosono:2011np} (only changing some notations so that we align with our own conventions throughout this paper). They go on to further study the mirror geometry in \cite{Hosono:2012hc}, prove their conjectured derived equivalence in \cite{hosono2016double}, and study infinite order birational automorphism groups in \cite{hosono2018movable}.

To begin with, one has the Reye Congruence Calabi-Yau threefold $\text{R}$, which can be realised as the following $\IZ_{2}$ quotient of a complete intersection Calabi-Yau threefold in $\IP^{4}\times\IP^{4}$ with hodge numbers $(2,52)$.
\begin{equation}
\text{R}\;\;\cong\;\;\cicy{\IP^{4}\\\IP^{4}}{1&1&1&1&1\\1&1&1&1&1}_{/\IZ_{2}}^{1,26}~.
\end{equation}
The mirror of the complete intersection of five degree (1,1) hypersurfaces in $\IP^{4}\times\IP^{4}$ is obtained using the Batyrev-Borisov mirror construction for complete intersections in toric varieties \cite{Batyrev:1994pg}. This produces a geometry with $(h^{1,1},h^{2,1})=(52,2)$, realised as a complete intersection in the appropriate toric variety $\IP_{\nabla^{*}}$. The mirror Reye Congruence $\MR$ is then constructed as the $\IZ_{2}$ quotient of this (52,2) geometry, with complex structure parameters set to the locus where the threefold has a $\IZ_{2}$ symmetry. With $U_{i},\,V_{i},\,1\leq i\leq4$ giving coordinates on a dense algebraic torus $(\IC^{*})^{8}$ in $\IP_{\nabla^{*}}$, the geometry $\MR$ is birational to the  $\IZ_{2}$ quotient of the locus
\begin{equation}
U_{i}+V_{i}\=1~,\qquad \frac{\varphi}{U_{1}U_{2}U_{3}U_{4}}+\frac{\varphi}{V_{1}V_{2}V_{3}V_{4}}\=1~,
\end{equation}
with the $\IZ_{2}$ action being $U_{i}\leftrightarrow V_{i}$. $\varphi$ is the complex structure parameter, and the fundamental period reads 
\begin{equation}
\varpi_{0}(\varphi)\=\sum_{n=0}^{\infty}\sum_{m_{1}+m_{2}=n}\left(\frac{(m_{1}+m_{2})!}{m_{1}!m_{2}!}\right)^{5}\varphi^{n}~.
\end{equation}
Equation (2.9) of \cite{Hosono:2011np} provides the Picard-Fuchs operator that annihilates $\varpi_{0}$. Remarkably, in addition to the MUM point at $\varphi=0$ from which BPS expansions for $\text{R}$ can be performed, there is a MUM point at $\varphi=\infty$. It was argued in \cite{Hosono:2011np} that BPS expansions about this additional MUM point gave GV invariants for a geometry we will denote\footnote{This was denoted $Y$ in \cite{Hosono:2011np,hosono2016double}. We are reserving $Y$ for the threefolds that we compute D4D2D0 indices for, so have forced ourselves to make this unpleasant change.} $\widetilde{Y}$. $\widetilde{Y}$ is the double cover of $H$, where
\begin{equation}
H=\left\{[y_{1}:y_{2}:y_{3}:y_{4}:y_{5}]\in\IP^{4}\;|\;\text{Det}\left[\sum_{i=1}^{5}y_{i}A_{i}\right]=0\right\}~,
\end{equation}
branched along a genus 26 curve $C$. The $A_{i}$ are suitably generic $5\times5$ symmetric matrices, so that $H$ is the locus in $\IP^{4}$ where $\sum y_{i}A_{i}$ has rank$\leq4$. The curve $C$ is the rank$\leq3$ locus, along which $H$ has an $A_{1}$ type singularity. $\widetilde{Y}$ is smooth and simply connected. It was proven in \cite{Hori:2016txh} that $\text{R}$ and $\widetilde{Y}$ are derived equivalent.

We are not able to compute D4D2D0 indices for any of the geometries just discussed. It is worth noting that a quotient geometry $\text{R}/\IZ_{5}$ appears in the tables of \cite{Candelas:2016fdy} (page 28).

There is a $\IZ_{5}$ symmetry of $\MR$. If we introduce additional coordinates $U_{5},\,V_{5}$ defined by $U_{1}U_{2}U_{3}U_{4}U_{5}=V_{1}V_{2}V_{3}V_{4}V_{5}=\varphi$ then this symmetry is generated by simultaneously effecting $U_{i}\mapsto U_{i+1\text{ mod }5}$, $V_{i}\mapsto V_{i+1\text{ mod }5}$. This is freely acting for $\varphi$ not on the discriminant locus. One can set about performing BPS expansions about $\varphi=0$ to obtain GV invariants for $\text{R}/\IZ_{5}$, and one is left to wonder what to associate to $\varphi=\infty$. Notice that for certain $A_{i}$ the double cover $\widetilde{Y}$ also has a freely acting $\IZ_{5}$ symmetry also. The $\IZ_{5}$ symmetry of $H$ is generated by $y_{i}\mapsto y_{i+1\text{ mod }5}$. Conveniently, Hosono and Takagi provide a set of matrices $A_{i}$ so that 
\begin{equation}
\sum_{i=1}^{5}y_{i}A_{i}\=\begin{pmatrix}
y_{2}&y_{1}&0&0&y_{5}\\
y_{1}&y_{3}&y_{2}&0&0\\
0&y_{2}&y_{4}&y_{3}&0\\
0&0&y_{3}&y_{5}&y_{4}\\
y_{5}&0&0&y_{4}&y_{1}
\end{pmatrix}
\end{equation}
provides a smooth $\widetilde{Y}$, and the rank$\leq4$ locus is symmetric under the $y_{i}\mapsto y_{i+1\text{ mod }5}$. We will use  $Y$ to denote the quotient $\widetilde{Y}/\IZ_{5}$. We conjecture that $Y$ is derived equivalent to $\text{R}/\IZ_{5}$. This is supported by the genus 0 BPS expansion (as the GV invariants divide by 5) and also the genus 1 expansion (from which we can read off the necessary $c_{2}^{Y}=8=c_{2}^{\widetilde{Y}}/5$). Subject to this assumption, we compute the GV invariants listed in \sref{sect:GVtables} by incorporating boundary conditions from Castelnuovo vanishing and constant terms in expansions about both MUM points.

The Picard-Fuchs equation has an apparent singularity at $\varphi=7/4$. There are hyperconifold singularities where 
\begin{equation}
\Delta_{1}=1-32\varphi\qquad \text{and}\qquad \Delta_{2}=1+11\varphi-\varphi^{2}
\end{equation}
vanish, where respectively there is a shrinking $S^{3}/\IZ_{10}$ and $S^{3}/\IZ_{5}$. This means that in \eqref{eq:Bgenus1} we will have $|G_{1}|=10$ and $|G_{2}|=5$

\subsubsection{Abelian D4D2D0 indices}\label{sect:HTresults}
The topological data for $Y$, obtained by dividing the data for $\widetilde{Y}$ given in \cite{Hosono:2011np} by 5, is
\begin{equation}
\kappa_{111}\=2~,\qquad c_{2}\=8~,\qquad \chi=-10~.
\end{equation}
We go on to compute, much the same as for the Pfaffian quotient in the previous section,
\begin{equation}\begin{aligned}
\chi_{\mathcal{D}_{1}}&\=\frac{1}{6}\kappa_{111}+\frac{1}{12}c_{2}\=1~,\qquad \chi(\mathcal{D}_{1})\=\kappa_{111}+c_{2}=10~, \\[5pt]
&\Delta_{0}\= \frac{\chi(\mathcal{D}_{1})}{24}=\frac{5}{12}~,\\[5pt]
&\Delta_{1}\= \frac{\chi(\mathcal{D}_{1})}{24}-\text{Fr}\left(\frac{1^{2}}{2\kappa_{111}}+\frac{1}{2}\right)=-\frac{1}{3}~.
\end{aligned}\end{equation}
From \eqref{eq:TorsionZm}, $\text{Tors}\left(H^{2}\right)\cong\IZ_{5}$ since $\widetilde{Y}$ is simply connected (a result of \cite{Hosono:2011np}). Then \eqref{eq:TheoremF1} provides
\begin{equation}
\Omega_{1,0}\left(\frac{5}{12}\right)\=25~.
\end{equation}
We arrive at
\begin{equation}\label{eq:HTZ5abelian}
\begin{aligned}
h_{1,\mu}(\tau)&\=\frac{25}{\eta(\tau)^{16}}\left(-\frac{E_{6}(\tau)}{12}\theta^{(2)}_{\mu}(\tau)-2E_{4}(\tau)D\left[\theta^{(2)}_{\mu}(\tau)\right]\right)~,\\[10pt]
h_{1,0}(\tau)&\=25\text{q}^{-5/12}\left(\underline{\,1\,}+136\text{q}+2081\text{q}^{2}+18152\text{q}^{3}+117028\text{q}^{4}+\,...\right)~,\\[5pt]
h_{1,1}(\tau)&\=25\text{q}^{1/3}\left(56+1136\text{q}+10912\text{q}^{2}+75072\text{q}^{3}+414304 \text{q}^{4}+\,...\right)~.
\end{aligned}
\end{equation}
Up to an overall rational factor $\frac{25}{49}$ this is the same as the result we obtained for the (1,8) model \eqref{eq:RPfaffianZ7abelian}. Therefore the ratio of Abelian MSW invariants for either of these models, with the same charges,  is a constant value as $\mu$ and $\hat{q}_{0}$ are varied. The theorem \eqref{eq:TheoremF2r} and the MNOP conjecture would then imply infinitely many identities relating GV invariants of either model. This may be of wider interest, although we do not at this time have any use for this surprising relation.

\subsubsection{Interplay with GV invariants}
This analysis proceeds very similarly to that of the (1,8) model. Here we are again unable to make any nontrivial tests of modularity or increase the maximal genus we can compute GV invariants for. We will content ourselves to compute a few GV invariants assuming modularity.

Attempting to compute $\Omega_{1,1}(-\frac{1}{3})$ with $k-k_{\text{min}}-1=1$ gives the incorrect $1400\neq1570=25*56$.

After computing $\Omega_{1,0}\left(\frac{5}{12}\right)$ and $\Omega_{1,1}\left(-\frac{1}{3}\right)$ using $k=k_{\text{min}}=2$, we arrive at
\begin{equation}
n^{(7)}_{4}\=75~,\qquad n^{(9)}_{5}\=5600~.
\end{equation}

Dubiously using $k=k_{\text{min}}-1$ to compute $\Omega_{1,0}\left(\frac{5}{12}-1\right)$, and incorporating the above value for $n^{(7)}_{4}$, leads to 
\begin{equation}
n^{(6)}_{4}\;\QQQeq\;-7700
\end{equation}
and once again we only needed a correct value of $n^{(6)}_{4}$ in order to extend the results of \sref{sect:GVtables} to genus 6. 

After using \eqref{eq:TheoremF2r} to compute $\Omega_{1,0}\left(\frac{5}{12}\right)$ and $\Omega_{1,0}\left(\frac{5}{12}-1\right)$, with respectively $k=k_{\text{min}}+1=3$ and $k=k_{\text{min}}=2$, \eqref{eq:HTZ5abelian} predicts
\begin{equation}
n^{(12)}_{6}\=-24600~,\qquad n^{(13)}_{6}\=175~.
\end{equation}

We use \eqref{eq:TheoremF2r} to compute $\Omega_{1,1}\left(-\frac{1}{3}\right)$, $\Omega_{1,1}\left(-\frac{1}{3}-1\right)$, and $\Omega_{1,1}\left(-\frac{1}{3}-2\right)$ with respectively $k=k_{\text{min}}+1=3$, $k=k_{\text{min}}=3$, and $k=k_{\text{min}}=3$. Then \eqref{eq:HTZ5abelian} provides
\begin{equation}
n^{(14)}_{7}\=13361000~,\qquad n^{(15)}_{7}\=-605200~,\qquad n^{(16)}_{7}\=12600~.
\end{equation}

\newpage

\section{Multiparameter Examples}\label{sect:Multiparameter}
Once again, we underline all polar terms. For these multiparameter models we have nothing to say about higher genus GV invariants. Assuming modularity, all of these cases are solved by computing their single polar terms. As we discussed following \eqref{eq:TheoremF1}, we propose to apply the result \eqref{eq:TheoremF1} (with $p\mapsto\bm{p}$) in these examples in spite of the fact that only the (2,12) model meets the Assumption $(\star)$ of \cite{Alexandrov:2023zjb} (while the rest of the models do not).
\subsection{The (2,29) and (2,20) models}
For both of these models we choose the homology class of our irreducible divisor to be $\bm{p}=(1,0)$. Due to the (not freely acting) $\IZ_{2}$ symmetry of the (2,29) model under the exchange of the two ambient $\IP^{2}$ factors before taking the $\IZ_{3}$ quotient, it does not make a difference for this model if we instead take $\bm{p}=(0,1)$. However, choosing $\bm{p}=(0,1)$ for the (2,20) model does give a genuinely different problem, one that we are unable to solve in this paper. 

For each of these models with this choice of divisor 
\begin{equation}
|\Lambda^{*}/\Lambda|\=|\text{Det}\left[\kappa_{ijk}p^{i}\right]|\=|\text{Det}\left[\kappa_{1jk}\right]|\=1~.
\end{equation}
In light of this there is only a single glue vector, $\bm{\mu}=(0,0)$. As a result the VVMF that generates MSW invariants has rank 1, and so is a standard modular form. This modular form must have weight $-1-\frac{b_{2}}{2}=-2$. The $\text{q}$-series starts at $\text{q}^{-1/2}$, which follows from 
\begin{equation}
\Delta_{(0,0)}=\frac{\chi(\mathcal{D}_{\bm{p}})}{24}\=\frac{\kappa_{ijk}p^{i}p^{j}p^{k}+c_{2,i}p^{i}}{24}\=\frac{\kappa_{111}+c_{2,1}}{24}\=\frac{12}{24}~.
\end{equation}
These considerations fix the modular form up to scale: it must be a multiple of $E_{4}\eta^{-12}$.

We use \eqref{eq:TheoremF1} to compute the first MSW invariant. Since
\begin{equation}
\chi_{\mathcal{D}_{\bm{p}}}\=\frac{1}{6}\kappa_{ijk}p^{i}p^{j}p^{k}+\frac{1}{12}c_{2,i}p^{i}\=1\quad\text{ and }\quad H^{2}(Y,\IZ)_{\text{Torsion}}\cong\IZ_{3}~,
\end{equation}
we find
\begin{equation}
\Omega_{1,(0,0)}\left(\frac{1}{2}\right)\=9~.
\end{equation}
The generating series of MSW invariants for both of the (2,29) and (2,20) models, with these choices of divisor, are the same function. This is
\begin{equation}
\begin{aligned}
h_{1,(0,0)}(\tau)&\=9E_{4}(\tau)\eta(\tau)^{-12}\\[5pt]
&\=9\text{q}^{-1/2}\left(\underline{\,1\,} + 252 \text{q} + 5130 \text{q}^2 + 54760 \text{q}^3 + 419895 \text{q}^4 + 2587788 \text{q}^5 + 
 13630694 \text{q}^6 +\,...\right)~.
\end{aligned}\end{equation}
It may be of interest that this equals $\frac{9}{25\eta(\tau)}$ times the MSW generating function of $\IP^{4}[5]/\IZ_{5}$, given in equation \eqref{eq:QuinticZ5abelian}.

\subsection{The (2,12) model}
We choose to work with the irreducible divisor $\bm{p}=(1,0)$, but the same result would be obtained if we instead chose $(0,1)$. This is the only multiparameter model we study where the divisor $\mathcal{D}_{\bm{p}}$ meets the technical condition $(\star)$ of \cite{Alexandrov:2023zjb}, Appendix A.

Our VVMF will this time have rank 2, because 
\begin{equation}
|\Lambda^{*}/\Lambda|\=|\text{Det}\left[\kappa_{ijk}p^{i}\right]|\=|\text{Det}\left[\kappa_{1jk}\right]|\=2~.
\end{equation}
Our two glue vectors $\bm{\mu}$ are $(0,0)$ and $(0,1)$. From \eqref{eq:Deltamu} we compute
\begin{equation}
\Delta_{(0,0)}\=\frac{11}{24}~,\qquad \Delta_{(0,1)}\=-\frac{7}{24}~.
\end{equation}
The MSW generating function will be of rank two and weight -2. The first component has a single polar term, and the second component will have no polar terms. Since there is only one polar term this problem becomes one-dimensional, and will be solved once we compute the first MSW index. Using \eqref{eq:TheoremF1}, and
\begin{equation}
\chi_{\mathcal{D}_{\bm{p}}}\=1~,\qquad H^{2}(Y,\IZ)_{\text{Torsion}}\cong\IZ_{5}~,
\end{equation}
we have that 
\begin{equation}
\Omega_{1,(0,0)}\left(\frac{11}{24}\right)\=25~.
\end{equation}
The multiplier system (from equation \eqref{eq:MultiplierSystem}) that $h_{1,\bm{\mu}}$ should have is
\begin{equation}
M(T)\=\begin{pmatrix}\text{e}^{13\pi\ii/12}&0\\0&\text{e}^{7\pi\ii/12}\end{pmatrix}~,\qquad
M(S)\=\frac{1}{\sqrt{2}}\begin{pmatrix}-1&-1\\-1&\+1\end{pmatrix}~.
\end{equation} 
It turns out that this is the multiplier system for $\eta(\tau)^{-17}\theta_{\alpha(\bm{\mu})}^{(2)}(\tau)$, where we introduce
\begin{equation}
\alpha\left((0,0)\right)=0~,\qquad \alpha\left((0,1)\right)=1~.
\end{equation}
$\eta(\tau)^{-17}\theta_{\alpha}^{(2)}(\tau)$ transforms as a VVMF with weight -8. Note that multiplying by $E_{4}$ or $E_{6}$, or acting with the Serre derivative $D$, gives a VVMF with a new weight but the same multiplier system. We obtain a weight -2 VVMF with the correct leading term as follows:
\begin{equation}
\begin{aligned}
h_{1,\bm{\mu}}(\tau)&\=\frac{25}{\eta(\tau)^{17}}\left(-\frac{E_{6}(\tau)}{12}\theta^{(2)}_{\alpha(\bm{\mu})}(\tau)-2E_{4}(\tau)D\left[\theta^{(2)}_{\alpha(\bm{\mu})}(\tau)\right]\right)~,\\[10pt]
h_{1,(0,0)}(\tau)&\=25\text{q}^{-11/24}\left(\underline{\,1\,}+ 137 \text{q} + 2219 \text{q}^2 + 20508 \text{q}^3 + 139755 \text{q}^4 + 779254 \text{q}^5+\,...\right)~,\\[5pt]
h_{1,(0,1)}(\tau)&\=25\text{q}^{7/24}\left(56 + 1192 \text{q} + 12160 \text{q}^2 + 88424 \text{q}^3 + 514888 \text{q}^4 + 2564184 \text{q}^5+\,...\right)~.
\end{aligned}
\end{equation}
This is $\frac{1}{\eta(\tau)}$ times the result we obtained for the $\IZ_{5}$ quotient of Hosono-Takagi's double quintic symmetroid threefold.

\subsection{The (3,18) and (3,15) models}
Finally we turn to the (3,18) and (3,15) models. We choose the divisor $\bm{p}=(1,0,0)$. For the (3,18) model we could equally well take $(0,1,0)$ or $(0,0,1)$ and get the same final result. For the (3,15) model we would get the same result if we chose $(0,1,0)$, but $(0,0,1)$ is a different problem altogether which we do not attempt to solve. The rank of the VVMF will be 2, since
\begin{equation}
|\Lambda^{*}/\Lambda|\=|\text{Det}\left[\kappa_{ijk}p^{i}\right]|\=|\text{Det}\left[\kappa_{1jk}\right]|\=2~.
\end{equation}
For representative glue vectors we take $\bm{\mu}$ to be either $(0,0,0)$ or $(0,0,1)$ (this is not a unique choice). From \eqref{eq:Deltamu} we have
\begin{equation}
\Delta_{(0,0,0)}\=\frac{1}{2}~,\qquad \Delta_{(0,0,1)}\=-\frac{1}{4}~.
\end{equation}
Once again, this problem is one-dimensional and solved by computing a single abelian MSW invariant. We have 
\begin{equation}
\chi_{\mathcal{D}_{\bm{p}}}\=1~,\qquad H^{2}(Y,\IZ)_{\text{Torsion}}\cong\IZ_{3}~,
\end{equation}
and so \eqref{eq:TheoremF1} provides
\begin{equation}
\Omega_{1,(0,0,0)}\left(\frac{1}{2}\right)\=9~.
\end{equation}
$h_{1,\bm{\mu}}$ should have weight $-1-\frac{b_{2}}{2}=-\frac{5}{2}$, and multiplier system (from equation \eqref{eq:MultiplierSystem}) 
\begin{equation}
M(T)\=\begin{pmatrix}-1&0\\0&\ii\end{pmatrix}~,\qquad
M(S)\=\frac{\text{e}^{\pi\ii/4}}{\sqrt{2}}\begin{pmatrix}-1&-1\\-1&\+1\end{pmatrix}~.
\end{equation}
This is the multiplier system for $\eta(\tau)^{-18}\theta_{\alpha(\bm{\mu})}^{(2)}(\tau)$, where
\begin{equation}
\alpha\left((0,0,0)\right)=0~,\qquad \alpha\left((0,0,1)\right)=1~.
\end{equation}
$\eta(\tau)^{-18}\theta_{\alpha(\bm{\mu})}^{(2)}(\tau)$ has weight $-17/2$. Similarly to the (2,12) model then, we can obtain
\begin{equation}
\begin{aligned}
h_{1,\bm{\mu}}(\tau)&\=\frac{9}{\eta(\tau)^{18}}\left(-\frac{E_{6}(\tau)}{12}\theta^{(2)}_{\alpha(\bm{\mu})}(\tau)-2E_{4}(\tau)D\left[\theta^{(2)}_{\alpha(\bm{\mu})}(\tau)\right]\right)~,\\[10pt]
h_{1,(0,0,0)}(\tau)&\=9\text{q}^{-11/24}\left(\underline{\,1\,}+ 138 \text{q} + 2358 \text{q}^2 + 23004 \text{q}^3 + 165117 \text{q}^4 + 967374 \text{q}^5+\,...\right)~,\\[5pt]
h_{1,(0,0,1)}(\tau)&\=9\text{q}^{7/24}\left(56 + 1248 \text{q} + 13464 \text{q}^2 + 103136 \text{q}^3 + 631488 \text{q}^4 + 3298752 \text{q}^5+\,...\right)~.
\end{aligned}
\end{equation}
This is $\frac{9}{25\eta(\tau)}$ times the result we obtained for the (2,12) model.

\newpage
\section{Discussion and outlook}
We have provided nine new examples of modular generating functions of Abelian D4D2D0 indices, including five multiparameter cases, four of which remain subject to an assumption we make on the polar term.

Future work should provide a better understanding of the problematic examples discussed above equation \eqref{eq:bad}. These could be studied more closely along the lines of this paper if we had more GV invariants. A lack of available GV invariants, even in spite of the substantial advances documented in \cite{BonnData}, is eventually always a problem and new ways of computing polar terms would hopefully circumvent this.

It was already anticipated by \cite{Alexandrov:2023zjb} that certain multiparameter examples may be interesting in light of the modular bootstrap for elliptic fibrations \cite{Klemm:2012sx,Alim:2012ss,Huang:2015sta}. Some of the multiparameter models that we have considered are quotients of elliptically fibred threefolds, and so this avenue might be fruitfully pursued using our new examples. It may be the case that the modularity associated to the elliptic fibration interacts in an interesting way with the modularity associated to MSW indices. Along similar lines, the instanton numbers for some of the multiparameter geometries we have investigated possess an infinite Coxeter symmetry \cite{Lukas:2022crp,Candelas:2021lkc,Kuusela:2023vgi} that may similarly provide nice interplay.

It remains to study the geometries investigated in this paper in cases with more than one unit of D4 charge, so that the generating function is mock modular, as has been performed in \cite{Alexandrov:2023ltz}. This also would require more GV invariants.

\vfill
\section*{Acknowledgements}
I am very grateful to Johanna Knapp for collaboration, conversation, and facilitating an excellent working environment in Melbourne. I thank Philip Candelas, Xenia de la Ossa, and Pyry Kuusela for collaboration on previous projects and wide-ranging discussions on black holes, elliptic genera, and theta functions.  I thank Emanuel Scheidegger for fruitful discussion and inspiring questions, including during his generous hosting of me at the BICMR. From this stay in China, I thank the organisers of the very interesting conference \textit{Topological Strings on Calabi-Yau Manifolds -- Maths and Physics} where I could present this work, and Xiaoni Tan for logistical assistance. I thank Adam Monteleone for helpful conversations on derived equivalences. I am very pleased to acknowledge pleasant and productive stays during programs at the Matrix Research Institute, Creswick. I am furthermore grateful to Luca Cassia, Mohamed Elmi, Robert Pryor, Shinobu Hosono, and Bryan Wang for a number of interesting conversations around the matters of this work and other topics. Joseph McGovern is supported by a University of Melbourne establishment grant.

\appendix
\section{Tables of GV invariants, computed without assuming MSW modularity}\label{sect:GVtables}
\vskip-10pt
\subsection{$\IZ_{5}$ quotient of the Quintic, $\IP^{4}[5]/\IZ_{5}$}
\vskip-20pt
\begin{minipage}{.45\linewidth}
\begin{table}[H]
\footnotesize{\begin{tabular}{| c || l | l | l |}
 \hline $\phantom{\bigg|}k\phantom{\bigg|}$ &\hfil $n^{(0)}_{k}$ &\hfil $n^{(1)}_{k}$ &\hfil $n^{(2)}_{k}$   \\\hline 
1 & 575 & 750 & 10 \\
2 & 121850 & 749650 & 316180 \\
3 & 63441275 & 996355600 & 1812388645 \\
4 & 48493506000 & 1485713351625 & 6832687291550 \\
5 & 45861177777525 & 2360745222311890 & 21386162464746280 \\
6 & 49649948423604400 & 3905048810312630500 & 60300725772067744370 \\
7 & 59018210114169131850 & 6641344898623706083650 & 
159076086322903496882380 \\
8 & 75126432187495320710000 & 11526755459840114914978125 & 
400786642257411458505334750 \\
9 & 100768102083397048729021250 & 20318695348931590786593466250 & 
976395092762568245382984038375 \\
\hline   
\end{tabular}\\[10pt]
\begin{tabular}{| c || l | l | l |}
 \hline $\phantom{\bigg|}k\phantom{\bigg|}$ &\hfil $n^{(3)}_{k}$ &\hfil $n^{(4)}_{k}$ &\hfil $n^{(5)}_{k}$   \\\hline 
1 & 0 & 0 & 0 \\
2 & 6605 & -50 & 0 \\
3 & 614019320 & 24204855 & -411100 \\
4 & 8454561591200 & 3245851807350 & 335376611250 \\
5 & 62482623318387100 & 73547811444806780 & 37415860873266590 \\
6 & 335260750296254643675 & 854728982385312743250 & 
1113446999293082406000 \\
7 & 1478729514546933367264780 & 6823074320028253950291680 & 
17599173386900553095722050 \\
8 & 5713442876756111478138384000 & 42686825299540747760264603450 & 
188810123308116593206813525925 \\
9 & 20066844116093245982572929494250 & 
224965680778877016475984526332550 & 
1554620412233924437357116215173950 \\
\hline   
\end{tabular}\\[10pt]
\begin{tabular}{| c || l | l | l |}
 \hline $\phantom{\bigg|}k\phantom{\bigg|}$ &\hfil $n^{(6)}_{k}$ &\hfil $n^{(7)}_{k}$ &\hfil $n^{(8)}_{k}$   \\\hline 
1 & 0 & 0 & 0 \\
2 & 0 & 0 & 0 \\
3 & 20195 & -100 & 0 \\
4 & 5659153300 & 180837000 & -19994300 \\
5 & 7939819376947330 & 635259149779950 & 16737062529500 \\
6 & 767600162108855476270 & 277471791066259667935 & 
50751903422009320250 \\
7 & 26800395547581653089999415 & 24570895165088774373417820 & 
13554556528840296607908120 \\
8 & 525864713996231860938258450860 & 950526985732429408456132354500 & 
1129570844990003922419953775500 \\
9 & 7063192000309086783572794733829590\!\! & 
21846585899244676317890567776136150\!\! & 
46879050731802968075938206207601875\!\! \\
\hline   
\end{tabular}\\[10pt]
\begin{tabular}{| c || l | l |}
 \hline $\phantom{\bigg|}k\phantom{\bigg|}$ &\hfil $n^{(9)}_{k}$ &\hfil $n^{(10)}_{k}$ \\\hline 
1 & 0 & 0 \\
2 & 0 & 0 \\
3 & 0 & 0 \\
4 & 1317930 & -41150 \\
5 & -178612452360 & 47212463805 \\
6 & 4404305636366815360 & 156953956651213430 \\
7 & 4438049154836488965109000 & 840110455805420692507800 \\
8 & 884720532073095769813844309260 & 454547508223416968855445391030 \\
9 & 70459942662058756197493625575447620 & 74442134047236071014645216834582425 \\
\hline   
\end{tabular}
}
\end{table}
\end{minipage}
\vskip-10pt
\begin{table}[H]
\begin{center}
\capt{\textwidth}{tab:QuinticZ5Numbers}{GV invariants for the $\IZ_{5}$ quotient of the quintic threefold. The conifold gap condition, regularity at the orbifold point, constant term, and the Castelnuovo bound only allow us to expand to genus 10. It may be of interest that $n^{(1)}_{1}>n^{(0)}_{1}$.}
\end{center}
\end{table}

\newpage

\subsection{$\IZ_{3}$ quotient of the Bicubic, $\IP^{5}[3,3]/\IZ_{3}$}
\vskip-20pt
\begin{minipage}{.45\linewidth}
\begin{table}[H]
\footnotesize{\begin{tabular}{| c || l | l | l | l |}
 \hline $\phantom{\bigg|}k\phantom{\bigg|}$ &\hfil $n^{(0)}_{k}$ &\hfil $n^{(1)}_{k}$ &\hfil $n^{(2)}_{k}$ &\hfil $n^{(3)}_{k}$   \\\hline 
1 & 351 & 54 & 0 & 0 \\
2 & 17604 & 18306 & 162 & 0 \\
3 & 2141442 & 5827014 & 814545 & 1188 \\
4 & 379816128 & 2051710641 & 971556336 & 51684048 \\
5 & 83262630861 & 764607449610 & 824228247186 & 182228884866 \\
6 & 20886988169844 & 295680170663946 & 593100643960080 & 321345364590549 \\
7 & 5752151300274003 & 117337498455503898 & 387461098072391913 & 405461363439976596 \\
8 & 1696280856142054320 & 47460584475061944453 & 237574270278361366560 & 418707568368082416330 \\
9 & 527083572658852629315 & 19479816270442546690932 & 139308884883240104134116 & 378117918242281386340212 \\
\hline   
\end{tabular}\\[10pt]
\begin{tabular}{| c || l | l | l |}
 \hline $\phantom{\bigg|}k\phantom{\bigg|}$ &\hfil $n^{(4)}_{k}$ &\hfil $n^{(5)}_{k}$ &\hfil $n^{(6)}_{k}$   \\\hline 
1 & 0 & 0 & 0 \\
2 & 0 & 0 & 0 \\
3 & 15 & 0 & 0 \\
4 & 253152 & -972 & 0 \\
5 & 10108036467 & 53329320 & -492723 \\
6 & 61989938556828 & 3978481806297 & 50096328498 \\
7 & 179505307950892326 & 35379506324522070 & 2880034261539471 \\
8 & 346230234722519249238 & 146594736156533810400 & 31919837327756766684 \\
9 & 516178348980696220839042 & 392321404995925930952196 & 171372524054889420881940 \\
\hline   
\end{tabular}\\[10pt]
\begin{tabular}{| c || l | l | l |}
 \hline $\phantom{\bigg|}k\phantom{\bigg|}$ &\hfil $n^{(7)}_{k}$ &\hfil $n^{(8)}_{k}$ &\hfil $n^{(9)}_{k}$   \\\hline 
1 & 0 & 0 & 0 \\
2 & 0 & 0 & 0 \\
3 & 0 & 0 & 0 \\
4 & 0 & 0 & 0 \\
5 & 1458 & 0 & 0 \\
6 & -112622568 & 3255660 & -4644 \\
7 & 75185909775432 & 345720294045 & 2479109922 \\
8 & 3380473998240899949 & 152773576290297936 & 2342948602615704 \\
9 & 42744163756609489532478 & 5859503926986268013226 & 410009966496890836800 \\
\hline   
\end{tabular}
}
\end{table}
\end{minipage}
\vskip-10pt
\begin{table}[H]
\begin{center}
\capt{\textwidth}{tab:BicubicZ3Numbers}{GV invariants for the $\IZ_{3}$ quotient of the bicubic threefold. The conifold gap condition, regularity at the K-point (infinity), constant term, and the Castelnuovo bound only allow us to expand to genus 9.}
\end{center}
\end{table}

\newpage

\subsection{$\IZ_{7}$ quotient of the R{\o}dland model}
\vskip-20pt
\begin{minipage}{.45\linewidth}
\begin{table}[H]
\footnotesize{\begin{tabular}{| c || l | l | l |}
\hline $\phantom{\bigg|}k\phantom{\bigg|}$ &\hfil $n^{(0)}_{k}$ &\hfil $n^{(1)}_{k}$ &\hfil $n^{(2)}_{k}$   \\\hline 
1 & 84 & 106 & 0 \\
2 & 1729 & 9731 & 2597 \\
3 & 83412 & 1189690 & 1548666 \\
4 & 5908448 & 162847656 & 561111558 \\
5 & 515627728 & 23784778992 & 165479914726 \\
6 & 51477011901 & 3619073252171 & 43608601458779 \\
7 & 5641036903908 & 566456134227334 & 10707782743676536 \\
8 & 661894028378002 & 90513043864684029 & 2505000342181522444 \\
9 & 81831403277082228 & 14692917426093647214 & 565787798626063356392 \\
\hline   
\end{tabular}\\[10pt]
\begin{tabular}{| c || l | l | l |}
\hline $\phantom{\bigg|}k\phantom{\bigg|}$ &\hfil $n^{(3)}_{k}$ &\hfil $n^{(4)}_{k}$ &\hfil $n^{(5)}_{k}$   \\\hline 
1 & 0 & 0 & 0 \\
2 & 21 & 0 & 0 \\
3 & 282100 & 2870 & 0 \\
4 & 425899488 & 73987368 & 1449231 \\
5 & 316671278140 & 199905031018 & 41043809856 \\
6 & 165288447125522 & 246295464565871 & 156541429038015 \\
7 & 69737863611873408 & 198387291602887002 & 275064061320436234 \\
8 & 25527142974989788328 & 122013764358829703549 & 308374855684567279377 \\
9 & 8442139465760064106140 & 62231778734823700282532 & 256500030949655724914284 \\
\hline 
\end{tabular}  
}
\end{table}
\end{minipage}
\vskip-10pt
\begin{table}[H]
\begin{center}
\capt{\textwidth}{tab:RodlandZ7PfaffianNumbers}{Above are GV invariants for the $\IZ_{7}$ quotient of R{\o}dland's Pfaffian threefold. Note $n^{(1)}_{1}>n^{(0)}_{1}$.}
\end{center}
\end{table}

\vskip-40pt
\begin{minipage}{.45\linewidth}
\begin{table}[H]
\footnotesize{\begin{tabular}{| c || l | l | l | l | l | l |}
\hline $\phantom{\bigg|}k\phantom{\bigg|}$ &\hfil $n^{(0)}_{k}$ &\hfil $n^{(1)}_{k}$ &\hfil $n^{(2)}_{k}$ &\hfil $n^{(3)}_{k}$ &\hfil $n^{(4)}_{k}$ &\hfil $n^{(5)}_{k}$   \\\hline 
1 & 28 & 18 & 0 & 0 & 0 & 0 \\
2 & 175 & 463 & 7 & 0 & 0 & 0 \\
3 & 1820 & 11526 & 3248 & 0 & 0 & 0 \\
4 & 28294 & 345024 & 321426 & 28245 & 35 & 0 \\
5 & 530992 & 10778248 & 22523634 & 7781872 & 344386 & 378 \\
6 & 11403315 & 352208877 & 1355627203 & 1134004599 & 219069333 & 7349629 \\
7 & 268281804 & 11824000122 & 74410696766 & 120414777910 & 59050816048 & 8607660880 \\
8 & 6755563416 & 405319921505 & 3848162314080 & 10540208130242 & 10243982279181 & 3762792168023 \\
9 & 179169428732 & 14113739316490 & 190766199128806 & 809447642951500 & 1352699340656188 & 980390268886546 \\
\hline   
\end{tabular} 
}
\end{table}
\end{minipage}
\vskip-10pt
\begin{table}[H]
\begin{center}
\capt{\textwidth}{tab:RodlandZ7GrassmannianNumbers}{Above are GV invariants for the $\IZ_{7}$ quotient of R{\o}dland's Grassmannian threefold.}
\end{center}
\end{table}
To compute the GV invariants in \tref{tab:RodlandZ7PfaffianNumbers} and \tref{tab:RodlandZ7GrassmannianNumbers} we expand a single B-model free energy about two different MUM points ($\varphi=0$ and $\varphi=\infty$). We impose the conifold gap condition; the constant term contributions in both expansions about 0 and $\infty$; and the Castelnuovo bounds for both geometries. These considerations only allow us to expand as high as genus 5.

\newpage

\subsection{$\IZ_{5}$ quotient of the Hosono-Takagi model}
\vskip-20pt
\begin{minipage}{.45\linewidth}
\begin{table}[H]
\footnotesize{\begin{tabular}{| c || l | l | l |}
\hline $\phantom{\bigg|}k\phantom{\bigg|}$ &\hfil $n^{(0)}_{k}$ &\hfil $n^{(1)}_{k}$ &\hfil $n^{(2)}_{k}$   \\\hline 
1 & 110 & 88 & 0 \\
2 & 3830 & 10920 & 2080 \\
3 & 233110 & 1806540 & 1590820 \\
4 & 21322480 & 330946550 & 750087075 \\
5 & 2455996570 & 64415464108 & 290194401190 \\
6 & 324701179500 & 13042591099243 & 100703783697240 \\
7 & 47154769689380 & 2714327362683188 & 32628936131965760 \\
8 & 7335485654525500 & 576387496496436593 & 10085276331992030630 \\
9 & 1202660670835792580 & 124298318900133467068 & 3012093129341767122550 \\
\hline   
\end{tabular}\\[10pt]
\begin{tabular}{| c || l | l | l |}
\hline $\phantom{\bigg|}k\phantom{\bigg|}$ &\hfil $n^{(3)}_{k}$ &\hfil $n^{(4)}_{k}$ &\hfil $n^{(5)}_{k}$   \\\hline 
1 & 0 & 0 & 0 \\
2 & 10 & 0 & 0 \\
3 & 215250 & 1570 & 0 \\
4 & 417960730 & 57124730 & 938035 \\
5 & 399276936750 & 196521594350 & 33494657920 \\
6 & 269595086591590 & 307606088617415 & 159359084509985 \\
7 & 147954130068172870 & 316701550590837000 & 351272641013964080 \\
8 & 70724470876231474060 & 250482960584090792325 & 497771824466038411950 \\
9 & 30631935324967527180560 & 165111821196548486009530 & 526934266333548120001620 \\
\hline 
\end{tabular}  
}
\end{table}
\end{minipage}
\vskip-10pt
\begin{table}[H]
\begin{center}
\capt{\textwidth}{tab:HTZ5DQSNumbers}{Above are GV invariants for the $\IZ_{5}$ quotient of Hosono-Takagi's threefold subvariety of the double quintic symmetroid.}
\end{center}
\end{table}

\vskip-40pt
\begin{minipage}{.45\linewidth}
\begin{table}[H]
\footnotesize{\begin{tabular}{| c || l | l | l | l | l | l |}
\hline $\phantom{\bigg|}k\phantom{\bigg|}$ &\hfil $n^{(0)}_{k}$ &\hfil $n^{(1)}_{k}$ &\hfil $n^{(2)}_{k}$ &\hfil $n^{(3)}_{k}$ &\hfil $n^{(4)}_{k}$ &\hfil $n^{(5)}_{k}$   \\\hline 
1 & 10 & 16 & 0 & 0 & 0 & 0 \\
2 & 65 & 238 & 5 & 0 & 0 & 0 \\
3 & 295 & 3001 & 1210 & 0 & 0 & 0 \\
4 & 3065 & 54024 & 66665 & 7460 & 0 & 0 \\
5 & 29715 & 905336 & 2616590 & 1177075 & 61840 & 60 \\
6 & 377115 & 16967201 & 88967385 & 96235145 & 22425380 & 766775 \\
7 & 4862130 & 315204632 & 2735270040 & 5767910725 & 3440264885 & 546451975 \\
8 & 69723305 & 6098433011 & 79647092230 & 284786224085 & 339337810075 & 140415307660 \\
9 & 1031662155 & 118674731165 & 2215776214620 & 12331891528815 & 25464777595225 & 21216255805620 \\
\hline   
\end{tabular} 
}
\end{table}
\end{minipage}
\vskip-10pt
\begin{table}[H]
\begin{center}
\capt{\textwidth}{tab:HTZ5RCNumbers}{GV invariants for the $\IZ_{5}$ quotient of Hosono-Takagi's Reye congruence threefold. $n_{1}^{(1)}>n_{1}^{(0)}$.}
\end{center}
\end{table}
To compute the GV invariants in \tref{tab:HTZ5DQSNumbers} and \tref{tab:HTZ5RCNumbers} we expand a single B-model free energy about two different MUM points ($\varphi=0$ and $\varphi=\infty$). We impose the conifold gap condition; the constant term contributions in both expansions about 0 and $\infty$; and the Castelnuovo bounds for both geometries. These considerations only allow us to expand as high as genus 5.

We remark that, as is the case for the BPS computations performed in \cite{Hosono:2011np}, the relation \eqref{eq:BtoA_FreeEnergy} in this example includes an additional numerical factor $8^{2g-2}$ required to correctly normalise the Yukawa coupling for the geometry at the second MUM point if the naive $\varpi_{0}$ obtained as the power series solution with leading coefficient $1$ is used.

\newpage

\section{Homotopy-homology relations, K{\"a}hler parameters, and Wall data}\label{sect:QuotientData}
We will study compact Calabi-Yau threefolds $Y$, obtained as quotients of simply connected threefolds $\widetilde{Y}$ (themselves compact and Calabi-Yau) by a freely acting discrete group $G$. To begin our discussion, we relax the assumptions in the body of this paper so that $G$ is not necessarily $\IZ_{M}$, and $h^{1,1}$ may be different for $Y$ and $\widetilde{Y}$. 

We have a quotient map
\begin{equation}
\rho:\,\widetilde{Y}\;\mapsto\;Y\cong\widetilde{Y}/G~.
\end{equation}
$\widetilde{Y}$ is the universal cover of $Y$. The threefold $Y$ has a second cohomology $H^{2}(Y,\IZ)$ with integral generators $e_{i}$, $1\leq i\leq h^{1,1}$, of the torsion-free part. The K{\"a}hler form $\omega$ of $Y$ and the Neveu-Schwarz B-field $B_{2}$ are combined into the complexified K{\"a}hler form of $Y$, which can be expanded as
\begin{equation}
B_{2}+\ii\,\omega \= t^{i}e_{i} ~.
\end{equation}
These coordinates $t^{i}$ parametrise the A-model moduli space. Similarly, $\widetilde{Y}$ has its own set of K{\"a}hler parameters $\widetilde{t}^{i}$.

Based on arguments by Aspinwall and Morrison \cite{Aspinwall:1994uj}, using results of \cite{EilenbergMacLane}, one can relate the K{\"a}hler parameters $t^{i}$ and $\widetilde{t}^{i}$ of $Y$ and $\widetilde{Y}$. The homology groups of quotient manifolds have already been studied in detail in \cite{Braun:2007xh}. We review the aspects of the Aspinwall-Morrison argument relevant to us, in order to explain how we obtain the Wall data that we used in this paper. If we allow for quotients such that $h^{1,1}$ differs for $Y$ and $\widetilde{Y}$ then a minor modification to the argument presented in \cite{Aspinwall:1994uj} must be incorporated.

$\pi_{n}$ and $H_{n}$ will denote homotopy and homology groups. For each $n$, there is a Hurewicz map
\begin{equation}\label{eq:Hurewicz}
\nu_{n}:\,\pi_{n}(Y)\mapsto H_{n}(Y,\IZ)~.
\end{equation}
And similarly for $\widetilde{Y}$ we have maps $\widetilde{\nu}_{n}:\pi_{n}(\widetilde{Y})\mapsto H_{n}(\widetilde{Y},\IZ)$. Fixing a generator $u_{n}\in H_{n}\left(S^{n}\right)$, the map $\nu_{n}$ takes a homotopy class $[f]$ (of maps $S^{n}\mapsto Y$) to the pushforward $f_{*}(u_{n})\in H_{n}(Y)$. 
\begin{equation}
\nu_{n}(f)\=f_{*}(u_{n})~.
\end{equation}
This map is used to define what were called in \cite{EilenbergMacLane} the spherical subgroups $\Sigma^{n}(Y)$ of $H_{n}(Y)$. These are the images of each $\pi_{n}(Y)$ under $\nu_{n}$,
\begin{equation}
\Sigma^{n}(Y)\;\cong\;\nu_{n}\left(\pi_{n}(Y)\right)~.
\end{equation}
$\Sigma^{n}$ is the set of homology classes generated by spheres. An exact sequence is provided by \cite{EilenbergMacLane} that relates homotopy and homology groups of $Y$:
\begin{equation}\label{eq:exact0}
0\rightarrow\Sigma^{2}\left(Y\right)\rightarrow H_{2}(Y,\IZ)\rightarrow H_{2}\left(\pi_{1}(Y),\IZ\right) \rightarrow 0~.
\end{equation}
We should like to massage this relation, and extract an exact sequence relating the homology groups of the spaces $\widetilde{Y},\,Y,$ and the group $G$. 

We remark that $\pi_{2}(Y)\cong\pi_{2}(\widetilde{Y})$, as more generally the homotopy groups $\pi_{i\geq2}$ of a space and its universal cover are isomorphic. Moreover, since $\pi_{1}(\widetilde{Y})\cong0$ there is an isomorphism $\pi_{2}(\widetilde{Y})\cong H_{2}(\widetilde{Y},\IZ)$ (indeed this is implied by \eqref{eq:exact0} if we replace $Y$ by $\widetilde{Y}$). Then the Hurewicz map $\widetilde{\nu}_{2}$ provides $\Sigma^{2}(\widetilde{Y})\cong H_{2}(\widetilde{Y},\IZ)$.

Crucially however, $\Sigma^{2}(Y)$ is not isomorphic to $\Sigma^{2}(\widetilde{Y})$ in general (but this was the case for the example in \cite{Aspinwall:1994uj}). Note that $\nu_{n}$ takes a homotopy class $f$ to the homology class $f_{*}(u_{n})$, but this map is not in general injective because non-homotopic spheres inside $Y$ can still be homologous. The group action $G$ may identify homology classes of $\widetilde{Y}$, in which case $h^{1,1}(Y)\leq h^{1,1}(\widetilde{Y})$. This leads to an identification
\begin{equation}
\Sigma^{2}(Y)\cong H_{2}(\widetilde{Y},\IZ)_{G}~,
\end{equation}
where we introduce $G$-coinvariant second homology. This is the set of equivalence classes under the $G$-action in the second homology of $\widetilde{Y}$. Now, from simply-connectedness of $\widetilde{Y}$ we learn that the group $\pi_{1}(Y)$, i.e. $\pi_{1}(\widetilde{Y}/G)$, is itself $G$. One uses these identifications to rewrite \eqref{eq:exact0} in such a way as to relate homologies:
\begin{equation}\label{eq:exact}
0\;\mapsto\; H_{2}(\widetilde{Y},\IZ)_{G}\;\stackrel{\rho_{*}}{\mapsto}\; H_{2}(Y,\IZ) \;\mapsto\; H_{2}(G,\IZ)\;\mapsto\; 0~.
\end{equation}
The group $H_{2}(G,\IZ)$ is sometimes referred to as the Schur multiplier of $G$. The pushforward map $\rho_{*}$ identifies $H_{2}(\widetilde{Y},\IZ)_{G}$ with a subgroup of $H_{2}(Y,\IZ)$. There is a nonsingular pairing
\begin{equation}\label{eq:pairing}
H_{2}(\widetilde{Y},\IZ)_{\text{Free}}\times H^{2}(\widetilde{Y},\IZ)_{\text{Free}}\mapsto\IZ~,\qquad \int_{\epsilon^{i}}e_{j}\=\delta^{i}_{j}~,
\end{equation} 
where $\epsilon^{i}$ and $e_{j}$ respectively generate the torsion-free parts of the second integral homology and cohomology. There is a similar such pairing for the quotient $Y$, and compatibility of \eqref{eq:exact} with \eqref{eq:pairing} tells us how the pullback $\rho^{*}$ acts on the integral cohomology generators. This is necessary information for us to compute Wall data. Since the quotient map $\rho$ effects a degree-$|G|$ covering of $Y$, we have that $|G|\int_{Y} V=\int_{\widetilde{Y}}\rho^{*}(V)$ for $V\in H^{6}(Y,\IZ)$. So once we understand how to perform the pullbacks, it becomes possible to express the topological data of $Y$ in terms of the data for $\widetilde{Y}$.

From the naturality axiom of the Chern classes, $\rho^{*}(c_{a}(Y))=c_{a}(\widetilde{Y})$. This is how one shows the well-known fact that the Euler characteristic always divides upon taking the quotient by a freely acting symmetry group, irrespective of the details of \eqref{eq:exact}:
\begin{equation}
\chi(Y)=\int_{Y}c_{3}(Y)\=\frac{1}{|G|}\int_{\widetilde{Y}}\rho^{*}(c_{3}(Y))\=\frac{1}{|G|}\int_{\widetilde{Y}}c_{3}(\widetilde{Y})\=\frac{1}{|G|}\chi(\widetilde{Y})~.
\end{equation}
For the triple intersection and second Chern numbers, 
\begin{equation}\label{eq:WallDivisionGeneral}
\begin{aligned}
\kappa^{(Y)}_{ijk}&\=\int_{Y}e_{i}\wedge e_{j}\wedge e_{k}\=\frac{1}{|G|}\int_{\widetilde{Y}}\rho^{*}(e_{i}\wedge e_{j}\wedge e_{k})\\[5pt]
&\=\frac{1}{|G|}\int_{\widetilde{Y}}\rho^{*}(e_{i})\wedge\rho^{*}(e_{j})\wedge\rho^{*}(e_{k})~,\\[10pt]
c_{2,i}^{(Y)}&\=\int_{Y}c_{2}(Y)\wedge e_{i}\=\frac{1}{|G|}\int_{\widetilde{Y}}\rho^{*}(c_{2}(Y)\wedge e_{i})\=\frac{1}{|G|}\int_{\widetilde{Y}}\rho^{*}(c_{2}(Y))\wedge \rho^{*}(e_{i})\\[5pt]
&\=\frac{1}{|G|}\int_{\widetilde{Y}}c_{2}(\widetilde{Y})\wedge \rho^{*}(e_{i})~.
\end{aligned}\end{equation}
One can use \eqref{eq:WallDivisionGeneral} to express the triple intersection and second Chern numbers of $Y$ in terms of the same data for $\widetilde{Y}$, but in order to do this one must know the explicit images $\rho^{*}(e_{i})$ as combinations of the $\widetilde{e}_{i}$. These are obtained in each example by studying \eqref{eq:exact} and \eqref{eq:pairing}.

We will now turn to a non-exhaustive set of examples, to illustrate the implications of \eqref{eq:exact} in different cases. We will consider genus 0 curves $\widetilde{C}\subset\widetilde{Y}$ and their smooth\footnote{We do not take care here to make our argument work for singular curves. However, to obtain our conclusions on the relations between $\widetilde{t}_{i}$ and $t$, we only require our following arguments with some smooth rational curve $C$ on $Y$. We do not consider the possibility of a Calabi-Yau threefold $Y$ not containing any smooth $C$.} images $C\subset Y$ under the quotient map. Since a smooth rational curve cannot have an unramified cover (as follows from the Riemann-Hurwitz formula), and $G$ acts freely on $\widetilde{Y}$ by assumption, we learn that no genus 0 curves on $\widetilde{Y}$ are fixed by the $G$-action and the preimage of $C$ is $|G|$ disjoint curves $\widetilde{C}$. The degree-vector of a curve $C$ is the set of intersections of that curve with a generating set $e_{i}$ of $H^{2}(Y,\IZ)_{\text{Free}}$,
\begin{equation}
\text{deg}(C)_{i}=\int_{[C]}e_{i}~.
\end{equation}
The homology class of the curve $C$ is expanded as $[C]=\text{deg}(C)_{i}\epsilon^{i}$, similarly we have $[\widetilde{C}]=\text{deg}(\widetilde{Y})_{i}\widetilde{\epsilon}^{i}$. The area of $C$ is $t^{i}\text{deg}(C)_{i}$ and the area of a single curve $\widetilde{C}$ is $\widetilde{t}^{i}\text{deg}(\widetilde{C})_{i}$. Since these areas must be equal (as these are identical curves), we read off relations between $\widetilde{t}^{i}$ and $t$.

\underline{Example 0, $H_{2}(G)\cong0$ and $h^{1,1}(Y)=h^{1,1}(\widetilde{Y})$ :}

All examples studied in this paper belong to this class, to which we pay the most attention.

In this simplest case we necessarily have $H_{2}(\widetilde{Y},\IZ)_{G}\cong H_{2}(\widetilde{Y},\IZ)\cong H_{2}(Y,\IZ)$. As a consequence of \eqref{eq:exact}, the map $\rho_{*}$ takes each generator $\widetilde{\epsilon}^{i}$ of $H_{2}(\widetilde{Y},\IZ)$ to a generator $\epsilon^{i}$ of $H_{2}(Y,\IZ)$,
\begin{equation}\label{eq:ex0pushforward}
\rho_{*}(\widetilde{\epsilon}^{i})\=\epsilon^{i}~.
\end{equation} 
As a result, taking the quotient by $G$ sends each curve in $\widetilde{Y}$ to a curve in $Y$ with the same degree:
\begin{equation}
\text{deg}(C)_{i}=\int_{[C]}e_{i}=\int_{\rho_{*}([\widetilde{C}])}e_{i}=\int_{\rho_{*}\left(\text{deg}(\widetilde{C})_{j}\widetilde{\epsilon}^{j}\right)}e_{i}=\int_{\text{deg}(\widetilde{C})_{j}\epsilon^{j}}e_{i}=\text{deg}(\widetilde{C})_{i}
\end{equation}
The complexified area of $\widetilde{C}$ is $\text{deg}(\widetilde{C})_{i}\widetilde{t}^{i}$, and the complexified area of $C$ is $\text{deg}(C)_{i}t^{i}$. Since these must be equal, we obtain for these examples
\begin{equation}
t^{i}=\widetilde{t}^{i}~.
\end{equation}
Comparing \eqref{eq:ex0pushforward} with \eqref{eq:pairing}, we get the pullback
\begin{equation}\label{eq:ex0pullback}
\rho^{*}(e_{i})=\widetilde{e}_{i}.
\end{equation}
This allows us to compute the Wall data. For the triple intersection number, we have
\begin{equation}
\begin{aligned}
\kappa^{(Y)}_{ijk}&\=\int_{Y}e_{i}e_{j}e_{k}\=\frac{1}{|G|}\int_{\widetilde{Y}}\rho^{*}(e_{i}e_{j}e_{k})\=\frac{1}{|G|}\int_{\widetilde{Y}}\rho^{*}(e_{i})\rho^{*}(e_{j})\rho^{*}(e_{k})\=\frac{1}{|G|}\int_{\widetilde{Y}}\widetilde{e}_{i}\widetilde{e}_{j}\widetilde{e}_{k}\\[5pt]
&\=\frac{1}{|G|}\kappa^{\widetilde{Y}}_{ijk}~.
\end{aligned}
\end{equation}
For the second Chern numbers,
\begin{equation}
\begin{aligned}
c_{2,i}^{(Y)}&\=\int_{Y}c_{2}(Y)e_{i}\=\frac{1}{|G|}\int_{\widetilde{Y}}\rho^{*}(c_{2}(Y)e_{i})\=\int_{\widetilde{Y}}\rho^{*}(c_{2}(Y))\rho^{*}(e_{i})\=\frac{1}{|G|}\int_{\widetilde{Y}}c_{2}(\widetilde{Y})e_{i}\\[5pt]
&\=\frac{1}{|G|}c_{2,i}^{\widetilde{Y}}~.
\end{aligned}
\end{equation}

\underline{Example 1, $\IP^{4}[5]_{/\IZ_{5}\times\IZ_{5}}$ :}

This example is illustrative of the fact that the Wall data of $Y$ cannot always be obtained by dividing the Wall data of $\widetilde{Y}$ by $|G|$. This particular example was studied in \cite{Aspinwall:1994uj}, and there is much more that is interesting about this model than we will cover. The salient point for us is that $H_{2}(\IZ_{5}\times\IZ_{5},\IZ)\cong\IZ_{5}$, which follows from a K{\"u}nneth-type computation. This means that the fourth term in the sequence \eqref{eq:exact} is nonzero, unlike in Example 0. Both $Y$ and $\widetilde{Y}$ have $h^{1,1}=1$. 

From \eqref{eq:exact}, we can be sure that $\rho_{*}$ is not surjective. However, that still allows for two possibilities. Either we have $H_{2}(Y,\IZ)\cong\IZ$ or $H_{2}(Y,\IZ)\cong\IZ\oplus\IZ_{5}$. To distinguish between these two cases, one must study the geometry carefully. There is necessarily a homology class $\upsilon$ not represented by a sphere, and either $5\upsilon=0$ or $5\upsilon$ generates $H_{2}(\IP^{4}[5],\IZ)$. It was shown in \cite{Aspinwall:1994uj} that the latter is the case. As such
\begin{equation}\label{eq:ex1pushforward}
H_{2}(Y,\IZ)\cong\IZ~,\qquad\text{ and }\qquad \rho_{*}(\widetilde{\epsilon})=5\epsilon~.
\end{equation}
Consider once again a genus 0 curve $C$ on $Y$, whose preimage is 25 disjoint curves $\widetilde{C}\subset\widetilde{Y}$. We wish to relate the degree of $C$ to the degree of a single curve $\widetilde{C}$ (not the full set of 25 taken together).
\begin{equation}
\text{deg}(C)=\int_{[C]}e\=\int_{\rho_{*}([\widetilde{C}])}e\=\int_{\rho_{*}\left(\text{deg}(\widetilde{C})\widetilde{\epsilon}\right)}e\=\int_{5\text{deg}(\widetilde{C})\epsilon}e\=5\text{deg}(\widetilde{C})~.
\end{equation}
We now demand that the area of $C$ equals the area of a single curve $\widetilde{C}$. This means that $\text{deg}(C)t=\text{deg}(\widetilde{C})\widetilde{t}$, and therefore
\begin{equation}\label{eq:5t}
\widetilde{t}\=5t~.
\end{equation}
This was noted in \cite{Aspinwall:1994uj}, which provided the very important observation that the A-model moduli space of $\IP^{4}[5]_{/\IZ_{5}\times\IZ_{5}}$ furnishes a five-fold cover of the moduli space of $\IP^{4}[5]$. Their equation (12), $\exp(2\pi\ii\, \widetilde{t}\,)=\exp(2\pi\ii\, t)^{5}$, provides the above relation \eqref{eq:5t}.

Comparing \eqref{eq:pairing} with \eqref{eq:ex1pushforward} provides us with the pullback map, which differs to that of Example 0.
\begin{equation}
\rho^{*}(e)\=5\widetilde{e}~.
\end{equation}
We now turn to the triple intersection and second Chern numbers. For $\IP^{4}[5]_{/\IZ_{5}\times\IZ_{5}}$ one obtains, as has already been done in \cite{Aspinwall:1994uj},
\begin{equation}\begin{aligned}
\kappa_{111}&\=\frac{1}{|G|}\int_{\IP^{4}[5]}\rho^{*}(e)\wedge\rho^{*}(e)\wedge\rho^{*}(e)\=\frac{1}{5^{2}}\int_{\IP^{4}[5]}5e\wedge 5e\wedge 5e\\[5pt]
&\=\frac{1}{5^{2}}5^{3}\big(5\big)\=25~,\\[5pt]
c_{2}&\=\frac{1}{|G|}\int_{\IP^{4}[5]}\rho^{*}(c_{2}(\IP^{4}[5]_{/\IZ_{5}\times\IZ_{5}}))\wedge\rho^{*}(e)\=\frac{1}{5^{2}}\int_{\IP^{4}[5]}c_{2}(\IP^{4}[5])\wedge 5e\\[5pt]
&\=\frac{1}{5^{2}}5\big(50\big)\=10~.
\end{aligned}\end{equation}
Note that when the dust settles the triple intersection number, as compared to that of the quintic, has been multiplied by five. However, the second Chern number has been divided by 5. Moreover, the second Chern numbers of $\IP^{4}[5]_{/\IZ_{5}\times\IZ_{5}}$ and $\IP^{4}[5]_{/\IZ_{5}}$ are both equal to 10.

For a much more involved example wherein $H_{2}(G,\IZ)\cong0$ and both $H_{2}(\widetilde{Y},\IZ)_{G}$ and $H_{2}(Y,\IZ)$ have torsion, one can see the examples and discussion in \cite{Braun:2008sf,Braun:2007xh,Braun:2007vy}. Note well that if $H_{2}(G,\IZ)\neq0$, one must still study the non-spherical part of $H_{2}(Y,\IZ)$ before one is able to take the above steps to compute the Wall data.

We also remark that, while in Examples 0 and 1 we have compared the areas of curves $\widetilde{C}$ and $C$ in order to obtain the relation between $t$ and $\widetilde{t}$, one could also consider the classical part of the quantum volume of $Y$ and $\widetilde{Y}$. Since $\text{Vol}(Y)=\frac{1}{|G|}\text{Vol}(\widetilde{Y})$, we can anticipate in any example that 
\begin{equation}
\kappa^{(Y)}_{ijk}t^{i}t^{j}t^{k}\=\frac{1}{|G|}\kappa^{\widetilde{Y}}_{ijk}\widetilde{t}^{i}\widetilde{t}^{j}\widetilde{t}^{k}~.
\end{equation}
Knowledge of the relation between the $\kappa_{ijk}$ thereby enables one to relate $t$ and $\widetilde{t}$.

\underline{Example 2, $H_{2}(G)\cong0$, $h^{1,1}(Y)<h^{1,1}(\widetilde{Y})$ :}

In these cases, the $h^{1,1}(\widetilde{Y})$ K{\"a}hler parameters $\widetilde{t}^{i}$ are not all independent. Since $G$ acts nontrivially on the cohomology of $\widetilde{Y}$, exchanging some of the generators $\widetilde{e}^{i}$, the corresponding subset of the $\widetilde{t}^{i}$ must all be set equal so that the K{\"a}hler form $\widetilde{t}^{i}\widetilde{e}^{i}$ is left invariant. 

For concreteness, we will consider two examples at once. These have been discussed in \cite{Candelas:2015amz} (Appendix A.1 therein).
\begin{equation}
\begin{aligned}
\widetilde{Y}^{(1)}&\cong\cicy{\IP^{1}\\\IP^{4}\\\IP^{4}}{1&1&0&0&0&0 \\ 1 & 0 & 1 & 1 & 1 & 1 \\ 0 & 1 & 1 & 1 & 1 & 1}^{h^{1,1}=3~,\,h^{2,1}=47}~,\quad &Y^{(1)}\cong \widetilde{Y}^{(1)}/\IZ_{2}~,\\[5pt]
\widetilde{Y}^{(2)}&\cong\cicy{\IP^{1}\\\IP^{3}\\\IP^{3}}{0&0&0&2 \\ 1 & 1 & 1 & 1  \\ 1 & 1 & 1 & 1}^{h^{1,1}=3~,\,h^{2,1}=47}~,\quad &Y^{(2)}\cong \widetilde{Y}^{(2)}/\IZ_{2}~.
\end{aligned}
\end{equation}
We have used the CICY notation \cite{Candelas:1987kf} to display complete intersections of polynomials in the ambient spaces defined as the product of the projective spaces on the left, with the degrees of the intersecting polynomials given by the entries of the matrices. An adjunction computation \cite{Hosono:1994ax} gives us the topological data of the covering manifolds,
\begin{equation}
\begin{aligned}
\kappa^{(\widetilde{Y}^{(1)})}_{111}&\=\kappa^{(\widetilde{Y}^{(1)})}_{112}\=\kappa^{(\widetilde{Y}^{(1)})}_{113}\=0~,\qquad\qquad &\kappa^{(\widetilde{Y}^{(2)})}_{111}&\=\kappa^{(\widetilde{Y}^{(2)})}_{112}\=\kappa^{(\widetilde{Y}^{(2)})}_{113}\=0~, \\[5pt]
\kappa^{(\widetilde{Y}^{(1)})}_{122}&\=\kappa^{(\widetilde{Y}^{(1)})}_{133}\=4~,&\kappa^{(\widetilde{Y}^{(2)})}_{122}&\=\kappa^{(\widetilde{Y}^{(2)})}_{133}\=4~,\\[5pt]
\kappa^{(\widetilde{Y}^{(1)})}_{123}&\=6~,&\kappa^{(\widetilde{Y}^{(2)})}_{123}&\=6~,\\[5pt]
\kappa^{(\widetilde{Y}^{(1)})}_{223}&\=\kappa^{(\widetilde{Y}^{(1)})}_{233}\=10~,&\kappa^{(\widetilde{Y}^{(2)})}_{223}&\=\kappa^{(\widetilde{Y}^{(2)})}_{233}\=6~,\\[5pt]
\kappa^{(\widetilde{Y}^{(1)})}_{222}&\=\kappa^{(\widetilde{Y}^{(1)})}_{333}\=5~,&\kappa^{(\widetilde{Y}^{(2)})}_{222}&\=\kappa^{(\widetilde{Y}^{(2)})}_{333}\=2~,\\[5pt]
c^{(\widetilde{Y}^{(1)})}_{2,1}&\=24~,&c^{(\widetilde{Y}^{(2)})}_{2,1}&\=24~,\\[5pt]
c^{(\widetilde{Y}^{(1)})}_{2,2}&\=c^{(\widetilde{Y}^{(1)})}_{2,3}\=50~,& c^{(\widetilde{Y}^{(2)})}_{2,2}&\=c^{(\widetilde{Y}^{(2)})}_{2,3}\=44~,\\[5pt]
\chi(\widetilde{Y}^{(1)})&\=-88~, & \chi(\widetilde{Y}^{(2)})&\=-88~.
\end{aligned}
\end{equation}
Any triple intersection numbers not displayed are obtained from the above by permuting indices.

In both cases, the $\IZ_{2}$ symmetry of the threefold descends from the $\IZ_{2}$ symmetry of the ambient space given by exchanging the lower two projective space factors. This means that in both examples the $\IZ_{2}$ action exchanges the integral cohomology generators $e_{2}$ and $e_{3}$. So although each of $\widetilde{Y}^{(1)},\,\widetilde{Y}^{(2)}$ has $h^{1,1}=3$, we will set some K{\"a}hler parameters equal before taking the quotient. In both examples there are two independent K{\"a}hler parameters,
\begin{equation}
\widetilde{t}^{1} \text{ and } \widetilde{t}^{2}\=\widetilde{t}^{3}~.
\end{equation}
The Hodge numbers of $Y^{(1)}$ and $Y^{(2)}$ are both $(h^{1,1},h^{2,1})=(2,24)$.

The second group cohomology $H_{2}(\IZ_{2},\IZ)$ is trivial, and so \eqref{eq:exact} guarantees that $\rho_{*}$ gives an isomorphism between $H_{2}(Y,\IZ)$ and $H_{2}(\widetilde{Y},\IZ)_{G}$ (crucially this is not the same as $H_{2}(\widetilde{Y},\IZ)\cong\IZ^{3}$). Consider a genus 0 curve $\widetilde{C}$ on either $\widetilde{Y}^{(1)}$ or $\widetilde{Y}^{(2)}$. From \eqref{eq:exact} we learn that the image curve $C$ has degree
\begin{equation}
\text{deg}(C)_{1}\=\text{deg}(\widetilde{C})_{1}~,\qquad\text{deg}(C)_{2}\=\text{deg}(\widetilde{C})_{2}+\text{deg}(\widetilde{C})_{3}~~,
\end{equation}
and therefore
\begin{equation}
t^{1}\=\widetilde{t}^{1}~,\qquad t^{2}\=\widetilde{t}^{2}~.
\end{equation}
The pullback will map $H^{2}(Y,\IZ)$ to $H^{2}(\widetilde{Y},\IZ)^{G}$, the $G$-invariant subspace of $H^{2}(\widetilde{Y},\IZ)$.
\begin{equation}
\rho^{*}(e_{1})\=\widetilde{e}_{1}~,\qquad\rho^{*}(e_{2})\=\widetilde{e}_{2}+\widetilde{e}_{3}~.
\end{equation}
To see that this is the correct pullback relation, note that $\widetilde{\epsilon}_{2}$ and $\widetilde{\epsilon}_{3}$ lie in the same equivalence class in $H_{2}(\widetilde{Y},\IZ)_{G}$. Note that $H^{2}(\widetilde{Y},\IQ)^{G}$ is spanned by $\widetilde{e}_{1}$ and $\widetilde{e_{2}}+\widetilde{e}_{3}$ with rational coefficients. It must be the case that $\rho^{*}(e_{2})=a \widetilde{e}_{1}+b(\widetilde{e}_{2}+\widetilde{e}_{3})$. To preserve \eqref{eq:pairing}, we need to have $\int_{\widetilde{\epsilon}_{1}}\rho^{*}(e_{1})=0$ and $\int_{\widetilde{\epsilon}_{2}}\rho^{*}(e_{2})=\int_{\widetilde{\epsilon}_{3}}\rho^{*}(e_{2})=1$, which forces $a=0$ and $b=1$ as claimed.
  
The triple intersection and second Chern numbers can now be computed via \eqref{eq:WallDivisionGeneral}. For each triple $i,j,k$ one expands $\rho^{*}(e_{i})\rho^{*}(e_{j})\rho^{*}(e_{k})$ to obtain
\begin{equation}
\begin{aligned}
\kappa^{(Y^{(1)})}_{111}&\=\frac{1}{|G|}\kappa^{(\widetilde{Y}^{(1)})}_{111}\=0~,\\[5pt]
\kappa^{(Y^{(1)})}_{112}&\=\frac{1}{|G|}\left(\kappa^{(\widetilde{Y}^{(1)})}_{112}+\kappa^{(\widetilde{Y}^{(1)})}_{113}\right)\=0~,\\[5pt]
\kappa^{(Y^{(1)})}_{122}&\=\frac{1}{|G|}\left(\kappa^{(\widetilde{Y}^{(1)})}_{122}+2\kappa^{(\widetilde{Y}^{(1)})}_{123}+\kappa^{(\widetilde{Y}^{(1)})}_{133}\right)\=10~,\\[5pt]
\kappa^{(Y^{(1)})}_{222}&\=\frac{1}{|G|}\left(\kappa^{(\widetilde{Y}^{(1)})}_{222}+3\kappa^{(\widetilde{Y}^{(1)})}_{223}+3\kappa^{(\widetilde{Y}^{(1)})}_{233}+\kappa^{(\widetilde{Y}^{(1)})}_{333}\right)\=35~,\\[5pt]
c_{2,1}^{(Y^{(1)})}&\=\frac{1}{|G|}c_{2,1}^{(\widetilde{Y}^{(1)})}\=12~,\\[5pt]
c_{2,2}^{(Y^{(1)})}&\=\frac{1}{|G|}\left(c_{2,2}^{(\widetilde{Y}^{(1)})}+c_{2,3}^{(\widetilde{Y}^{(1)})}\right)\=50.
\end{aligned}
\end{equation}

The same contractions relate the numbers for $Y^{(2)}$ to those of $\widetilde{Y}^{(2)}$, yielding
\begin{equation}
\kappa^{(Y^{(2)})}_{111}\=0~,\quad \kappa^{(Y^{(2)})}_{112}\=0~,\quad \kappa^{(Y^{(2)})}_{122}\=10~,\quad \kappa^{(Y^{(2)})}_{222}\=20~,\quad c_{2,1}^{(Y^{(2)})}\=12~,\quad c_{2,2}^{(Y^{(2)})}\=44~.
\end{equation}

\subsection{Remarks on Wall's theorem and non-simply connected threefolds}
\subsection{The Hori-Knapp model}
Our choice of the specific $Y^{(1)},\,Y^{(2)}$ of Example 1 were not merely taken to illustrate how to use \eqref{eq:exact} and \eqref{eq:pairing} to compute topological data for quotients satisfying some specific properties (namely $H_{2}(G,\IZ)=0$ and $h^{1,1}(Y)<h^{1,1}(\widetilde{Y})$). The topological data that we have computed for these models allows us to address an open problem in the literature.

Consider the Calabi-Yau geometries discussed in \cite{Hori:2016txh}. Their nonabelian GLSM realised, in two different phases (in their labelling, phases $I_{+}$ and $IV$), a pair of Calabi-Yau threefolds with hodge numbers $(h^{1,1},h^{2,1})=(2,24)$. One of these (phase $IV$) had trivial fundamental group, and so this geometry was known to be distinct to the quotient geometries $Y^{(1)},\,Y^{(2)}$ of \cite{Candelas:2015amz}. Nonetheless, it was left as open to determine if their phase $I_{+}$ really did provide a new Calabi-Yau threefold, or if this geometry was in the same diffeomorphism class as one of $Y^{(1)},\,Y^{(2)}$. Now that we have the full set of Wall data for the two quotient models we have taken from \cite{Candelas:2015amz}, we can compare this to the Wall data of the phase $I_{+}$ geometry of \cite{Hori:2016txh} and use the methods of \cite{Gendler:2023ujl} to check if the phase $I_{+}$ geometry lies in a distinct family of Calabi-Yau threefolds. 

Strictly speaking, Wall's theorem \cite{Wall:1966rcd} states that the homotopy type of a compact, simply connected Calabi-Yau threefold with torsion-free homology is determined by the Wall Data (triple intersection numbers, second Chern numbers, and Hodge numbers). At the time of writing, this is not known to apply to non-simply connected threefolds (which we are studying presently). Nonetheless, as argued in \cite{Gendler:2023ujl} it is anticipated on physical grounds that one can drop the assumption of simply-connectedness. We note that there is no $\text{GL}(2,\IZ)$ transformation with determinant $\pm1$ that takes the Wall data for phase $I_{+}$ of \cite{Hori:2016txh} to the Wall data that we have computed here for either of $Y^{(1)},\,Y^{(2)}$. We determine this by comparing the GCDs of the sets of triple intersection numbers and second Chern numbers for each geometry. To use the language of \cite{Gendler:2023ujl}, these geometries are not Wall-equivalent. Therefore, we expect that the geometry associated to phase $I_{+}$ is distinct from either of $Y^{(1)},\,Y^{(2)}$, assuming that homotopic non-simply connected threefolds must have the same Wall data. 

\subsection{Non-homotopic threefolds with the same Wall data}
As we will now demonstrate, non-simply connected threefolds with the same Wall data need not be homotopic. We guess that if Wall's theorem can be generalised beyond the simply connected case, then the Wall data must be extended to include the fundamental group.
 
Consider for the sake of nuisance two further geometries from \cite{Candelas:2015amz} (Table 25 therein),
\begin{equation}
\IP^{7}[2,2,2,2]_{/\IZ_{8}}~,\quad \text{ and } \quad \IP^{7}[2,2,2,2]_{/\IQ_{8}}~.
\end{equation}
These are of the type considered in our discussion of Example 0, with $H_{2}(\IZ_{8},\IZ)\cong H_{2}(\IQ_{8},\IZ)\cong0$. $h^{1,1}=1$ for both quotients and their shared universal cover $\IP^{7}[2,2,2,2]$. Both quotient groups have the same order, $|G|=8$. Nonetheless both quotient geometries have different fundamental groups, respectively $\IZ_{8}$ and $\IQ_{8}$ (the quaternion group). Therefore the manifolds cannot be homotopy equivalent. They have the same Wall data, obtained by dividing the triple intersection, second Chern, and Euler numbers of $\IP^{7}[2,2,2,2]$ by 8: $\kappa_{111}=2,\,c_{2}=8,\,\chi=-16,\,h^{1,1}=1$.

\bibliographystyle{JHEP}
\bibliography{D4D2D0}

\end{document}